\documentclass[letterpaper,11pt]{article}

\pdfoutput=1
\usepackage{color}
\usepackage{epsfig, palatino}
\usepackage{epic}
\usepackage{dsfont}
\usepackage{colonequals}
\usepackage{mathrsfs}
\usepackage{ae} 
\usepackage[T1]{fontenc}
\usepackage[ansinew]{inputenc}
\usepackage{amsmath}
\usepackage{amssymb}
\usepackage{mathtools}
\usepackage{graphicx}
\usepackage[normalem]{ulem}
\usepackage{xcolor}
\definecolor{darkblue}{cmyk}{0.9,0.9,0,0}
\usepackage[colorlinks=true,linkcolor=darkblue,citecolor=darkblue,urlcolor=darkblue]{hyperref}
\usepackage{cite}
\usepackage{hyperref}
\usepackage{varioref}
\usepackage{makeidx}
\usepackage[english]{babel}
\usepackage{simplewick}
\usepackage{array}
\usepackage{multirow}
\usepackage{float}
\usepackage{sidecap}

\usepackage{subcaption}

\usepackage[font={small}]{caption}

\newcommand{\comment}[1]{}

\newcommand{\begBvR}[1]{\begin{#1}} 
\newcommand{\beq}{\begBvR{equation}}
\newcommand{\eeq}{\end{equation}}

\newcommand{\eeqq}{\end{equation*}}

\newcommand\eeqaa{\end{eqnarray*}}

\newcommand\eeqa{\end{array}}

\newcommand{\eea}{\end{eqnarray}}

\newcommand{\im}{{\rm Im}\;}

\renewcommand{\Re}{\operatorname{Re}}
\renewcommand{\Im}{\operatorname{Im}}

\newcommand{\nn}{\nonumber}

\newcommand{\neqa}{\nonumber\end{eqnarray}} 
\newcommand{\la}[1]{\label{#1}}

\renewcommand{\d}{\partial}

\newcommand{\<}{{\langle}}
\renewcommand{\>}{{\rangle}}

\newcommand{\cC}{{\cal C}}

\newcommand{\re}{\relax{\rm I\kern-.18em R}}

\renewcommand{\sp}{p\hspace{-.40em}/}

\definecolor{darkgreen}{rgb}{0.0, 0.45, 0.0}
\definecolor{mathematicablue}{RGB}{94,130,182}

\newcommand{\Blue}[1]{{\color{blue}#1\color{black}}}

\def\XXint#1#2#3{{\setbox0=\hbox{$#1{#2#3}{\int}$}
\vcenter{\hbox{$#2#3$}}\kern-.5\wd0}}

\def\su2{{SU(2)}}

\def\eps{{\epsilon}}

\def\[{\left[}
\def\]{\right]}

\def\l{\lambda}

\def\({\left(}
\def\){\right)}
\def\[{\left[}
\def\]{\right]}

\def\<{\langle}
\def\>{\rangle}

\def\i2{\frac{i}{2}}

\def\spi{\relax{\rm \pi\kern-0.5em /}}
\def\sA{\relax{\rm A\kern-0.5em /}}
\def\sp{\relax{\rm p\kern-0.5em /}}
\def\sd{\relax{\rm \d\kern-0.5em /}}
\def\sk{\relax{\rm k\kern-0.5em /}}
\def\sn{\relax{\rm n\kern-0.5em /}}
\def\sl{\relax{\rm l\kern-0.5em /}}
\def\sP{\relax{\rm P\kern-0.7em /}}
\def\sBethe{\relax{\rm \Bethe\kern-0.5em /}}
\def\cN{{\cal N}}

\def\One{1\hskip-.16cm1}

\def\cC{{\cal C}}
\def\cD{{\cal D}}

\def\cO{{\cal O}}

\def\cN{{\cal N}}

\def\cP{{\cal P}}

\def\cT{{\cal T}}
\def\2F1{\,_2{\rm F}_1}

\def\Re{{\rm Re}}

\makeindex

\def \la {\langle}
\def \ra {\rangle}
\def \pa {\partial}

\def \eps {\epsilon}

\def\be#1\ee{\begin{align}#1\end{align}}

\begin{document}

\thispagestyle{empty}

\renewcommand{\thefootnote}{\fnsymbol{footnote}}
\setcounter{page}{1}
\setcounter{footnote}{0}
\setcounter{figure}{0}

\begin{flushright}
CERN-TH-2020-095
\end{flushright}

\begin{center}
$$$$

{\Large\textbf{\mathversion{bold}
An Analytical Toolkit for the S-matrix Bootstrap 
}\par}
\vspace{1.0cm}

\vspace{1.0cm}

\textrm{Miguel Correia,$^{a,b}$ Amit Sever,$^{a,c}$ Alexander Zhiboedov$^a$}
\\ \vspace{1.2cm}
\footnotesize{\textit{ 
$^a$CERN, Theoretical Physics Department, CH-1211 Geneva 23, Switzerland\\
$^b$Institute of Physics, Ecole Polytechnique Federal de Lausanne (EPFL) CH-1015 Lausanne, Switzerland\\
$^c$School of Physics and Astronomy, Tel Aviv University, Ramat Aviv 69978, Israel
}
\vspace{4mm}
}

\par\vspace{1.5cm}

\textbf{Abstract}\vspace{2mm}
\end{center}

\noindent We revisit analytical methods for constraining the nonperturbative $S$-matrix of unitary, relativistic, gapped theories in $d \geq 3$ spacetime dimensions. We assume extended analyticity of the two-to-two scattering amplitude and use it together with elastic unitarity to develop two natural expansions of the amplitude. One is the threshold (non-relativistic) expansion and the other is the large spin expansion. The two are related by the Froissart-Gribov inversion formula. When combined with crossing and a local bound on the discontinuity of the amplitude, this allows us to constrain scattering at finite energy and spin in terms of the low-energy parameters measured in the experiment. Finally, we discuss the modern numerical approach to the $S$-matrix bootstrap and how it can be improved based on the results of our analysis.

\noindent

\numberwithin{equation}{section}

\setcounter{page}{1}
\renewcommand{\thefootnote}{\arabic{footnote}}
\setcounter{footnote}{0}

\setcounter{tocdepth}{2}

 \def\nref#1{{(\ref{#1})}}

\newpage

\tableofcontents

\parskip 5pt plus 1pt   \jot = 1.5ex

\newpage

\section{Introduction}

The idea of bootstrapping the $S$-matrix of a unitary, relativistic, gapped theory in $d\geq 3$ was actively pursued in the 60's. While many interesting results have been derived \cite{Eden:1966dnq, Martin:1967zzd,Eden:1971fm,Gribov:2003nw}, no nonperturbative physical $S$-matrices have been computed. 
The main reason being that no solid, nonperturbative calculation scheme was ever put forward without relying on some unreliable approximations. In addition to that, analytic properties of the multi-point amplitudes were never fully understood. Moreover, even at the level of the two-to-two scattering amplitude, the region of analyticity that is usually assumed in the bootstrap analysis has not been rigorously established. 

While the problem of analytic properties of multi-point scattering amplitudes is still widely open, the question of finding a good calculation scheme has recently acquired an interesting twist with the development of the conformal bootstrap \cite{Rattazzi:2008pe,Poland:2018epd}. In this context, by exploring bounds on the OPE data in the space of solutions to the CFT bootstrap equations, it was found that sometimes the physical theories of interest saturate the bootstrap bounds \cite{ElShowk:2012ht,Kos:2015mba,Agmon:2019imm} and are in this sense solvable.\footnote{It is still an open question if the islands observed in the exclusion plots shrink to zero size or not.} Remarkably, a similar phenomenon was observed for 2d $S$-matrices \cite{Paulos:2016but,Cordova:2018uop,Cordova:2019lot}, where some previously known two-dimensional scattering amplitudes of physical theories were found to saturate the bootstrap bounds. We can loosely call such theories {\it bootstrap-solvable}. A fundamental, open question in the $S$-matrix theory program therefore is: {\it Are there bootstrap-solvable $S$-matrices in $d \geq 3$?}

A distinguishing feature characteristic to scattering in $d \geq 3$ dimensions is a remarkable connection between scattering and particle production. Here, by scattering we mean a non-trivial $2 \to 2$ amplitude and by particle production, we mean a non-zero $2 \to n$ amplitude with $n>2$. While scattering without particle production is a commonplace in $d=2$ \cite{Dorey:1996gd}, in $d \geq 3$ it is widely believed that scattering implies production \cite{Aks}.\footnote{This result is sometimes called the Aks theorem. However, since it relies on some unproven assumptions that we discuss in detail below, its status is still not completely solid.} The underlying reason is an elegant interplay between analyticity, elastic unitarity and crossing symmetry.

The simplest non-trivial S-matrix element is a $2\to2$ connected scattering amplitude $T(s,t)$ as a function of Mandelstam invariants $s$ and $t$. In a gapped theory, this amplitude is subject to an exact non-linear equation called {\it elastic unitarity}. This equation originates from the fact that when we scatter the lightest particles in the theory at energies below the first multi-particle threshold ($s_0$), only two particles can be produced in the final state.\footnote{As such elastic unitarity is absent in theories with massless particles. Similarly, there is no elastic unitarity in CFTs.} As a result, unitarity of the $S$-matrix becomes a non-linear equation satisfied by $T(s,t)$ for $4m^2 < s < s_0$. It is the purpose of the present paper to revisit the implications of elastic unitarity, when combined with crossing and analyticity, on the structure of the nonperturbative amplitude $T(s,t)$.

One motivation for our analysis is recent numerical investigations of higher-dimensional scattering in \cite{Paulos:2017fhb,Guerrieri:2018uew}. In these works elastic unitarity was not imposed and it was observed that various bootstrap bounds tend to be saturated by purely elastic functions.
It is therefore an interesting, open question how to efficiently implement elastic unitarity and particle production in the current $S$-matrix bootstrap program. An obvious way to tackle the problem is to include higher-point amplitudes in the bootstrap analysis explicitly. However, due to unknown and complicated analytic properties of higher-dimensional amplitudes it is not clear if it is feasible in $d \geq 3$.
Another possible way to make progress, which we will follow in the present paper, is to focus on how to 
implement structures of the amplitude that are dictated by elastic unitarity into the current numerical approach of \cite{Paulos:2017fhb,Guerrieri:2018uew}. In this way we hope to be able to zoom in closer on the physical higher-dimensional $S$-matrices and, if we are lucky, maybe eventually solve them. Here, we discuss various possibilities of doing that 
and will report the numerical results in \cite{numerics}. We start by laying out our assumptions that serve as the basis for the further analytic study.

\subsection{Assumptions}

We assume that in the far past and in the far future states of the system are described by a set of free particles. The Hilbert space therefore is taken to be the Fock space of free particles.\footnote{To the best of our knowledge this assumption, known as {\it asymptotic completeness}, does not follow from the non-zero gap and Wightman axioms, see e.g. \cite{Haag:1992hx}.} For simplicity we assume that the spectrum contains a single scalar particle of mass $m$ together with its multi-particle states. Under these conditions it has been rigorously proven that the amplitude satisfies:
\begin{enumerate}
\item {\bf Crossing symmetry}: for scattering of identical particles we have
    \be
    T(s,t) = T(t,s) = T(u,t)\ ,\qquad ~s+t+u = 4m^2 \ .
    \ee
    \item {\bf Real analyticity}:  for scattering of identical particles we have
    \be
    T(s^*,t^*) = T^*(s,t) \ .
    \ee
  
\end{enumerate}

 Our extra assumptions for the connected two-to-two scattering amplitude $T(s,t)$, which have not been rigorously established,  are the following:

\begin{enumerate}

\item[$3.$]{\bf Extended analyticity}: $T(s,t)$ is an analytic function for complex $s$ and $t$ in some region $\cD$, except for potential poles in $0<s<4m^2$ and a cut starting at $s=4m^2$, as well as images of these singularities under crossing. We will often assume that the region of analyticity $\cD$ extends to the full complex $s$-plane ($t$-plane)  for some finite region in $t$-plane ($s$-plane), but many of our arguments can be adopted to the situation when $\cD$ is bounded in both variables simultaneously.\footnote{If $\cD = \mathbb{C}^2$ one says that the function is maximally analytic. Maximal analyticity is not necessary for the present paper.} As usual in this paper $T(s,t)$ stands for the analytic continuation of the amplitude from the physical regime to the principle sheet -- without going through the multi-particle cuts.

\item[$4.$]{\bf Polynomial Boundedness}: for fixed $t$
\be
    | T(s,t)| < |s|^{J_0(t)}\ ,\qquad  |s|\to \infty\ ,\qquad  (s,t) \in \cD\ . 
    \ee
The formula above assumes that $\cD$ includes $s=\infty$ for fixed $t$. More generally, we will assume that the amplitude is polynomially bounded on the principal sheet (away from the bound states poles and multi-particle thresholds).
\end{enumerate}

We also have two extra technical assumptions:
\begin{enumerate}
  \item[$5.$] {\bf Continuity}: partial waves $f_J(s)$ are real analytic functions in the elastic region $4m^2<s<s_0$.\footnote{This is closely related to what is called absence of pathologies a-la A.~Martin \cite{Martin:1966zz}. We discuss it in more detail in section \ref{sec:continuity}.}
  \item[$6.$] {\bf ${\mathbb Z}_2$ symmetry and no bound states}: for simplicity we assume no bound states in the spectrum. We also assume that we have a single stable particle of mass $m$ which is odd under ${\mathbb Z}_2$ symmetry. Therefore only an even number of particles can be produced in the scattering of two particles.
\end{enumerate}

\subsection{Plan of the Paper}

In section \ref{sec:amplbasics} we review basics of the two-to-two scattering amplitude. We consider scattering of identical scalar particles and briefly review known analyticity results. We discuss unitarity and elastic unitarity in terms of $T(s,t)$. We introduce partial wave expansion and the Froissart-Gribov formula. We briefly discuss continuity properties of $T(s,t)$.

In section \ref{aceu} we consider analytic continuation of elastic unitarity. We first review the derivation of the Mandelstam equation for the double spectral density $\rho(s,t)$, or equivalently double discontinuity of $T(s,t)$, based on analytic continuation of elastic unitarity in one of the Mandelstam variables ($s$ or $t$). We then discuss its relation to analytic continuation of partial waves in spin $J$ via the Froissart-Gribov formula. We also consider analytic continuation of elastic unitarity in energy $s$. Finally, we exhibit the structure of the Landau-Karplus curves along which the double spectral density develops a nontrivial support in the elastic region.

We then analyze various implications of elastic unitarity:

\begin{itemize}
    \item In section \ref{productionsec} we discuss positivity properties of double spectral density $\rho(s,t)$. We review the argument that scattering implies production in $d \geq 3$. This result follows from the combination of crossing symmetry and positivity of $\rho(s,t)$.

\item In section \ref{THboot} we introduce the notion of the threshold expansion for partial waves, as well as for the first and second discontinuities of the scattering amplitude. The threshold expansion of partial waves and the discontinuity of the scattering amplitude is a consequence of elastic unitarity and it describes low-energy or non-relativistic scattering. The Mandelstam equation then maps it to the expansion of double spectral density close to the boundary of its nontrivial support, the so-called Kaprlus-Landau curve.

\item In section \ref{LJboot} we map the threshold expansion in the $t$-channel to the large $J$ expansion of partial wave coefficients in the $s$-channel. It comes from ``inversion'' of the threshold expansion via the Froissart-Gribov formula. 
Turning to the ${1 \over J}$ corrections we find that remarkably the computations can be sometimes done exactly in ${1 \over J}$.

\item In section \ref{sec:boundinelasticity} we turn the large $J$ results into finite $J$ predictions plus an error estimate. This error estimate is deducted 
from a local bound on the discontinuity of the amplitude 
in the region $s,t>4m^2$. 
No such rigorous bound is known. We discuss natural error estimates and perform the finite spin, finite energy computations in a simple toy model.
\end{itemize}

In section \ref{sec:numboot} we consider the modern numerical approach to the $S$-matrix bootstrap following \cite{Paulos:2017fhb}. We discuss why and what should be improved in the existing approach. We suggest several ways in which 
this approach can be improved. In particular, we discuss various ways to implement elastic unitarity numerically. In section \ref{eq:commentCFT} we briefly comment on the relation between the present analysis and similar ideas in the conformal bootstrap. Finally, in section \ref{sec:conclusions} we conclude and present some future directions. Several appendices contain technical details that should be helpful in understanding the details of our arguments and calculations.

\section{Amplitude Basics}
\label{sec:amplbasics}

In this section we briefly review the basic kinematics of the two-to-two scattering of identical, scalar particles to set the conventions for the further analysis.
As usual we write the $S$-matrix as
\beq
\label{eq:hatOne}
\hat S\equiv\hat\One+i\,\hat T\ ,
\eeq
where $\hat T$ is zero in the theory of a free massive scalar. We are interested in the matrix elements that describe two-to-two scattering
\be
S_{2,2}(p_3, p_4 | p_1, p_2) \equiv \la p_3,p_4 |  \hat S |p_2,p_1 \ra\ ,
\ee
where the initial and final states are characterized by the on-shell momenta
\beq
p^2 = {\vec p\,}^2- (p^0)^2  = - m^2\ ,\qquad  {\vec p\,}^2 =\sum_{i=1}^{d-1}(p^i)^2\ ,\qquad p^0 > 0\ .
\eeq

As in  (\ref{eq:hatOne}) we can separate the contribution of the disconnected and connected parts of the $S$-matrix
\be\label{2to2}
S_{2,2}(p_3, p_4 | p_1, p_2) &= S_{1,1}(p_3|p_1) S_{1,1}(p_4|p_2) + S_{1,1}(p_4|p_1) S_{1,1}(p_4|p_2)  + S_{2,2}^{c} (p_3, p_4 | p_1, p_2).
\ee
The disconnected part is given by an overlap of the one particle states which is uniquely fixed by Lorentz symmetry
\be
S_{1,1}(p|q) &= \One_{1,1}(p|q) = 2(2 \pi)^{d-1}\sqrt{\vec p^2 + m^2}\times \delta^{d-1} (\vec p - \vec q) \ , 
\ee
The connected part is the main object of our interest
\be
S_{2,2}^c (p_3, p_4 | p_1, p_2) &\equiv i \la p_3,p_4 |  \hat T |p_2,p_1 \ra =   i (2 \pi)^{d} \delta^{d} \left(p_1 + p_2 - p_3 - p_4 \right) T(s,t) \ .
\ee
In the formula above we introduced Mandelstam invariants
\be\label{stu}
s &=- (p_1 + p_2)^2 =  4 ( m^2 + \vec p^2)\ , \nn \\
t  &=- (p_1 - p_3)^2 =- 2 \vec p^2 (1-\cos \theta)\ , \nn \\
u &=- (p_1 - p_4)^2 =  -2 \vec p^2 (1+\cos \theta) \ ,
\ee
where in the last equality we wrote their form in the center-of-mass frame, $p_1=(m,\vec p)$, $p_2=(m,-\vec p)$. Here,
\be
\cos \theta = 1 + {2 t \over s - 4 m^2} \ ,
\ee
is cosine of the scattering angle. Only two of the Mandelstam invariant are independent while the third is related to the other two through the relation
\be
s+t+u = 4 m^2 \ .
\ee
From above, the dimension of the amplitude is
\be
[T(s,t)] = m^{4-d} \ .
\ee

\subsection{Analyticity}\label{sec-analytics}

In the discussion above scattering matrix elements were defined for physical momenta that correspond to actual scattering. The starting point of the $S$-matrix considerations is the statement that the physical matrix element $T(s,t)$ is a boundary value of an analytic function of the relativistic invariants regarded as complex variables
\be\label{analyticboundary}
T(s,t) = \lim_{\eps \to 0} T(s + i \eps, t) \ ,
\ee
where we assumed $s$ and $t$ to be real and $-(s-4m^2)< t< 0$, $4m^2<s$. This corresponds to scattering in the $s$-channel $1,2 \to 3,4$. 
Using the basic principles of QFT correlation functions, reduction formulas that relate them to the $S$-matrix elements, and techniques of analytic completion one can establish various analytic properties of the scattering amplitudes as functions of complex $s$ and $t$, see e.g. \cite{Itzykson:1980rh, Sommer:1970mr, Martin:1967zzd} for a pedagogical exposition.

The first result of this type is subtracted dispersion relations in $s$ for fixed $-t_0 < t \leq 0$ in the physical $s$-channel region \cite{Sommer:1970mr,Bogoliubov}. In this case one has to understand analytic properties of $T(s,t)$ as a function of complex $s$. For the case of $\pi^0 \pi^0 \to \pi^0 \pi^0$ scattering, for which our treatment applies directly, one can show that $T(s,t)$ is an analytic function of $s$ with two cuts: $s>4 m^2$ and $u>4 m^2$, with $t_0 = 28 m^2$. See Table 1 of \cite{Sommer:1970mr} for the processes for which a fixed-$t$ dispersion relation has been proven and the corresponding values of $t_0$.\footnote{The cases for which a dispersion relation has not been proven, baryon-baryon scattering for instance \cite{Sommer:1970mr}, still enjoy a domain of analyticity that connects the $s$- and $u$-channel cuts. Thus, the property of crossing (see below) can still be established for these cases \cite{Bros:1965kbd}.}

Another well-known property, originally due to Lehmann \cite{Lehmann:1958ita}, 
concerns analytic properties of $T(s,t)$ as a function of $t$ for fixed physical $s>4 m^2$. Lehmann showed that $T(s, \cos \theta)$ is analytic inside an ellipse, the so-called Lehmann ellipse, in the $\cos \theta$ complex plane with foci at $\cos \theta = \pm 1$ and semi-major axis $\cos \theta_{sL} > 1$ which depends on the details of the theory, energy and masses of particles, as well as the scattering process. Lehmann also showed that the absorptive part or discontinuity of the amplitude ${\rm Disc}_s T(s, \cos \theta)$ is analytic in a larger ellipse, the so-called large Lehmann ellipse, with a semi-major axis $\cos \theta_{LL} = 2 \cos^2 \theta_{sL} -1$.

The third class of results concerns analyticity of $T(s,t)$ when both $s$ and $t$ are complex. Bros, Epstein and Glaser \cite{Bros:1964iho} showed that any point $(s, \cos \theta)$ in the physical region is surrounded by an analyticity neighborhood whose precise form is not known in general, see e.g. \cite{Martin:1967zzd} for details. An explicit domain of simultaneous analyticity in both variables was derived by Lehmann \cite{Lehmann66} for elastic processes that obey a fixed-$t$ dispersion relation by continuing the Lehmann ellipse to complex $s$. For cases with single variable dispersion relations in all three channels (as in $\pi \pi \to \pi \pi$ scattering), Mandelstam \cite{Mandelstam64} derived domains of the form $|s\,t| < b$ for any complex $s$ and $t$ outside the single variable dispersion relation cuts. For pion scattering the largest domain occurs for $b = 256 m_\pi^4$. These domains have the drawback that when $s \to \infty$ the Lehmann domain shrinks to the line $-t_0<t<0$ and the Mandelstam domain shrinks to the point $t=0$.

Based on the results above, the analyticity domain was further enlarged using unitarity by A. Martin \cite{Martin:1965jj}. The final result is that the amplitude is analytic within $|t|<R$ and $s$ cut-plane. For scattering of identical particles that we consider in the present paper $R=4m^2$. From this result the standard bounds on high energy behavior of the amplitudes follow. It also follows that for $|t|<4 m^2$ the scattering amplitude admits fixed $t$ dispersion relations with at most two subtractions.

Finally, let us introduce the notion of maximal analyticity (or Mandelstam analyticity) which states that the scattering amplitude is analytic in the $(s,t)$ complex planes with only singularities on the principal sheet being unitarity cuts $s,t,u>4m^2$ and bound state poles for $0<s,t,u<4m^2$. This property is consistent, at least for the scattering of lightest particles in the theory, with expectations from unitarity and analysis of perturbation theory.
However it is important to keep in mind that it has not been proven. The original attempts to prove maximal analyticity in perturbation theory \cite{EdenM,TaylorM} were later found to have a loophole \cite{AnomM} which, to the best of our knowledge, has never been closed even for the scattering of lightest particles. In this paper we freely assume analyticity beyond what has been rigorously proven but we do not assume maximal analyticity.

There are two other properties that we will use. One is crossing
\be
T(s,t) = T(t,s) = T(u,t)\ ,
\ee
which states that in particular that scattering in different channels is described by different boundary values of a single analytic function. For the two-to-two scattering it has been proven in \cite{Bros:1965kbd}. Beyond the two-to-two scattering only a partial progress has been achieved \cite{Bros:1985gy}. Another property is real analyticity
\be\label{eq:realanalyticity}
T(s^*,t^*) = T^*(s,t) \ .
\ee
This was established within axiomatic quantum field theory in \cite{Olive62}. For the $S$-matrix argument see e.g. \cite{Eden:1966dnq}.

Finally, let us also mention in a related context the result by A.~Martin \cite{MartinMot} who showed that maximal analyticity in the form of the Mandelstam representation together with knowledge of the double spectral density in the elastic region fix the scattering amplitude completely. Similarly, knowledge of the amplitude at fixed energy in the elastic region as a function of the scattering angle is believed to fix it almost completely, see e.g. \cite{Martin:2020jlu}.

\subsection{Unitarity and Elastic Unitarity}

In terms of the $T$-matrix, the unitarity of the $S$-matrix $\hat S\cdot\hat S^\dagger=\hat\One$ reads
\be\label{unitarity}
{1 \over i } \la p_3,p_4 | \hat T - \hat T^{\dagger} |p_2,p_1 \ra &= \la p_3, p_4 | \hat T\cdot \hat T^{\dagger} |p_2,p_1 \ra\\
&=\sum_{n=1}^{\infty} \int d\mu(q_1,\dots,q_{2n})\la p_3, p_4 | \hat T|\{q_i\}_{i=1}^{2n}\> \,\<\{q_i\}_{i=1}^{2n}|\hat T^{\dagger} |p_2,p_1 \ra \nn\ ,
\ee
where in the second line we have inserted a complete basis of asymptotic states and the Lorentz invariant measure is 
\beq
d\mu(p_1,\dots, p_n)\equiv {1\over n!}\prod_{i=1}^n d\mu(p_i)\ ,\qquad d \mu(p)\equiv {1\over (2 \pi)^{d-1}}\theta(p^0)\,  \delta(p^2+m^2)\, d^d p\ .
\eeq 

Due to the momentum conservation, a $2n$-particle intermediate state can only contribute if $s>(2n m)^2$. Otherwise, there is no enough energy to create an on-shell state with $2n$ particles. In particular, for $4m^2<s<16m^2$, only two particle states are possible. Hence, in that regime we can replace the sum in (\ref{unitarity}) by the first term and the equation closes on the $2\to2$ transitions only. This is the elastic unitarity regime. Explicitly, we have
\be
\label{eq:elun}
2 T_s (s,t) = {1 \over 2} \int {d^{d-1} \vec{q'} \  \over (2 \pi)^{d-1} (2 E_{\vec{q'}})}  \int {d^{d-1} \vec{q''}\over (2 \pi)^{d-1} (2 E_{\vec{q''}})} (2 \pi)^d \delta^{d}(p_1 + p_2 - q' - q'') T^{(+)}(s,t') T^{(-)}(s,t'')\ ,
\ee
where we have introduced the notations
\be
\label{eq:PMdef}
T^{(\pm)} \equiv \lim_{\eps \to 0} T(s \pm i \eps , t) \ ,\qquad T_s (s,t)={\rm Disc}_{s}T(s,t) &\equiv {1 \over 2 i} \left(T^{(+)}(s,t) - T^{(-)}(s,t) \right)\ ,
\ee
and $t' = -(\vec p_1 - \vec{q'})^2$, $t''=- (\vec{q''} - \vec p_4)^2$. In writing the above we used real analyticity of the amplitude (\ref{eq:realanalyticity}). Finally, the overall factor of one half is a symmetry factor for identical bosons.

\begin{figure}[t]
\centering
\includegraphics[width=0.9\textwidth]{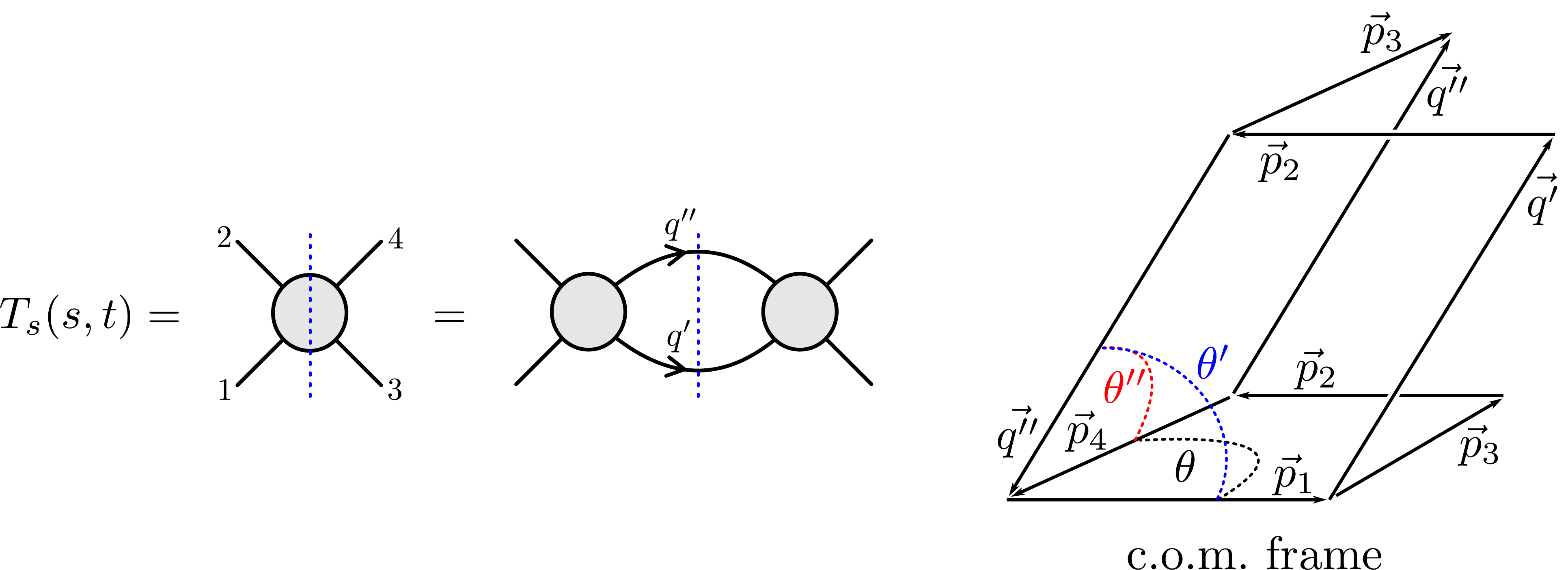}
\caption{\small In the elastic strip $4m^2<s<16m^2$ the discontinuity of the amplitude comes from two intermediate particle exchange only. This result in the exact elastic unitarity equation (\ref{eq:elauni1}). The corresponding phase space integration kernel ${\cal P}_d(\cos\theta,\cos\theta',\cos\theta'')$ in (\ref{eq:dlips}) is proportional to a step/delta function which has a simple geometrical origin. In the center of mass frame we have three $(d-1)$-dimensional vectors, $\vec p_1=-\vec p_2$, $\vec p_3=-\vec p_4$, and $\vec{q'}=-\vec{q''}$. The geometrical angles between these three vectors, $\{\theta,\theta',\theta''\}$ are therefore restricted to the range $\theta_1+\theta_2\ge\theta_3$, where $\theta_{1,2,3}$ are any permutation of $\{\theta,\theta',\theta''\}$.}
\label{stepfunction}
\end{figure}

We will now reduce the integral to an integration over the two scattering angles by performing all the kinematical integrations explicitly. For that aim, we first go to the center of mass frame where $\vec{q'} = - \vec{q''}\equiv p \,\vec n$, where $\vec n$ is a unit $d-1$ vector and $p=\sqrt{s-4m^2}/2$. In these variables the elastic unitarity constraint (\ref{eq:elun}) becomes 
\be
\label{eq:unitampl}
2 T_s (s,t) ={p^{d-2} \over (2 \pi)^{d-2} (2 E_{p})^2 } {E_p \over 4 p} \int d^{d-2} \Omega_{\vec n}\,T^{(+)}(s,t')\, T^{(-)}(s,t'')\ ,
\ee
where ${E_p \over 2p}={\sqrt s\over2\sqrt{ s-4m^2}}$ is the Jacobian coming from the energy conservation delta-function. The integrand only depends on the two scatttering angles
\be\label{cosines}
z' = \cos \theta' = {\vec p_1 \cdot \vec n \over |\vec p_1| }\qquad\text{and}\qquad  z'' = \cos \theta'' =  {\vec p_3\cdot \vec n \over |\vec p_3|}\ ,
\ee
in terms of which we can write the measure as 
\be\label{angularkin}
\int d^{d-2} \Omega_{\vec n} \equiv
\int\limits_{-1}^1 dz'\int\limits_{-1}^1 dz''\,{\cal P}_d(z,z',z'')\qquad\text{where}\qquad z = \cos \theta = {\vec p_1 \cdot \vec p_3 \over |\vec p_1| | \vec p_3|}=1+\frac{2t}{s- 4 m^2}\ ,
\ee
is the cosine of the external scattering angle. We find that (see appendix \ref{AppendixA} for more details) 
\be
\label{eq:dlips}
\mathcal{P}_3(z,z',z'') &=2
\sqrt{1-z^2} \, \delta(1 - z^2 - z'^2 - z''^2 + 2 z z' z'')\ ,  \\
\mathcal{P}_{d>3}(z,z',z'') &={2\pi^{d-3\over2}
\over\Gamma\({d-3\over2}\)}(1 - z^2)^{4 - d \over 2}
{\Theta(1 - z^2 - z'^2 - z''^2 + 2 z z' z'')\over(1 - z^2 - z'^2 - z''^2 + 2 z z' z'')^{5-d \over 2}} \ .\nn
\ee
Using that $E_k = \sqrt{s}/2$
, we can write \eqref{eq:unitampl} covariantly as
\be
\label{eq:elauni1}
\boxed{T_s (s,t) = {(s - 4 m^2)^{d-3\over2} \over8(4\pi)^{d-2} \sqrt{s} }
\int\limits_{-1}^1 dz'\int\limits_{-1}^1 dz''\,{\cal P}_d(z,z',z'')\, T^{(+)}(s, t(z'))\, T^{(-)}(s, t(z''))\ , \quad 4 m^2 \leq s \leq 16 m^2}\ ,  
\ee
where $t(x)\equiv-(s-4m^2)(1-x)/2$, not to be confused with the external momentum transfer $t$, which is held fixed. The step/delta function in the phase space integration kernel ${\cal P}_d(z,z',z'')$ has a simple geometrical origin, see figure \ref{stepfunction}. 

For $s>16m^2$ and general $t$ the unitarity constraint involves scattering elements with more than two particles. To get a constraint on the two particle amplitude, we note that $\hat T\cdot\hat T^\dagger$ on the right hand side of (\ref{unitarity}) is a positive semi-definite matrix. Hence, for any state $\Psi$ we have that 
\beq
\la \Psi | \hat T|\{q_i\}_{i=1}^{2n}\> \,\<\{q_i\}_{i=1}^{2n}|\hat T^{\dagger} |\Psi \ra=|\la\Psi | \hat T|\{q_i\}_{i=1}^{2n}\>|^2\ge0\ ,
\eeq
and hence, if we drop all the contributions with more than two particles in (\ref{unitarity}) we get an inequality for the $2\to2$ scattering matrix
\be\label{unitarity2}
{1 \over i }&\int d\mu(p_1,p_1)\,d\mu(p_3,p_4)\,\psi(p_1,p_2)\,\psi^*(p_3,p_4)\times \la p_3,p_4 | \hat T - \hat T^{\dagger} |p_2,p_1 \ra\\
\ge&\int d\mu(p_1,p_1)\,d\mu(p_3,p_4)\,\psi(p_1,p_2)\,\psi^*(p_3,p_4)
\times\int d\mu(q_1,q_2)\,\la p_3, p_4 | \hat T|q_1,q_2\> \,\<q_1,q_2|\hat T^{\dagger} |p_2,p_1 \ra\ge0 \nn\ .
\ee
For example, if we pick a wave function that consists of two particles with a specific momenta then we have the the amplitude in the forward limit where $p_3=p_1$ and $p_4=p_2$. For this choice of wave function, the unitarity constraint (\ref{unitarity2}) becomes\footnote{The unitarity relation in the forward limit is nothing but the optical theorem. In $d$ dimensions it takes the form
\be
T_s(s,0) = {\rm Im}[T(s,0)] = \sqrt{s (s-4 m^2)}\, \sigma_{tot}(s)\ ,
\ee
where $\sigma_{tot}(s)$ is the total cross-section, of dimension $[\sigma(s)] = L^{d-2}$.}
\be
\label{eq:nonelauni1}
T_s (s,0) &\ge {(s - 4 m^2)^{d-3\over2} \over 8(4\pi)^{d-2} \sqrt{s} } \int\limits_{-1}^1 dz'\int\limits_{-1}^1 dz''\,{\cal P}_d(1,z',z'')\, T^{(+)}(s, t(z')) \, T^{(-)}(s, t(z''))\ \nn \\
&\propto \int\limits_{-1}^1 dz' (1 - z'^2)^{d - 4 \over 2} | T^{(+)}(s, t(z')) |^2 \, ,\quad s \geq 16 m^2\ ,
\ee
where we used that $\mathcal{P}_d(1,z',z'') \propto \delta(z' - z'')$, see \eqref{eq:PPlocal} for the precise formula. We also used that $T^{(-)}(s,t(z'))= \left(T^{(+)}(s,t(z')) \right)^*$ for $-1 \leq z' \leq 1$.

\subsection{Partial Wave Expansion}

Unitarity of the $S$-matrix implies the non-linear integral relations that the $2\to2$ $T$-matrix has to satisfy, (\ref{eq:elauni1}) and (\ref{unitarity2}). To simplify these complicated constraints we choose a wave function $\Psi$ that diagonalizes the $T$-matrix and therefore also the integral kernel in (\ref{eq:elauni1}), (\ref{unitarity2}). This can be done using the Lorentz symmetry of the problem. Namely, we decompose the amplitude $T(s,t)$ in a complete basis of intermediate states which transform in irreducible representations of the $SO(1,d-1)$ symmetry. These representations are characterised by their energy and the little group $SO(d-1)$ angular momentum in the center of mass frame, $E$ and $J$. For two particle states the $SO(1,d-1)$ quantum numbers are enough to characterize the states and we have
\beq
\<p_1,p_2|p,J,\vec m\>\propto \delta^d(p-p_1-p_2)Y^{(d)}_{J,\vec m}(\hat p_1)\ ,
\eeq
where $p^2=E^2$, $Y^{(d)}_{J,\vec m}$ are the $d$-dimensional spherical harmonics, and the energies dependant pre-factor will not be relevant for us.\footnote{For $d=4$ the factor is ${\sqrt{E_{\vec p}\over|\vec p_1|E_{\vec p_1}E_{\vec p_2}}}$, see \cite{Weinberg:1995mt} for details.} We can now insert a complete basis to these states to decompose the S-matrix element $\<p_3,p_4|\hat T|p_1,p_2\>$ in all possible spins. Since the operator $\hat T$ is both, translation and $SO(1,d-1)$ invariant, due to the Wigner-Eckart theorem we have that 
\beq
f_J(p^2)\propto{\<p,J,\vec m|\hat T|p,J,\vec m\>\over \<p,J,\vec m|p,J,\vec m\>}\ ,
\eeq
where the convention-dependent proportionality factor is independent of the energy and and the angular momentum $\vec m$. These functions are the so-called partial wave coefficients, in terms of which the amplitude takes the form
\beq
\label{eq:nJd}
T(s,t)=\sum_{J=0}^{\infty} n_J^{(d)}f_J(s)P_J^{(d)}\(\cos\theta\)\ ,
\eeq
where the sum runs over all (even) spins and $n_J^{(d)}$ are convention-dependent normalization factors. Here, $P_J^{(d)}(\cos\theta)$ are the partial waves. They represents the angular dependence of the amplitude due to the exchange of all the states with spin $J$. 
A simple way of determining these functions is to go to the center of mass frame and act with the $SO(d-1)$ quadratic Casimir on the two outgoing particle, while holding the momentum of the two incoming particles fixed. This equation takes the form
\beq\label{casimir}
\[(1-z^2)^{\frac{4-d}{2}} \frac{d}{dz} (1-z^2)^{\frac{d-2}{2}} \frac{d}{dz} +J(J+d-3)\]P_J^{(d)}(z)=0\ ,
\eeq
where $z=\cos\theta$ 
is cosine of the scattering angle (\ref{angularkin}). This second order differential equation has two independent solutions. Spin $J$ unitary representations are composed of states with angular momentum in the plane of scattering ranging between $-J$ and $J$. Hence, the corresponding solution of (\ref{casimir}) is a degree $J$ polynomial of $\cos\theta$ that is given by
\beq
P^{(d)}_J(z) = {}_2F_1\(-J,J+d-3,\frac{d-2}{2},\frac{1-z}{2}\)\ .
\eeq

The partial wave coefficients can be extracted from the amplitude using the orthogonality relation of these polynomials
\be
{1\over2}\int\limits_{-1}^1 d z\, (1-z^2)^{{d-4 \over 2}} P^{(d)}_J(z) P^{(d)}_{\tilde J}(z) ={\delta_{J \tilde J}\over {\cal N}_d\, n_J^{(d)}}\ .
\ee
Here we have chosen the convention 
\beq
\label{eq:nJd2}
{\cal N}_d={(16\pi)^{2-d\over2}\over\Gamma\({d-2\over2}\)}\ ,\qquad n_J^{(d)}=\frac{(4\pi)^{d\over2}(d+2J-3) \Gamma (d+J-3)}{\pi\,\Gamma \({d-2\over2}\) \Gamma (J+1)}\ ,
\eeq
for which the unitarity constraint presented below takes a simple form. In this convention we have
\beq
\label{eq:pwprojection}
f_J(s)={{\cal N}_d\over2} \int\limits_{-1}^1dz\,(1-z^2)^{d-4\over2}P_J^{(d)}(z)\,T\(s,t(z)\)\ .
\eeq

Because the S-matrix is diagonal in the spin basis, so does the unitary constraint. We consider first the elastic regime $4m^2<s<16m^2$ where this constraints takes the form (\ref{eq:elauni1}). Using (\ref{eq:pwprojection}), we project both sides to a fixed spin $J$. On the left hand side we find the discontinuity of the partial wave coefficient. Real analyticity (\ref{eq:realanalyticity}) of $T(s,t)$ leads to real analyticity of $f_J$
\be\label{eq:realanalytPW}
f_J(s^*) = f_J^*(s)\ .
\ee
Hence, the discontinuity of the partial wave is equal to the imgionary part ${1\over2i}\(f_{J}(s+i\epsilon)-f_{J}(s-i\epsilon)\)={\rm Im}f_{J}(s)$. 
On the right hand side, it is useful to first represent the kernel as a sum over partial waves of $z_1$, $z_2$ and $z$. Because this kernel represents the angular integration in (\ref{eq:unitampl}), its partial wave decomposition must also be diagonal in spin. It takes the form (see appendix C) 
\be
\label{eq:PPP}
{\cal P}_d(z,z',z'') =(4\pi)^{d-2}{\cal N}_d^2\, (1 - z'^2)^{d - 4 \over 2} (1 - z''^2)^{d - 4 \over 2}\sum_{J=0}^\infty n_J^{(d)} P_J^{(d)}(z) P_J^{(d)}(z') P_J^{(d)}(z'')\ .
\ee
Using (\ref{eq:pwprojection}) for the three integrals and real analyticity (\ref{eq:realanalytPW}), we arrive at the elastic unitarity constraint
\be\label{eq:EUfJ}
\boxed{2{\rm Im }f_J(s) = {(s-4 m^2)^{d-3\over2}\over\sqrt s}|f_J(s)|^2}\ ,
\ee
or equivalently
\beq
\label{eq:partialwave}
|S_J(s)| = 1\ ,\qquad\text{with}\qquad S_J(s)\equiv1+i{(s-4m^2)^{d-3\over2}\over  \sqrt s}f_J(s)\ .
\eeq
Here $1$ can be traced to back to $\hat \One$ in (\ref{eq:hatOne}). In this way the trivial unitary $S$-matrix $\hat S = \hat\One$ becomes $S_J = 1$ in the partial wave basis.

The solution to this is 
\be
\label{eq:elunPW}
f_J(s) = {\sqrt{s} \over (s-4 m^2)^{d-3\over2}} i (1 - e^{2 i \delta_J(s)} )\ ,
\ee
with $\delta_J(s)$ being real for $4 m^2 <s<16 m^2$ and is called the scattering phase.

Similarly to the above, for $s>16 m^2$ we chose $\psi(p_1,p_2)=\<p_1,p_2|p,J,\vec m\>$ in (\ref{unitarity2}). In that way we arrive at the same equation, but with an inequality instead of an equality
\be
\label{UfJ}
2{\rm Im }f_J(s)\ge{(s-4 m^2)^{d-3\over2}\over\sqrt s}|f_J(s)|^2\ .
\ee
or equivalently, $|S_J(s)|\le1$, ${\rm Im}[\delta_J(s)] \geq 0$.

We close this section with a short discussion on the range of convergence of the partial wave sum (\ref{eq:nJd}) for fixed physical $s$ as a function of $\cos \theta$.\footnote{\label{foot:Neumann} The convergence of the partial wave expansion can be seen using Neumann's argument \cite{Whittaker}. Consider a function $f(z)$ analytic inside some region $\cC$ which includes the $[-1, 1]$ interval. We can then write
\be
\label{eq:pwT}
f(z) = \oint\limits_{\gamma} {dt \over 2 \pi i} {f(t) \over t-z} = \oint\limits_{\gamma} {dt \over 2 \pi i} \sum_{J=0}^{\infty} n_J^{(d)} P^{(d)}_J(z) \left(\mathcal{N}_d (t^2-1)^{d-4\over2} Q^{(d)}_J(t) f(t) \right) = \sum_{J=0}^{\infty} n_J^{(d)} f_J P^{(d)}_J(z) ,
\ee
where $\gamma \in \cC$ is some contour that wraps the interval $ [-1,1]$ counterclockwise and contains $z$ inside the integration contour. To exchange the summation and integration, we also used that given $z$, ${1 \over t-z} =\mathcal{N}_d (t^2-1)^{d-4\over2} \sum_{J=0}^\infty n_J^{(d)} P^{(d)}_J(z) Q^{(d)}_J(t)$ converges uniformly in $t$ as long as $t$ is outside the ellipse with foci at $-1$ and $1$ that passes through $z$. We also used the relation between $Q_J^{(d)}(z)$ and $P_J^{(d)}(z)$ which will be explained below, see (\ref{eq:Qdisc}). Therefore, the partial wave expansion (\ref{eq:pwT}) converges when $z$ is inside the ellipse with foci at $-1$ and $1$ and $\in \cC$. Note therefore that the size of the domain of convergence of the partial wave expansion can be less that the analyticity domain $\cC$.} It is a well-known fact that the amplitude $T(s,\cos\theta)$ is analytic inside the small Lehmann-Martin ellipse and its absorptive part, $T_s(s, \cos\theta)$ is analytic inside the large Lehmann-Martin ellipse. These ellipses have foci at $\cos \theta = \pm 1$ and semi-major axis $z_{{\rm small}}$ and $z_{{\rm large}}$. Correspondingly, inside these ellipses the sum (\ref{eq:nJd}) and its discontinuity converge. 

In the case of scattering of identical lightest particles which is our main interest we have
\be
z_{\rm small} = 1 + {8 m^2 \over s-4m^2}\ ,\qquad z_{\rm large} = 2 z_{\rm small}^2 - 1 
= 1+{32 m^4 \over (s- 4m^2)^2} \ .
\ee
In section \ref{sec:PositivityDSD} we will see that extended analyticity, elastic unitarity and crossing imply that the partial wave expansion converges in a larger region.

\subsection{Froissart-Gribov Formula}

The Froissart-Gribov formula is a representation of the partial wave coefficients in terms of the discontinuity of the amplitude. It has multiple applications and, in particular, it allows us to analytically continue partial wave coefficients in spin. Correspondingly, in section \ref{sec:acJ} we will use the Froissart-Gribov formula to analytically continue elastic unitarity in spin. It will also allow us to better understand the analytic structure of the amplitude in the Mandelstam invariants and relate the threshold expansion, see section \ref{THboot}, to the large spin expansion, see section \ref{LJboot}.

Let us introduce the Gegenbauer $Q$-functions. These are given by the second linearly independent solution of the second order Casimir equation (\ref{casimir}). They are uniquely fixed by their asymptotic behavior
\beq\label{Qasimp}
\lim_{|z|\to\infty}Q^{(d)}_J(z)={c^{(d)}_J\over z^{J+d-3}} + \dots  \ ,
\eeq
where $c^{(d)}_J$ is a normalization constant.    
The corresponding $Q$-function is
\beq
Q^{(d)}_J(z)={c_J^{(d)}\over z^{J+d-3}} {}_2F_1\(\frac{J+d-3}{2},\frac{J+d-2}{2},J+\frac{d-1}{2}, {1 \over z^2}\)\ .
\eeq
Our convention is 
\beq
c_J^{(d)}=\frac{\sqrt{\pi}\Gamma(J+1)\Gamma(\frac{d-2}{2})}{2^{J+1}\Gamma(J+\frac{d-1}{2})}\ .
\eeq

The $Q$-function has a cut running between $z=-1$ and $z=1$. The fact that there are only two independent solutions to the Casimir equation means that the discontinuity of $Q$ can be expressed in terms of $Q$ and $P$. The precise relation takes the form
\beq\label{eq:Qdisc}
{\rm Disc}_{z}(z^2- 1)^{{d-4 \over 2}} Q^{(d)}_J(z)=-{\pi\over2}(1-z^2)^{{d-4 \over 2}}P^{(d)}_J(z)\ ,\qquad z\in[-1,1]\ ,
\eeq
or equivalently (for integer $J$)
\be
\label{eq:Qfunc}
Q^{(d)}_J(z)={1\over2}\int\limits_{-1}^1dz'\({1-z'^2\over z^2-1}\)^{d-4\over2}{P^{(d)}_J(z')\over z-z'}\ .
\ee

We can then plug \eqref{eq:Qdisc} into the partial wave coefficient \eqref{eq:pwprojection} as
\be
\label{eq:fJNeumann}
f_J(s) = {\cal N}_d \oint\limits_{[-1,1]} {d z \over 2 \pi i} \left(z^2 - 1 \right)^{{d-4 \over 2}} Q^{(d)}_J(z) T(s,t(z))\ ,
\ee
\begin{figure}[t]
\centering
\includegraphics[width=0.7\textwidth]{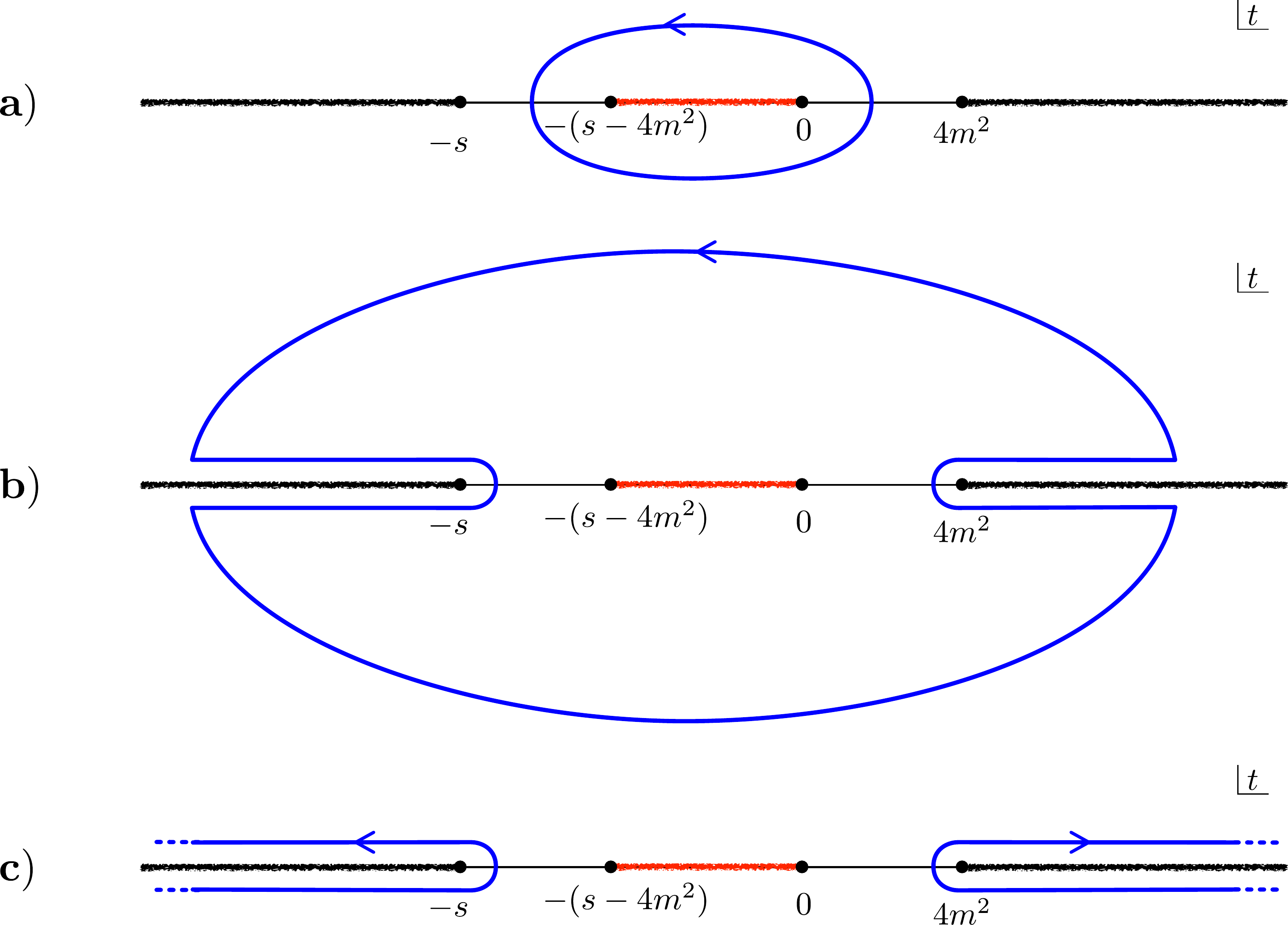}
\caption{\small {\bf a}. The partial wave projection integral (\ref{eq:fJNeumann}) is a contour integral (in blue) that circles around the cut of the $Q$-function, between $t=0$ and $t=-(s-4m^2)$, (in red). {\bf b}. We partially open up the contour. Sometimes this representation for partial waves is called the truncated Froissart-Gribov formula. The advantage of this representation is that we only use a finite amount of extended analyticity that has not been rigorously proven. {\bf c}. We open the contour all the way to infinity and arrive at the usual Froissart-Gribov formula (\ref{FG1}) with two integrations of the discontinuity of the amplitude along the $t$-channel and $u$-channel cuts (in black).}
\label{GFfig}
\end{figure}
where the integral is counterclockwise around the interval $z \in [-1,1]$. By blowing up the contour, we get two integrals along the $t$- and the $u$-channel cuts, see figure \ref{GFfig}
\beq\label{FG1}
f_J(s) ={{\cal N}_d\over\pi} \[\int\limits_{z_1}^\infty\! d z  (z^2-1)^{{d-4 \over 2}}Q^{(d)}_J(z) T_t(s,t(z))+\int\limits^{-z_1}_{-\infty}\! d z(z^2-1)^{{d-4 \over 2}}Q^{(d)}_J(z) T_u(s,u(z))\]\ ,
\eeq
where
\beq\label{z1definition}
z_1\equiv z|_{t=4m^2}= 1 + {8 m^2 \over s- 4 m^2}\ ,
\eeq
and we have assumed that $s>4m^2$, so that the $t$ channel cut runs from $z_1=z_1>1$ to infinity. 
Here we have dropped the contributions of the arcs at infinity. This is justified for large enough spin $J>J_0(s)$ using (\ref{Qasimp}), where $J_0(s)$ is the Regge intercept
\be\label{Regge}
\lim_{|t| \to \infty} |T(s,t) | < |t|^{J_0(s)}\ .
\ee
We can now use crossing to simplify (\ref{FG1}). We change the integration variable for the $u$-channel integral from $z$ to $-z$. Crossing symmetry implies that $T_u\(s,u(z)\)=- T_t\(s,t(-z)\)$, where we have used that $z(u)=-z $. Under this change of variables
\beq
(z^2-1)^{{d-4 \over 2}} \to (-1)^{d-4} (z^2-1)^{{d-4 \over 2}}\ ,\qquad Q^{(d)}_J(z)\to Q^{(d)}_J(-z)=(-1)^{J + 3 - d} Q^{(d)}_J(z)\ .
\eeq
We get that $f_J=0$ for odd $J$. For even $J$ we get
\beq\label{eq:FGform}
 \boxed{f_J(s) ={2 \, {\cal N}_d\over\pi}\int\limits_{z_1}^\infty d z \,(z^2-1)^{{d-4 \over 2}}Q^{(d)}_J(z) T_t(s,t(z))\ ,\qquad {\rm Re} J >J_0(s)}\ . 
\eeq

As opposed to (\ref{eq:pwprojection}), the Froissart-Gribov representation of the partial waves (\ref{eq:FGform}) is suitable for analytic continuation in $J$. It follows from the Carlson theorem that this analytic continuation 
is the unique continuation that does not grow too fast at large $J$. The Froissart-Gribov integral (\ref{eq:FGform}) converges as long as ${\rm Re} J >J_0(s)$ thanks to (\ref{Regge}) and (\ref{Qasimp}).

This integral is written for $s>4m^2$. As $s$ approaches the threshold from above $s-4m^2\to0^+$, the lower end of the integral is pushed to infinity, $z_1\to\infty$. To analyze $f_J(s)$ in this limit, it is useful to use (\ref{Qasimp}) and to switch back to an integral over $t$. In that way one finds 
\be
\label{eq:thresholdFJFG}
f_J(s) = {2 \, {\cal N}_d\over\pi} c_J^{(d)} \({s-4m^2\over2}\)^{J} \int\limits_{4 m^2}^{\infty} {d t \over t^{J+1}} T_t(4 m^2 , t) \left( 1 + O\((s - 4 m^2)/t\) \right) \ .
\ee
This integral should be understood as follows. The large $t$ contribution is finite because $|{\rm Im}_tT(4m^4,t)|<|T(4m^2,t)|<t^{J_0(4m^2)}$ and $J>J_0(4m^2)$ by assumption. If the integrand diverges at some finite $t$, and in particular as $t-4m^2\to0^+$, then we should step back and write it as a contour integral of $T(4m^2,t)$ around the cut, which is manifestly finite.

\subsection{Further Continuity Assumption}
\label{sec:continuity}

Based on the standard QFT axioms, scattering amplitudes, cross sections and partial waves are distributions rather than continuous functions. This leads to various subtleties, sometimes known as {\it pathologies a-la Martin} \cite{Martin:1966zz}. For example, we can imagine total cross-section having singularities which are local in energy variable $s$, that are not detectable by the finite resolution experiments. As such it is hard to exclude them based on physical grounds. One way to produce such a singularity is to consider an infinite number of resonances that accumulate on the real axis, see \cite{Martin:1966zz}. 

To the best of our knowledge there is no known, first principle argument that can exclude these possibilities. One way to eliminate them is to simply assume that various cross sections are continuous functions of energy $s$. We adopt this practical approach add this to the list of our assumptions. More precisely, we assume that  boundary values of $T(s,t)$ (this includes both single and double discontinuity) are continuous functions. It is common in the literature to impose the condition that boundary values of $T(s,t)$ are uniformly continuous or H\"{o}lder continuous, see e.g. \cite{Kupsch:2008hq}, but we will not use it in the present analysis.

Another related common assumption is regarding finiteness of scattering lengths which are commonly measured in the experiments or using the lattice. They are defined as follows, see (\ref{eq:thresholdFJFG}),
\be\label{eq:scatteringlengths}
a_J = \lim_{s \to 4 m^2} { m^{d-4} f_J(s) \over ({s \over 4 m^2}- 1 )^{J}} \geq 0\ ,\qquad J \geq 2\ .
\ee
We will assume that the scattering lengths are finite for $J \geq 2$. Through the Froissart-Gribov formula these are related to the assumption of finiteness of the discontinuity $T_t(4m^2,t)$ at $s = 4m^2$ as well as convergence of the $J=2$ Froissart-Gribov integral (\ref{eq:thresholdFJFG}).

There is an interesting connection between the continuity of the amplitude and macrocausality \cite{Chandler:1969bd,Williams}. Macrocausality is a set of statements about scattering amplitudes when particles grouped according to space and time of interactions and then moved away from each other by large translation.  The notion relevant for the continuity of scattering amplitudes is what is called {\it strong asymptotic causality} in \cite{Chandler:1969bd} and it has not been proved within the field theory.

\section{Analytic Continuation of Elastic Unitarity}\label{aceu}

The elastic unitarity relations (\ref{eq:dlips}) was derived for energies in the elastic region, above the two particle threshold $4m^2<s<16m^2$, and for physical kinematics $4m^2-s<t<0$. In this section we analytically continue this relation in $t$, outside of the regime of real scattering angles. We also consider the double discontinuity of the amplitude and the closely related analytic continuation of elastic unitarity in spin.

\subsection{Mandelstam Kernel}\label{sec:Mandelstam}

The dependence on the scattering angle $z=\cos\theta$ enters the right-hand side of the elastic unitarity relation (\ref{eq:elauni1}) through the kernels (\ref{eq:dlips}). These kernels contain a delta or a step functions and are thus not suitable for analytic continuation. To overcome this difficulty, we use the analyticity of $T^{(\pm)}(s,t(z))$ inside the small Lehmann-Martin ellipse to express them as a counterclockwise Cauchy integral around $[-1,1]$
\be
\cT^{(\pm)}(s,z') =\oint\limits_{[-1,1]} {d \eta' \over 2 \pi i} {\cT^{(\pm)}(s, \eta') \over \eta' - z'} \ ,\qquad-1<z'<1\ ,\qquad \cT(s,z)\equiv T(s,t(z))\ .
\ee
We can now exchange the order on integrations in (\ref{eq:elauni1}) and perform the $z'$ and $z''$ integrals explicitly. In this way we arrive at 
\be
\label{eq:elasticun}
\cT_s (s,z) &={(s - 4 m^2)^{(d-3)/2} \over 8(4\pi)^{d-2} \sqrt{s} } \oint\limits_{[-1,1]} {d \eta' \over 2 \pi i} \oint\limits_{[-1,1]} {d \eta'' \over 2 \pi i}\,\cT^{(+)}(s, \eta') \cT^{(-)}(s, \eta'')\times K_d(z, \eta', \eta'')\ ,
\ee
where the new kernel is
\be\label{MandelstamK}
K_d (z, \eta', \eta'')\equiv\int{ d^{d-2}\Omega_{\vec n} \over (\eta' - z')(\eta'' - z'')}=
\int\limits_{-1}^1 dz'\int\limits_{-1}^1 dz'' {\cP_d(z,z',z'') \over (\eta' - z')(\eta'' - z'')}\ .
\ee

These integrals are evaluated in appendix \ref{AppendixA}. For $|\eta'|,|\eta''| > 1$ the result is
\be
\label{eq:kernels}
K_{d=3}(z, \eta' , \eta'') &= {2\pi \over \eta_+ - z} \left( {\eta' \over \sqrt{\eta'^2-1}} + {\eta'' \over \sqrt{\eta''^2-1}} \right) \ ,\\
K_{d \geq 4} (z, \eta', \eta'') &={4\pi^{d-1\over2} \over \Gamma({d-3\over 2})}\int\limits_{\eta_+}^\infty{d\eta\over \eta - z} {(\eta^2 - 1)^{4-d \over 2}\over (\eta - \eta_+)^{5-d\over2} (\eta - \eta_-)^{5-d\over2}}\ . \nn
\ee
where
\be\label{etapm}
\eta_{\pm}(\eta',\eta'')\equiv\eta' \eta'' \pm \sqrt{\eta'^2-1} \sqrt{\eta''^2-1}\ .
\ee
The Mandelstam kernel for $|\eta'|<1$ or $|\eta''|<1$ is obtained from (\ref{eq:kernels}) by analytic continuation. Note that in (\ref{MandelstamK}) $\cP_d(z,z',z'')$ is not analytic in $z$, see (\ref{eq:dlips}). On the other hand, the Mandelstam kernel $K_{d}(z, \eta', \eta'')$ is analytic in $z$ and therefore is suitable for analytic continuation.

Similarly, the representation of $\cT_s(s,z )$ in (\ref{eq:elasticun}) is now suitable for analytic continuation in $t$. That is because $t$ only enters through the Mandelstam kernel that is manifestly analytic in $z$.

As for the kernel $\cP_J^{(d)}$ (\ref{eq:PPP}), the Mandelstam kernel (\ref{MandelstamK}) is also diagonal in spin. To represent it in angular momentum basis, we start from the representation of $\cP_J^{(d)}$ in (\ref{eq:PPP}) and plug it into the definition (\ref{MandelstamK}). We then note that the $z'$ and $z''$ integration has the effect of converting the partial waves $P_J^{(d)}(z')$ and $P_J^{(d)}(z'')$ into $Q_J^{(d)}(z')$ and $Q_J^{(d)}(z'')$ correspondingly, (\ref{eq:Qfunc}). In that way we arrive at
\be
\label{eq:PQQ}
K_d (z, \eta', \eta'') =4 (4 \pi)^{d-2} {\cal N}_d^2\, (\eta'^2-1)^{d - 4 \over 2} (\eta''^2-1)^{d - 4 \over 2}\sum_{J=0}^\infty n_J^{(d)} P_J^{(d)}(z) Q_J^{(d)}(\eta') Q_J^{(d)}(\eta'')\ .
\ee

\subsection{The Double Spectral Density and Crossing}\label{sec:rho}

The combination of elastic unitary with crossing symmetry is very restrictive. In the next sections we will explore some of its consequences at length. With this aim in mind, we now represent elastic unitarity in a form that is more suitable for imposing crossing. By taking another discontinuity of (\ref{eq:elasticun}) with respect to $t$, the left-hand side becomes the double discontinuity of the amplitude\footnote{Stated in words, the double spectral density $\rho(s,t)$ is defined as a certain combination of boundary values of the analytic function $T(s,t)$ unambiguously specified by $i \eps$ in the definition above.} 
\be
\rho(s,t)&\equiv - {1 \over 4} \lim_{\eps \to 0}\[ T(s+i \eps, t+ i \eps) - T(s-i \eps, t+ i \eps) - T(s+i \eps, t-i \eps) + T(s-i \eps, t - i \eps)\]\nn\\
&=\ \,{\rm Disc}_t{\rm Disc}_sT(s,t) = {\rm Disc}_s{\rm Disc}_t T(s,t)=\rho(t,s) \ .\label{dsd}
\ee
This crossing-symmetric function is known as {\it the double spectral density}.

By taking the $t$ discontinuity of (\ref{eq:elasticun}) we arrive at
\be\label{rhoTT}
\rho(s,t) &={(s - 4 m^2)^{d-3\over2} \over 8(4\pi)^{d-2} \sqrt{s} } \oint\limits_{[-1,1]} {d \eta' \over 2 \pi i} \oint\limits_{[-1,1]} {d \eta'' \over 2 \pi i}\,\cT^{(+)}(s, \eta') \cT^{(-)}(s, \eta'')\times{\rm Disc}_t K_d(z , \eta', \eta'')\ ,
\ee
where $4m^2<s<16m^2$. Outside of the elastic region there are additional non-elastic contributions to the double discontinuity that will be considered elsewhere. Next, we deform the $\eta'$ and $\eta''$ integration to wrap the $t$ and $u$ channel cuts of $\cT^{(\pm)}(s,\eta)$. The discontinuity of the Mandelstam kernel in (\ref{rhoTT}) is analytic in $\eta'$ and $\eta''$ along the deformation. In that way, we end with real $\eta'$ and $\eta''$ that are positive on the $t$ channel cut and are negative on the $u$ channel cut. For $\eta'\eta''<-1$ the integral in the Mandelstam kernel (\ref{eq:kernels}) starts at $\eta_+(\eta',\eta'')<-1$ and can be chosen to run along the negative real $\eta$-axis. This choice make it manifest that the discontinuity of the kernel for $z>1$ and $\eta_+(\eta',\eta'')<-1$ is zero. We remain with\footnote{The result for $z<-1$ is obtained by analytic continuation and takes the same form as (\ref{eq:doubleSD}) with ${\rm Disc}_zK(-z, \eta', \eta'')$ instead of ${\rm Disc}_zK(z, \eta', \eta'')$.}
\be\label{eq:doubleSD}
\rho(s,t) = {(s - 4 m^2)^{d-3\over2} \over 4\pi^2 (4\pi)^{d-2} \sqrt{s} }\int\limits_{z_1}^{\infty} d \eta'  \int\limits_{z_1}^{\infty} d \eta''\,\cT_t^{(+)}(s, \eta') \cT_t^{(-)}(s, \eta'')\, {\rm Disc}_z K_d(z, \eta', \eta'')\  ,
\ee
where we have mapped the $u$-channel cut to the $t$-channel cut using $\cT^{(\pm)}_u(s,\eta)=\Blue{-}\cT^{(\pm)}_t(s,-\eta)$ and $K(z,-\eta', -\eta'')=K(z, \eta', \eta'')$. Here, the lower limit of integration is the point where the $t$ channel cut starts (\ref{z1definition}). 
\begin{figure}[t]
\centering
\includegraphics[width=0.55\textwidth]{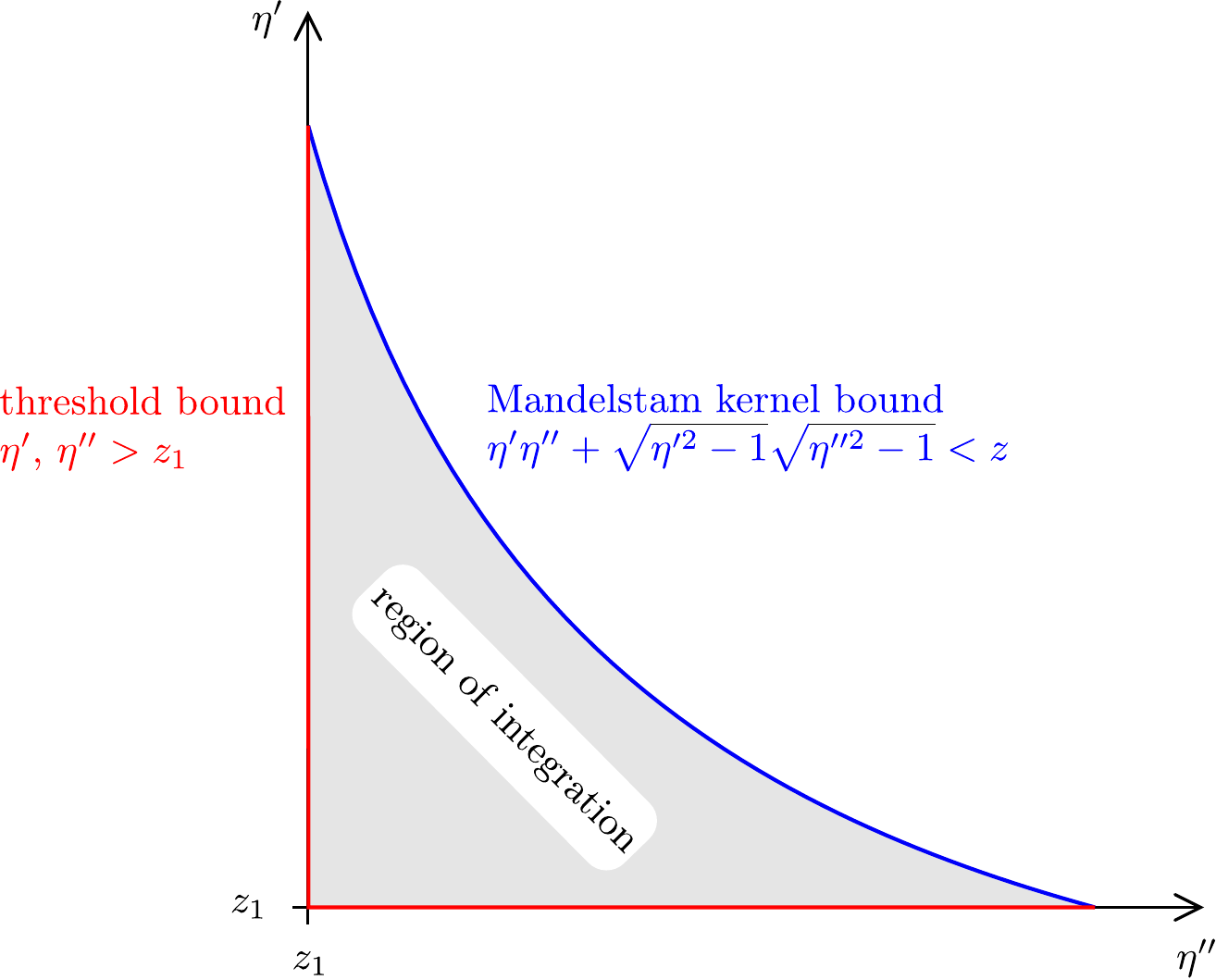}
\caption{\small The region of integration in equation (\ref{eq:doubleSD}). As $s$ or $t$ approaches the Landau curve from above, the integration region shrinks to zero. As a result, the double spectral density vanishes below the Landau curve $z=2 z_1^2-1$.}
\label{etarange}
\end{figure}
The discontinuity of the kernel, for $\eta' \eta'' > 0$ and $z>1$ is given by
\be\label{eq:kernelpositivity}
{\rm Disc}_z K_3(z, \eta', \eta'') &
= 4 \pi^2\delta(z-\eta_+){\sqrt{z^2 - 1}\over \eta_+ - \eta_-}\ , \nn \\
{\rm Disc}_z K_{d \geq 4}(z, \eta', \eta'') &= {4\pi^{d+1\over2} \over \Gamma({d-3\over 2})}  \Theta( z - \eta_+){(z^2 - 1)^{4-d \over 2}  \over (z-\eta_-)^{5-d\over2}(z-\eta_+)^{5-d\over2}}\ge 0\ .
\ee
This formula was first derived by Mandelstam in $d=4$ \cite{Mandelstam:1958xc}. In appendix \ref{AppendixA} we present the derivation for any dimension $d\ge3$.

\begin{figure}[t]
\centering
\includegraphics[width=0.65\textwidth]{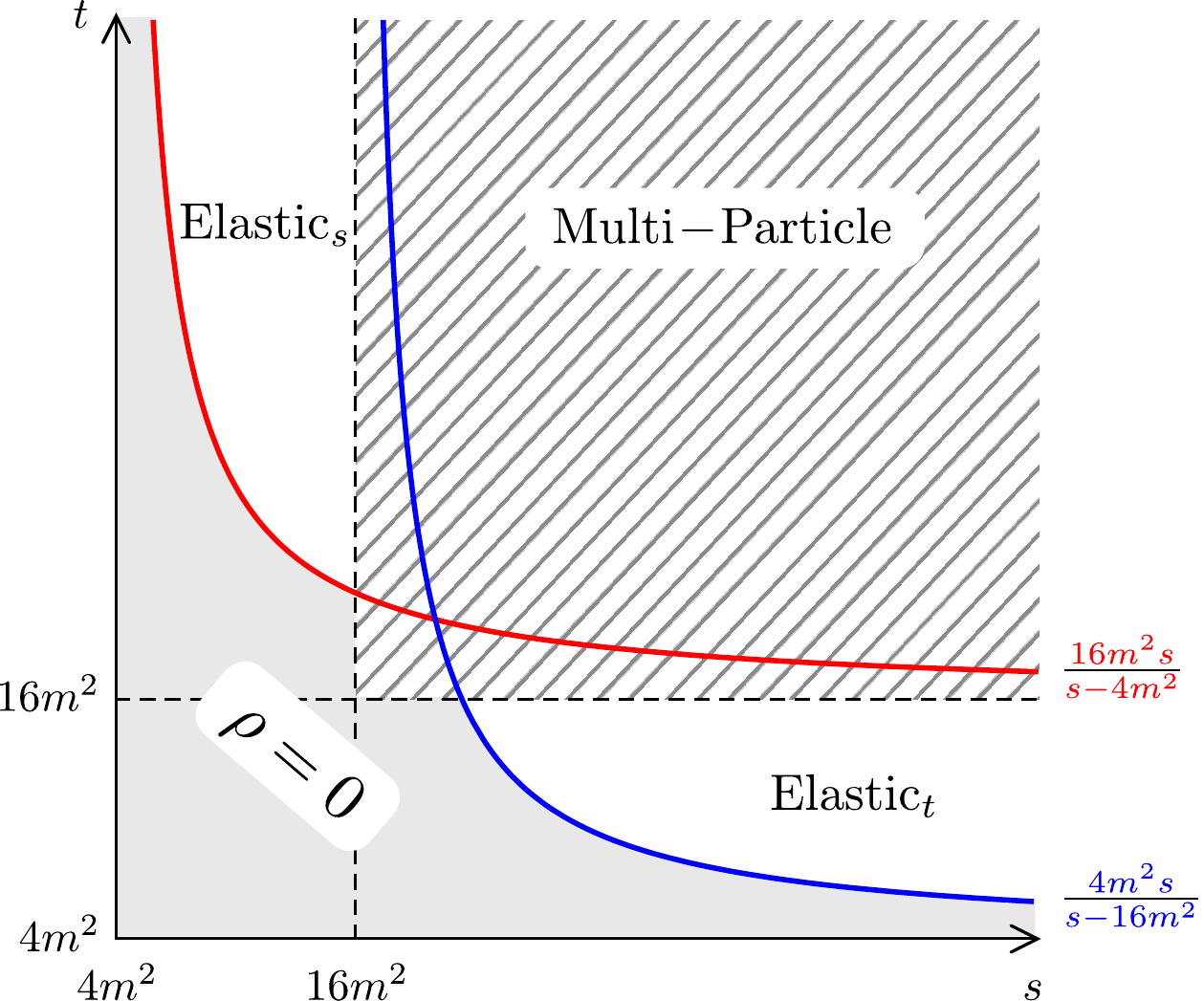}
\caption{\small The double discontinuity of the amplitude $\rho(s,t)$ in the real $(s,t)$ plane. In gray is {\it the Steinmann shadow} region where $\rho$ vanishes. This region extends inside the elastic bands, $\{(4m^2<s<16m^2,t),(s,4m^2<t<16m^2)\}$, and is bounded by the Landau curve $t_1(s)={16m^2 s\over s-4m^2}$ (in red) and its crossed curve $t_1^\text{cross}(s)={4m^2 s\over s-16m^2}$ (in blue). These two curves extend out of the elastic bands, where there are additional multi-particle contributions. They cross at $s=t=20m^2$.}
\label{fig:karplus1}
\end{figure}

We now discuss the region of support of the double discontinuity $\rho(s,t)$. Because the first discontinuity has only support for energies above the two particle threshold, so does the double discontinuity. Looking at (\ref{eq:doubleSD}), it is clear that the situation is more interesting. The discontinuities of the amplitude only start at $\eta',\eta''>z_1$. At the same time, the Mandelstam kernel has a non-zero support only for $z>\eta_+(\eta',\eta'')$. This constraint is an upper bound on the $\eta'$, $\eta''$ limits of integration, see figure \ref{etarange}. Hence, the double spectral density in the elastic region is non-zero only for
\be\label{Landau1}
z = 1 + {2 t \over s - 4m^2} > \eta_+ \left( z_1, z_1  \right) =1+{32m^2s\over(s-4m^2)^2}\ . 
\ee
In terms of $s$ and $t$, the boundary of this region is known as the Landau or Karplus curve and is given by (red in figure \ref{fig:karplus1})
\be
\label{eq:red}
t = {16m^2\,s \over s-4m^2}\equiv t_1(s) \ .
\ee

As $t$ is increased above $t_1(s)$, the range of the $\eta'$ and $\eta''$ integration opens up. For example, for any value of $t>t_1(s)$, the integral over $\eta'$ is bounded in the range
\beq\label{etapmax}
z_1\le\eta'\le z \eta''-\sqrt{(z ^2-1)({\eta''}^2-1)}\le z z_1-\sqrt{(z ^2-1)(z_1^2-1)}\ .
\eeq
Hence, as we increase $t$ inside the elastic region $4m^2<s<16m^2$ (where (\ref{eq:doubleSD}) is valid), more and more channels of $\cT^{(+)}_t(s,\eta')$ and $\cT^{(-)}_t(s,\eta'')$ kick in. Their corresponding contributions to $\rho(s,t)$ start at other Landau curves in the $(s,t)$ plane that are above the elastic one (\ref{eq:red}), see figure \ref{fig:karplus1}. In section \ref{sec:Karplus} we present a minimal and complete set of Landau curves in the elastic region $4m^2 < s < 16 m^2$ that appears in the physical $S$-matrices.

In general, the positions of the Landau curves look somewhat technical. However, they are all kinematic and therefore have a geometrical origin. For example, the constraint (\ref{Landau1}) takes a simple form in terms of the integral scattering angles $\eta'=\cosh\theta'$,   $\eta''=\cosh\theta''$, and is given by
\beq\label{thetacurve}
\theta'+\theta''\ge\theta\ .
\eeq
In physical kinematics, this constraint follows from a simple geometrical consideration that is described in figure \ref{stepfunction}. The analytic continuation to the non-physical kinetatical regime of positive $s$ and $t$ effect the range of the angles, but leaves this geometrical constraint unchanged. At the technical level, this is because the kinematical constraint only involves the cosines of the scattering angles.

We end this section with a comment regarding the region below the Landau curve, where the double discontinuity vanishes (the gray region in figure \ref{fig:karplus1}). For the two-to-two scattering, existence of this region is a direct consequence of elastic unitarity continued to $s,t>4m^2$. The precise shape of the region depends on the details of the unitarity kernel $K_d$, as well as on the analytic structure of the amplitude that enters into the elastic unitarity relation. 
A similar phenomenon occurs in the higher-point amplitudes as well. 
In this case one considers double discontinuity in the so-called overlapping channels, see e.g. \cite{Cahill:1972ye} for a detailed definition. This time it is possible to consider double discontinuity for physical kinematics directly, as opposed to the two-to-two case which requires continuation in one of the Mandelstam invariants. It then follows that for physical kinematics the double discontinuity vanishes for the overlapping channels. These are known as the {\it Steinmann relations} \cite{Steinmann,Cahill:1973qp}, and it is again a direct consequence of the multi-particle unitarity.  Steinmann relations are useful for constraining the analytic structure of amplitudes with six and more particles, see \cite{Caron-Huot:2016owq} for a recent discussion. Based on the two-to-two case one can try to analytically continue the relevant unitarity relations to the unphysical values of the relevant kinematical invariants, and find the extended region where the double discontinuity vanishes. It would be interesting to understand the precise shape of this region for the multi-particle case. This would require analytic continuation of multiparticle unitarity kernels, as well as relevant scattering amplitudes, analogous to the one made above.

\subsection{Analytic Continuation in Spin}\label{sec:acJ}

We will now argue that the elastic unitarity relation (\ref{eq:EUfJ}) holds in the complex $J$ plane, provided that the partial wave coefficients are analytically continued using the Froissart-Gribov representation (\ref{eq:FGform}). For that aim, we first rewrite the elastic unitarity condition (\ref{eq:EUfJ}) with the $i \eps$ prescription explicitly
\be\label{eq:generalFJ}
 f_J(s+ i \eps) - f_J(s- i \eps)=i{(s-4 m^2)^{d-3\over2}\over\sqrt s} f_J(s+i \eps) f_J(s - i \eps)\ ,\qquad4 m^2 < s < 16 m^2\ .
\ee
In our previous discussions $f_J(s) \equiv f_J(s + i \eps)$. For integer $J$ and real $s$ in that range, real analyticity (\ref{eq:realanalytPW}) leads to (\ref{eq:EUfJ}). The form (\ref{eq:generalFJ}) is however more suitable for analytic continuation.

Originally, (\ref{eq:generalFJ}) was derived for $J$ being integer and even, however using the Froissart-Gribov representation we can continue partial waves in spin $J$
\be
\label{eq:FGpw}
f_J(s \pm i \eps) &={2 \, {\cal N}_d\over\pi}\int\limits_{z_1}^\infty d z\,(z^2-1)^{{d-4 \over 2}}Q^{(d)}_J(z) \cT_t^{(\pm)}(s,z)\ ,\qquad {\rm Re}[J]>{\rm Re}[J_0(s)] \ .
\ee
This representation can in principle be continued to the whole complex $J$ plane, going beyond the ${\rm Re}[J]>{\rm Re}[J_0(s)]$ region. However, the continued partial waves are not guaranteed to coincide with the physical ones for $J<J_0$. Moreover, it is clear from (\ref{eq:realanalyticity}) and the reality properties of $Q_J^{(d)}(z)$ (\ref{eq:FGpw}), that real analyticity of partial waves (\ref{eq:realanalytPW}) is only guaranteed to hold for real $J>J_0$.

We can then use (\ref{eq:FGpw}) to separately analytically continue the left and right hand sides of (\ref{eq:generalFJ}). It follows from the Carlson theorem that these two analytic continuations have to agree in the whole complex $J$ plane. Namely, the two analytic continuations agree for real integer $J>J_0$ and do not grow too fast at large $|J|$, as can be seen from the large $J$ exponential decay of $f_J(s)$, see section \ref{sec:largespinF} for more details,
\beq
f_J(s)\sim \({\sqrt s+2m\over\sqrt s-2m}\)^{-J}\ .
\eeq

Let us relate the analytic continuation in $J$ of (\ref{eq:generalFJ}) to the Mandelstam equation (\ref{eq:doubleSD}). To get the former from the latter we integrate (\ref{eq:doubleSD}) with $\int_{2 z_1^2-1}^\infty d z (z^2-1)^{{d-4 \over 2}} Q_J(z)$, where $Q_J(z)$ are given in (\ref{eq:Qfunc}). We then use the following identity (for derivation see appendix \ref{Idappendix})
\be\label{eq:Atkinson}
\int\limits_{\eta_+}^\infty dz\,(z^2-1)^{{d-4 \over 2}} Q_J(z) {\rm Disc}_z K_d(z, \eta', \eta'') ={4 \pi^{d/2} \over \Gamma({d-2 \over 2})}\[(\eta'^2-1)^{{d-4 \over 2}} Q_J^{(d)}(\eta')\]\[(\eta''^2-1)^{{d-4 \over 2}} Q_J^{(d)} (\eta'')\]\ ,
\ee
which is valid for ${\rm Re}[J]>-1$, $d \geq 3$ and $|\eta_1|, |\eta_2|>1$. In practice, we will use this identity in \eqref{eq:doubleSD}, where $\eta'$ and $\eta''$ are real and positive.

\par
Finally, we can write
\be
\label{eq:dsdPW}
 f_J(s+ i \eps) - f_J(s- i \eps)  &=2i 
 {2\cN_d\over\pi}\!\!\!\int\limits_{2z_1^2-1}^\infty\!\!\!\! dz(z^2-1)^{{d-4 \over 2}}Q^{(d)}_J(z) \rho(s,t(z))\ ,\qquad{\rm Re}[J]>{\rm Re}[J_0(s)]\ .
\ee
By combining (\ref{eq:Atkinson}) with (\ref{eq:dsdPW}), it is easy to check that (\ref{eq:doubleSD}) becomes precisely (\ref{eq:generalFJ}).

Let us also mention that analytically continued elastic unitarity constrain the possible Regge limit behavior of the amplitude. In particular, the leading Regge singularity of the amplitude in the elastic region $4m^2 < s < 16m^2$ cannot we a pole located at some real Regge spin $\tilde J_0(s)$. We review the derivation of this and a slightly more general result, known as Gribov's theorem, in appendix \ref{sec:gribovstheorem}.

\subsection{Analytic Continuation in $s$}

As $s$ is increased above $16m^2$ there are new multi-particle cuts contributions to ${\rm Im}f_J(s)$ that are not captured by (\ref{eq:EUfJ}). Still, if we denote by $f^{\circlearrowleft}_J(s)$ the partial wave that was analytically continues on the second sheet through the elastic cut $4m^4<s<16m^2$, then the equation
\be\label{eq:generalFJ2}
 f_J(s) - f_J^{\circlearrowleft}(s)=i{(s-4 m^2)^{d-3\over2}\over\sqrt s} f_J(s) f_J^{\circlearrowleft}(s)
\ee
holds true when analytically continued away from the elastic strip in the full multi-sheet complex $s$ plane. In that sense, elastic unitarity can be analytically continued in $s$.

Note that there is a difference in the nature of the two-particle cut between odd and even dimensions. Due to the power of $(s-4m^2)$ in (\ref{eq:partialwave}) we get that
\beq\label{Scirc}
S_J^{\circlearrowleft}(s)\equiv1+i{(s-4m^2)^{d-3\over2}\over\sqrt s}f_J^{\circlearrowleft}(s)\times\left\{\begin{array}{ll}-1&d\text{-even}\\+1&d\text{-odd}\end{array}\right.\ .
\eeq
Correspondingly, for even dimension the continued elastic unitarity condition (\ref{eq:generalFJ2})  becomes
\be
\label{eq:evenDsJ}
S_J(s) S_J^{\circlearrowleft}(s) = 1\ ,\qquad d\text{-even}\ .
\ee
This relation implies that the elastic cut $4 m^2 < s < 16 m^2$ describes a two-sheeted Riemann surface in even $d$. Indeed, continuing (\ref{eq:evenDsJ}) around the elastic cut again we conclude that $S_J^{2 \times \circlearrowleft}(s) = S_J(s)$.  In odd $d$, because of the sign difference in (\ref{Scirc}), we do not have (\ref{eq:evenDsJ}) and the elastic cut is not two-sheeted. Similar conclusion holds for $T(s,t)$. As we will see below, the nature of the elastic unitarity cut is the one of a simple square-root for even $d$ and is infinitely-sheeted for odd $d$. It immediately follows from (\ref{eq:evenDsJ}) that a pole of $S_J^{\circlearrowleft}(s)$ (resonance) corresponds to a zero of $S_J(s)$ in even $d$. Similarly, in odd $d$, a pole of $S_J^{\circlearrowleft}(s)$ corresponds to a zero of $1-i {(s-4m^2)^{d-3\over2} \over \sqrt s} f_J(s)$ on the principal  sheet.

\subsection{Elastic Landau Curves}\label{sec:Karplus}

In this section we further study the kinematical properties of the double spectral density in the elastic strip. In general, discontinuities of the amplitude result from the exchange of intermediate on-shell states. Correspondingly, the double discontinuity of the amplitude receives its support from intermediate on-shell particles exchanged in both channels. As we have seen in section \ref{sec:rho}, the first double discontinuity in the elastic strip starts at the leading Landau curve (\ref{eq:red}). It comes from the exchange of two particle in the $s$-channel and four particles in the $t$-channel. As we increase $s$ or $t$, more and more processes become accessible. The curves in the $s-t$ plane where they start to contribute are higher Landau curves. We now derive an infinite set of higher Landau curves in the elastic strip that are required by elastic unitarity. 

\begin{figure}[t]
\centering
\includegraphics[width=0.98\textwidth]{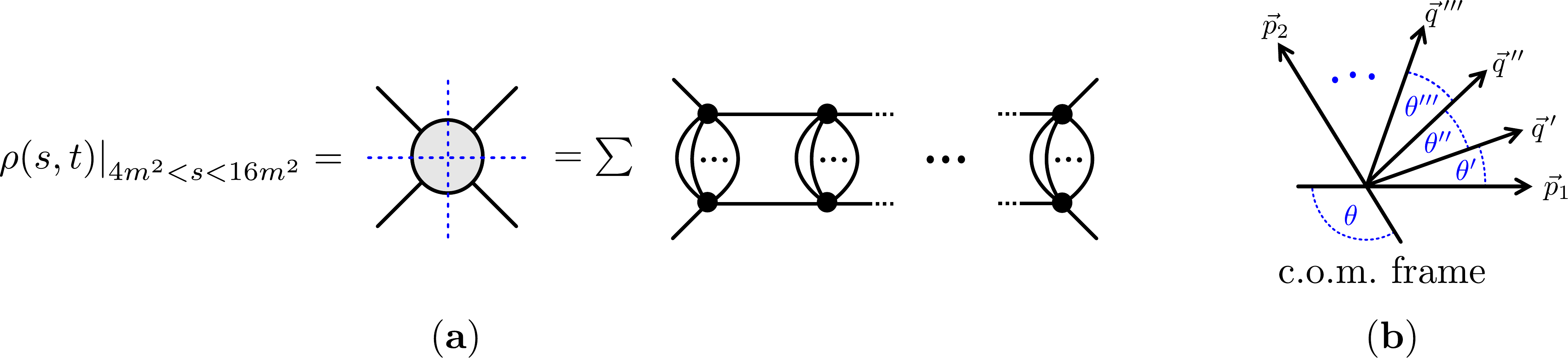}
\caption{\small {\bf a}. The Landau diagrams that contribute to the double spectral density in the elastic strip $4m^2<s<16m^2$ and $4m^2<t$. In this kinematical regime there can be only two particles in the $s$-channel. Hence, the corresponding Landau diagrams have a simple structure of iterative two-particle exchange in the $s$-channel and in between, any number of particles exchange in the $t$-channel. {\bf b}. Analogously to figure \ref{stepfunction}, the corresponding Landau curves originate from a simple geometric constraint on physical kinematics, see discussion after (\ref{thetacurve}).}
\label{LandauDiagrams}
\end{figure}

To each of these Landau curves one can associate a Landau diagram. In the elastic strip, all these diagrams have a simple iterative structure that is plotted in figure \ref{LandauDiagrams}. Correspondingly, the shape of the curve has a relatively simple geometrical origin, (for the case of the first Landau curve, see figure \ref{stepfunction} and the discussion around (\ref{thetacurve})). Here instead, we follow a shortcut and derive their shape directly from the Froissart-Gribov representation and the elastic unitarity constraint. 

The starting point is the kinematical structure of the discontinuity and the double discontinuity of the amplitude as function of one of the Mandelstam invariants
\be
&{\rm Disc}_tT(s,t)=\qquad T_t(s,t) =\sum_{n=1}^{\infty} \Theta(t - (2 n m)^2)\, T_t^{2 \to 2n}(s,t)\ ,\qquad 4m^2-t<s<0\ ,\label{thresholds}\\
&{\rm Disc}_s{\rm Disc}_tT(s,t)=\rho(s,t)=\sum_i \Theta(t - t^{(i)}(s))\,\rho^{(i)}(s,t)\ ,\qquad\qquad\  4m^2<s,t\ ,\label{thresholds2}
\ee
where the functions $t^{(i)}(s,t)$ are the Landau curves we are after. For that aim, we first have to analytically continue the discontinuities of the amplitudes, $T_t^{2 \to 2n}(s,t)$, to the regime of positive $s$ and $t$. As we do so, they develop new thresholds. Because $T_t(s,t)$ is a real analytic function, these new thresholds must coincide with the Landau curves $t^{(i)}(s)$. What is important for us here is that the multi-particle thresholds that are manifest in (\ref{thresholds}) are also present in the non-physical kinematical regime.

To derive the functional shape of the Landau curves in the elastic strip we impose the consistency of (\ref{thresholds2}) and (analytically continued) (\ref{thresholds}) with elastic unitarity. This can be done by either plugging them into the Mandelstam equation or by imposing elastic unitarity at the level of the partial waves. Here we follow the latter strategy and in appendix \ref{sec:Mandelstam_Karplus} we present the former, both leading to the same result.  

To impose consistency of (\ref{thresholds}) and (\ref{thresholds2}) with the partial waves elastic unitarity, we plug them into the Froissart-Gribov projection (\ref{eq:FGform}), that we quote here for convenience
\beq\label{FGRIm}
{\rm Re}f_J(s)+i\,{\rm Im}f_J(s)={2 \, {\cal N}_d\over\pi}\int\limits_{z_1}^\infty d z \,(z^2-1)^{{d-4 \over 2}}Q^{(d)}_J(z)\[{\rm Re}\,T_t(s,t(z))+i\rho(s,t(z))\]\ .
\eeq

Next, we take the large $J$ limit of (\ref{FGRIm}). Using the large $J$ decay of the $Q$ functions that can be schematically written as, see section \ref{sec:largespinF},
\beq\label{QlargeJ}
Q_J^{(d)}(z) \sim \lambda^{-J}(z)\ , \qquad \lambda(z) \equiv z+\sqrt{z^2 -1}=e^\theta\ ,\qquad 
z > 1 \ ,
\eeq
as well as the two-particle step function in (\ref{thresholds}),
we conclude that the leading large $J$ behavior of ${\rm Re}\,f_J(s)$ is
\beq\label{ReflargeJ}
\Re f_J(s) \sim \lambda^{-J}(z_1)\ .
\eeq

Similarly to (\ref{ReflargeJ}), the $n$-particle threshold of $T_t(s,t)$ and the $i$'th Landau curve threshold of $\rho(s,t)$ in (\ref{thresholds}) result in contributions to ${\rm Re}\,f_J(s)$ and ${\rm Im}\,f_J(s)$ that start at large $J$ as
\beq\label{nparticle}
\Re f_J(s) \sim \lambda^{-J}(z_n)\ ,\qquad \Im  f_J(s) \sim \left.\lambda^{-J}\(z\)\right|_{t=t^{(i)}(s)}\ ,
\eeq
where $z_n \equiv 1 + {8 n^2 m^2 \over s - 4m^2}$. Importantly, (\ref{QlargeJ}) receives only $1/J$-power corrections and no nonperturbative exponential corrections, see appendix \ref{sec:Q}. Hence, $\Im f_J(s)$ does not receive any exponential large $J$ behavior other than the ones that result from Landau curves thresholds (\ref{nparticle}). Similarly, the exponential large $J$ behavior of $\Re f_J(s)$ can only come from thresholds of $T_t(s,t)$ in (\ref{FGRIm}), but not from nonperturbative terms in the large $J$ expansion of $Q_J^{(d)}(z)$. 

We now plug these set of large $J$ exponential behaviors into elastic unitarity and derive the minimal set of Landau curves, $t^{(i)}(s)$, that are required to close the equation, together with new thresholds of $T_t(s,t)$ that are not present in physical kinematics. 

Elastic unitarity (\ref{eq:EUfJ}) can be written in the following schematic form
\be
\label{eq:elak}
\mathrm{Im} f_J(s) \propto [\mathrm{Re} f_J(s)]^2 + [\mathrm{Im} f_J(s)]^2 \ ,
\ee
where we omitted the pre-factor because it is irrelevant for the present discussion. First, we see that the multi-particle threshold exponents of $\Re f_J(s)$ in (\ref{nparticle}) result in the following exponents in $\Im f_J(s)$
\be\label{ImfJquadratic}
\Im f_J(s) \sim \left[ \lambda(z_n) \lambda(z_m) \right]^{-J}\ ,\qquad n,\,m \geq 1\ .
\ee
Using (\ref{nparticle}), these then lead to the Landau curves
\beq\label{nmLandauc}
\left.\lambda\(z\)\right|_{t=t^{(n,m)}(s)}=\lambda(z_n) \lambda(z_m)\ .
\eeq
For example, the leading Landau curve (\ref{eq:red}) corresponds to the case where $m=n=1$. This behavior must then also exist in $\Re f_J(s)$. To see that, recall that $T_t(s,t)$ is a real analytic function of $t$ for fixed $s>4m^2$. Hence, a threshold in the imaginary part of $T_t(s,t)$, namely in $\rho(s,t)$, must be accompanied by a corresponding threshold in ${\rm Re}\, T_t(s,t)$.

Having established the presence of the quadratic in $\lambda(z_n)$ terms in the large $J$ expansion of $\Re f_J(s)$  and $\Im f_J(s)$, we now come back to elastic unitarity (\ref{eq:elak}). The presence of these terms together with the linear one (\ref{nparticle}), now induces higher powers of $\lambda(z_n)$ in the large $J$ expansion of $\Im f_J(s)$ and, via real analyticity, of $\Re f_J(s)$ as well. These take the general form
\be
\label{eq:lambdasol}
\mathrm{Re} f_J(s) &\sim  \left[ \lambda(z_{n_1}) \cdots \lambda(z_{n_{L}}) \right]^{-J}\ ,\qquad n_{i}\geq 1\ ,\quad L\geq 1\ ,\\
\label{eq:lambdasol2}
\mathrm{Im} f_J(s) &\sim \left[ \lambda(z_{n_1}) \cdots \lambda(z_{n_{L}}) \right]^{-J}\ ,\qquad n_{i}\geq1\ ,\quad L\geq 2 \ .
\ee

This large $J$ structure implies existence of an infinite set of Landau curves labeled by a set of integers $\{N_1, N_2, N_3, \dots\}$ and given by the following equation
\be
\label{eq:Karplus_curve}
\lambda\(z(t_{\{N_1,N_2,N_3,\dots\}})\) = \lambda^{N_1} (z_1) \, \lambda^{N_2} (z_2) \, \lambda^{N_3} (z_3)  \cdots  \ .
\ee
In terms of the scattering angle, for which $z = \cosh \theta$ and $\lambda = e^{ \theta}$, these Landau curves take the form
\be
\theta(z) = N_1 \theta_1 + N_2 \theta_2 + N_3 \theta_3 + \dots\ ,
\ee
with $\theta_n = \arccos(z_n)$. This form is the generalization of (\ref{thetacurve}) for all the elastic Landau curves. In a direct analogy with figure \ref{stepfunction}, it also has a simple geometrical origin that is discussed in figure \ref{LandauDiagrams}.b. Using that $z = 1+ {2t \over s-4m^2}$, we can translate (\ref{eq:Karplus_curve}) into polynomial equations in $s$ and $t$ with real coefficients.\footnote{In this form they are familiar in the study of perturbative Feynman integrals, see e.g. \cite{Eden:1966dnq} for more details.} For example, we have
\be\label{fewcurves}
t_{\{2,0,\dots\}} = {16 m^2 s \over s - 4m^2}\ ,\qquad t_{\{3,0,\dots\}} = {36m^2(s + {4 m^2 \over 3})^2 \over (s - 4m^2)^2}\ ,\qquad t_{\{4,0,\dots\}} = {64m^2s(s+4 m^2)^2 \over (s - 4m^2)^3}\ .
\ee

The Landau curves have the following asymptotic behavior
\beq
\lim_{s \to 4 m^2} t_{\{N_1, N_2, \dots\}}(s) \sim (s - 4m^2)^{1 - \sum_{j=1}^\infty N_j}\ ,\qquad \lim_{s \to \infty} t_{\{N_1, N_2, \dots\}}(s) =\Big(2 m \sum_{j=1}^{\infty} j N_j\Big)^2 \ .
\eeq
These asymptotes are precisely the $t$-channel normal thresholds \eqref{thresholds}. This is evident from the Landau diagram interpretation of figure \ref{LandauDiagrams} and it would be interesting to see if it can be established rigorously.\footnote{For the case of two particles in the final state it was done by Mandelstam and we reviewed it in section \ref{sec:Mandelstam}. For the multi-particle case some limited results have been obtained \cite{Gribov:1962ft,Lardner1,Lardner2,Kim}.}

\section{Positivity of $\rho$ and Multi-Particle Production}\label{productionsec}

Drawing an analogy with the conformal bootstrap, one can expect that combining crossing symmetry of $\rho(s,t)$ with some sort of positivity property leads to nontrivial constraints on scattering amplitudes.

In this section we show that there is indeed a finite region in the elastic strips of the $(s,t)$-plane, 
where $\rho(s,t)$ is positive,
as noted in \cite{Mahoux}. In subsection \ref{sec:aks} we discuss a direct consequence of this positivity and crossing -- the necessity of multi-particle production $T_{2\to2n}$ for any $n$.

\subsection{Positivity of The Double Spectral Density}
\label{sec:PositivityDSD}

\begin{figure}[t]
\centering
\includegraphics[width=1\textwidth]{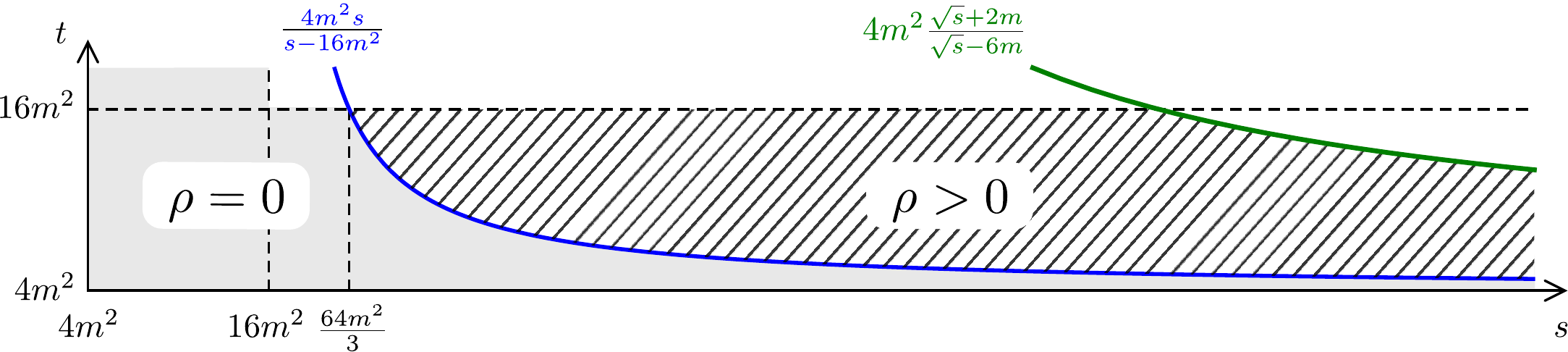}
\caption{\small The regime of positivity of the double spectral density $\rho(s,t)$ in the elastic strip $4m^2<t<16m^2$, (dashed region). This is the region above the first Karplus curve $s_1(t)$ (in blue) and below the curve $t=4m^2{\sqrt s+2m\over\sqrt s-6m}$ (in green). There is an identical positive region in the crossed strip of $4m^2<s<16m^2$, see (\ref{positiverho}).}
\label{positive_region}
\end{figure}

To establish positivity, we assume that the integral in the Mandelstam equation for the double spectral density (\ref{eq:doubleSD}) converges and study under what conditions the integrand 
is positive.

Consider first the discontinuity of the Mandelstam kernel ${\rm Disc}_z K_d(z, \eta', \eta'')$ given in (\ref{eq:kernelpositivity}). This function is strictly positive for $z>1$, 
which is the case for any $s,t>4m^2$. Next we turn our attention to $\cT_t^{(\pm)}(s,\eta')$. It is clear that $\cT_t(s,\eta')$ is non-negative in the region where the partial wave expansion converges\footnote{In this region we can drop the $(\pm)$ subscript of $\cT_t^{(\pm)}(s,\eta')$ and $\cT_t^{(\pm)}(s,\eta'')$ in (\ref{eq:doubleSD}) because both functions are equal.}
\be\label{eq:TttA}
\cT_t(s,\eta') =T_t(s,t') = \sum_J n_J^{(d)} \, {\rm Im} f_J(t') P_J^{(d)}\(1+{2s\over t'-4m^2}\)\ ,\qquad t'={\eta'-1\over2}(s-4m^2)\ , 
\ee
where $t'$ is not to be confused with the external $t$ in (\ref{eq:doubleSD}). Indeed, unitarity implies that ${\rm Im} f_J(t') \geq 0$ and $P_J^{(d)}\(1+{2s\over t'-4m^2}\)$ is positive for $s,t' > 4 m^2$. Consider the function $g\(1+{2s\over t'-4m^2}\)\equiv T_t(s,t')$ for fixed $t'$. Provided that $g(z)$ is analytic inside an ellipse in the complex $z$-plane with foci at $z=\pm1$, it then follows from Neumann's argument, see footnote \ref{foot:Neumann}, that the partial wave expansion (\ref{eq:TttA}) converges inside that ellipse. Under the assumption of extended analyticity and
for fixed $t'$, the first singularity is where $T_t(s,t')$ develops a discontinuity with respect to $s$. Namely, the partial wave expansion (\ref{eq:TttA}) converges below the first Landau curve, red in figure \ref{fig:karplus1},\footnote{As a consistency check, we can also study the convergence region of (\ref{eq:TttA}) using the large $J$ asymptotics of the partial wave coefficients. Using (\ref{eq:dsdPW}) and the shape of the first Landau curve we have arrived at the same conclusion. } 

\beq\label{convergenceregion}
t'<t_1(s)={16 m^2 s \over s - 4 m^2}\qquad\text{or equivalently}\qquad\eta'<1+{36m^2\over(s-4m^2)^2}
\ ,\qquad 4m^2<s<16m^2\ . 
\eeq

This region of convergence is called the large $s$-channel Martin-Lehmann ellipse \cite{Martin:1965jj}. Note that the standard argument of \cite{Martin:1965jj} refers to the convergence of (\ref{eq:dsdPW}) in the crossed region of $s<s_1(t')$, $4m^2<t'<16m^2$ and does not require the extended analyticity assumption. Here, we are saying that under this assumption, the partial wave expansion (\ref{eq:TttA}) converges in the union of the $t$- and $s$-channel large Martin-Lehmann ellipses.

The maximal value of the $\eta'$ integration is given in (\ref{etapmax}). Hence, we conclude that the double spectral density is non-negative for 
\beq\label{etappositive}
z z_1-\sqrt{(z ^2-1)(z_1^2-1)}\le1+{36m^2\over(s-4m^2)^2}\ .
\eeq
By solving this condition for $t$ we conclude that 
\be\label{positiverho}
\boxed{\rho(s,t) > 0\ ,\qquad 4 m^2 < s < 16 m^2\ ,\qquad{16 m^2 s \over s - 4 m^2} < t\le 4 m^2 {(3 s + 4 m^2)^2 \over (s - 4 m^2)^2}}\ .
\ee
where the lower bound comes from (\ref{eq:red}), and the upper bound from (\ref{etappositive}). This region of positivity was first derived in \cite{Mahoux} and is plotted in figure \ref{positive_region}.\footnote{Note that this is the regime netween the curves $t_{\{2,0,\dots\}}$ and $t_{\{3,0,\dots\}}$ in (\ref{fewcurves}).}

Note that at the upper bound of that region $\rho(s,t)$ is strictly positive. Hence, the region of positivity can always be extended from above. A question to which we do not know the answer is whether the region of positivity of the double spectral density can be extended to arbitrary large $s$ and $t$. Here we only list some known implications of such a scenario.
In \cite{Martin:1969cq} A.~Martin argued that positivity of the double spectral density for all $s$ and $t$ implies that the total cross-section at high energies satisfies the following bound
\be
\sigma_{tot}(s) \leq {C \over \log s}\ ,\qquad s \to \infty \ .
\ee
This immediately implies that for theories that saturate the Martin-Froissart bound $\sigma_{tot}(s) \sim \log^2 s$ \cite{Froissart:1961ux,Martin:1965jj}, the double spectral density is not positive-definite.

Similarly, in \cite{Martin:1969wv} A.~W.~Martin explored implications of the positive-definite double spectral density for scattering at finite angles. He showed that amplitudes with positive double spectral density {\it do not} exhibit diffraction peak in the near-forward scattering, which is
\be
T_t(s,t) \sim e^{s b(t) + \dots } \ , 
\ee
where $b(t)$ is a slowly-varying function. Such a diffraction peak is observed, for example, in the scattering of pions. Therefore, the double spectral density for pion scattering cannot be positive for arbitrary $s$ and $t$.

It would be interesting to understand positivity properties of $\rho(s,t)$ in physical theories more systematically.

\subsection{The Aks Theorem and Necessity of Particle Production}
\label{sec:aks}

We now review an elegant argument for scalar particles by Aks \cite{Aks}.\footnote{For the generalization to the case of spinning particles or particles in nontrivial representations of some global symmetry, see \cite{CheungAks}.} It states that scattering implies particle production. Namely, provided that $T_{2\to2}\ne 0$, also $T_{2\to n}\ne 0$ with $n>2$. The theorem applies to any crossing symmetric scalar scattering amplitude in $d \geq 3$ that satisfies extended analyticity in a finite region above the leading Landau curve.

To derive Aks's result, let us therefore assume that we have a nontrivial scattering amplitude $T(s,t)$, but $T_{2 \to n}$ are identically zero for $n>2$. This implies that elastic unitarity in the form of Mandelstam (\ref{eq:doubleSD}) holds for any $s\geq 4m^2$, whereas in the the theories with particle production elastic unitarity only holds below the first threshold $s_0>s\ge4m^2$. It then follow that $\rho(s,t)=0$ below the first $s$-channel Landau curve $4m^2<t<t_1(s)$ (\ref{eq:red}) for any $s$, see red curve in figure \ref{fig:karplus1}. This region however includes the crossed Mahoux-Martin region of positive $\rho(s,t)>0$ (\ref{positiverho}), $s<4 m^2 {(3 t + 4 m^2)^2 \over (t - 4 m^2)^2}$, see gray region in figure \ref{positive_region}.\footnote{Recall that positivity of $\rho(s,t)>0$ is a direct consequence of our assumption about nontriviality of scattering.} Therefore, we conclude that our assumptions of crossing symmetry and scattering without production are not consistent.

Note that in the case of an infinite $J=0$ scattering length, the derivation of the positivity of $\rho(s,t)$ in the Mahoux-Martin region does not always apply. That is because in that case, the integral in (\ref{eq:doubleSD}) fails to converge for $d \geq 5$. However, when the $J=0$ scattering length is infinite, $\rho(s,t)$ is already non-zero at the leading Landau curve and we reach the same conclusion that there must be particle production.

A way to relax the assumption of extended analyticity was explained in \cite{CheungToll} by exploiting polynomial boundedness and continuity assumptions. The idea is to use elastic unitarity to extend the region of analyticity of the amplitude. One starts with fixed $t<4m^2$ dispersion relations and then use elastic unitarity to first extend the analyticity domain of $T_s$ and then use dispersion relations to continue the scattering amplitude itself. The key point being that assuming no production we can use elastic unitarity to continue the discontinuity at arbitrary energy, which is necessary if we want to use this inside the dispersion relations.

Note that the argument above also implies that we must have four-particle production. That is because the crossed region of positivity starts at $s_1(16m^2)={64\over3}m^2<36m^2$, it is enough to assume that $T_{2\to4}=0$ to reach a contradiction. One can wonder if having $T_{2 \to 4}$ is enough to fix the problem, or $T_{2 \to 2 n}$ with $n>2$ are also necessary? To address this question, we can then proceed via crossing.\footnote{Note that crossing has been only rigorously proven within the standard QFT framework for $2\to 2$ amplitudes \cite{Bros:1965kbd}. For some further progress in the multi-particle case $2\to n$ see \cite{Bros:1985gy}.} By unitarity of the $4\to4$ amplitude, ${\rm Im} T_{4 \to 4} \sim |T_{2 \to 4}|^2$, non-vanishing $T_{2 \to 4}$ implies that we have a non-vanishing $T_{4\to 4}$ amplitude. 
Applying crossing this becomes $T_{2 \to 6}$. Continuing this recursion we conclude that all $T_{2 \to 2 n}$ should be non-zero. Therefore, not only scattering implies production but it requires all possible production (here we assumed $Z_2$ symmetry so that only an even number of particles is present in the final state).

An alternative argument that does not use crossing symmetry of higher-point amplitudes was put forward in \cite{Cheung:1968cko}. This argument relies on an unproven assumption that Landau curves can only asymptote to normal thresholds for the scattering of lightest particles, see section \ref{sec:Karplus}. This assumption is consistent with the perturbative analysis of \cite{EdenM,TaylorM,AnomM} and it would be interesting to establish it rigorously.

\subsection{Bounding Inelasticity}

\begin{figure}[t]
\centering
\includegraphics[width=0.9\textwidth]{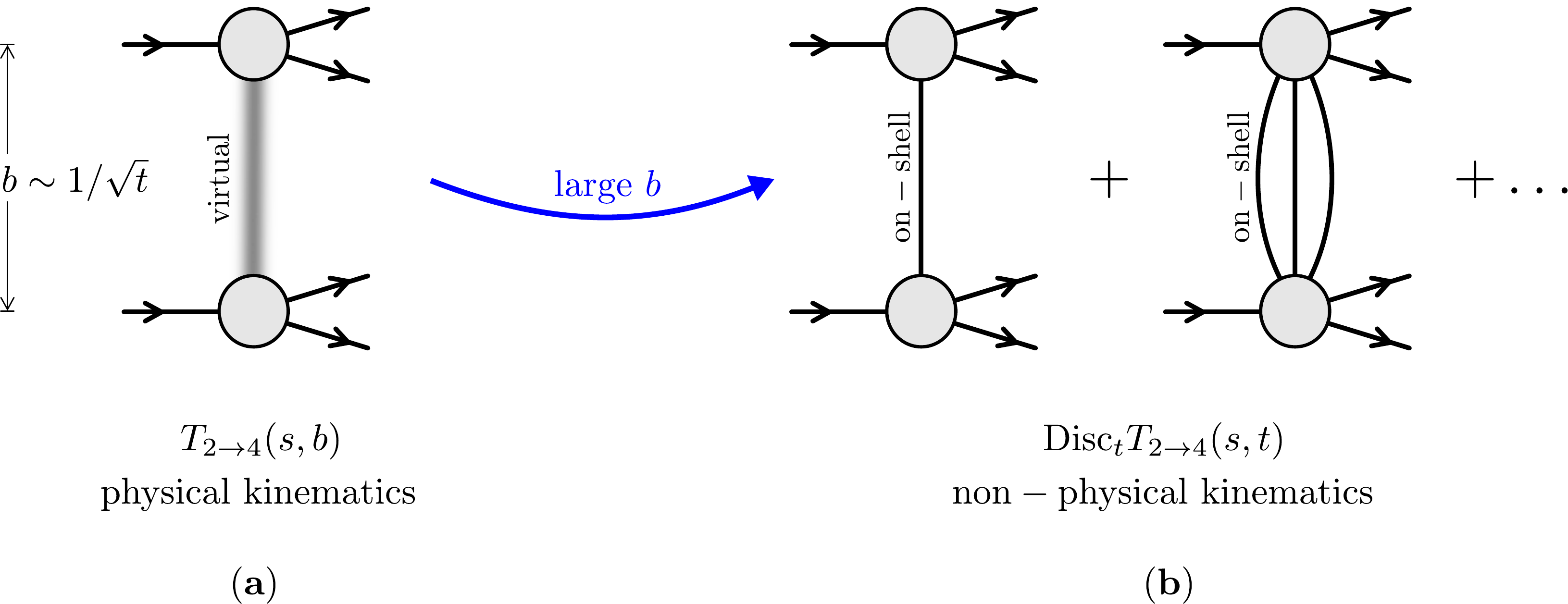}
\caption{\small {\bf a}. We consider a scattering experiment at fixed impact parameters $b$. This process is controlled by the exchanged momentum $t \sim {1 \over b^2} > 0$. {\bf b}. At large impact parameters the amplitude can be organized as a sum over on-shell particles exchange. The dominant contribution comes from the exchanged of the lightest on-shell particle. }
\label{T24}
\end{figure}

There is another, more intuitive way to think about the result of Aks and necessity of particle production in higher dimensions. Ideally, one would like to take a discontinuity of the $2\to4$ amplitude that is given by a product of two $2\to2$ amplitudes. For physical kinematics however, such a discontinuity only exist for the $3\to3$ setup. Instead, let us discuss impact parameter scattering, which is only possible in $d\ge3$. 

As reviewed for example in appendix E of \cite{Camanho:2014apa}, the effect of going to the impact parameter space is the same as continuing the conjugate momentum invariant to the unphysical kinematics. It follows that inelasticity in a gapped theory cannot be exactly zero at very large impact parameters. To see this, decompose the four particles in the final state into a pair of dipoles. Then consider a scattering in which the two dipoles in the final state, as well as the pair of incoming particles in the initial state, are separated by a finite distance $b$ in the transverse space, see figure \ref{T24}.a. Unitarity becomes, in the impact parameter space, the expansion in Yukawa potential suppressed terms 
$T_{2\to4}(s,b)\sim T_{2\to2}^2\times e^{-b m}+[\text{multi-particle}\sim e^{-nb m}]$. At large separation, this expansion is dominated by the one-particle exchange $e^{-b m}$, while the multi-particle corrections are further exponentially suppressed.\footnote{In terms of partial waves, large impact parameter scattering corresponds to the large spin limit and therefore we expect to have inelasticity at large spin which will be analyzed in detail in the sections below.} In that way, a non-trivial (analytically continued) four-point amplitude imply a non-trivial $2\to4$ amplitude. We would then like to bound the $2\to4$ amplitude from below.

There is a convenient way of bounding the integrated discontinuity of $T_{2\to4}$ in the kinematical regime of figure \ref{T24}.b from below. It is based on the discussion in the previous sections, see in particular sections \ref{sec:rho} and \ref{sec:PositivityDSD}, \ref{sec:aks}. It also highlights how the crossing of $\rho(s,t)$ discussed in the section above works microscopically. Let us start with the the square of $T_{2 \to 4}$ that appears in the discontinuity of the $2\to2$ amplitude, $T_s(s,t)$
\be\label{eq:inel24}
T^{\text{inel}, 2 \to 4}_s (s,t) \equiv {1 \over 2} {1 \over 4!} \int \prod_{i=1}^{4} {d^{d-1} \vec{q_i} \  \over (2 \pi)^{d-1} (2 E_{\vec{q_i}})}  (2 \pi)^d \delta^{d}(p_1 + p_2 - \sum_{i=1}^{4} q_i) T_{2 \to 4}^{(+)}(p_1,p_2 | q_i) T_{2 \to 4}^{(-)}(q_i | p_3, p_4)\ .
\ee
By construction, the unitarity integral in the right-hand side of (\ref{eq:inel24}) depends only on $s$ and $t$. For physical scattering we consider $s>16 m^2$ and $t<0$.

We would like next to analytically continue (\ref{eq:inel24}) to the unphysical Martin-Mahoux region discussed above
\be
\label{eq:MMcrossed}
4 m^2 < t<16m^2\ ,\qquad {16 m^2 t \over t - 4 m^2} < s < 4 m^2 {(3 t + 4 m^2)^2 \over (t-4 m^2)^2}\ .
\ee
We also would like to consider $16 m^2 < s <36 m^2$ to focus on the $T_{2 \to 4}$ amplitude. This condition together with (\ref{eq:MMcrossed}) imply ${36 m^2 \over 5}<t$. For the $2\to2$ case we reviewed the procedure in the sections above. For the multi-particle case it was discussed in \cite{Gribov:1962ft,Lardner1,Lardner2,Kim}.

After taking a discontinuity in $t$ and using crossing symmetry of the double spectral density, we arrive at the following schematic form
\be\label{eq:identityCr}
\rho(s,t)= {(t - 4 m^2)^{d-3\over2} \over 4\pi^2 (4\pi)^{d-2} \sqrt{t} }&\int\limits_{\bar z_1}^{\infty} d \eta'  \int\limits_{\bar z_1}^{\infty} d \eta''\,{\rm Disc}_s\cT_{2\to2}^{(+)}(t, \eta')\,{\rm Disc}_s\cT_{2\to2}^{(-)}(t, \eta'')\times {\rm Disc}_{\bar z} K_d(\bar z, \eta', \eta'')\nn\\
=&\int d {\rm LIPS}_4\times\, {\rm Disc}_t T_{2 \to 4}^{(+)}\, {\rm Disc}_t T_{2 \to 4}^{(-)}\times K_{\text{Mandelstam}}^{2 \to 4}\ ,
\ee
where in the formula above we switched to $\bar z = 1 + {2 s \over t-4m^2}$ and $\bar z_1= 1 + {8 m^2 \over t-4 m^2}$. Here in the right-hand side each ${\rm Disc} T_{2 \to 4}$ contains a delta-function that puts the exchanged particle in figure \ref{T24} on-shell. The phase space integral $d {\rm LIPS}_4$ should be understood in terms of the analytic continuation a-la Mandelstam, and we will discuss it in more detail elsewhere \cite{paper3}.

Note that since we are in the Mahoux-Martin region, the partial wave expansion of ${\rm Disc}_s\cT^{(\pm)}_{2\to2}(t,\eta)$ converges and 
both the Mandelstam kernel and the Legendre polynomials that enter into partial waves are positive. Therefore we can write
\be
\label{eq:boundAmpunph}
T_s(t,s) = \sum_{J=0}^{\infty} n_J^{(d)} {\rm Im} f_J(s) P_J(1+{2 t \over s - 4 m^2}) \geq \sum_{J=0}^{J_0} n_J^{(d)} {\rm Im} f_J(s) P_J(1+{2 t \over s - 4 m^2})\ .
\ee
By plugging (\ref{eq:boundAmpunph}) into (\ref{eq:identityCr}) and using positivity of the Mandelstam kernel ${\rm Disc}_{\bar z} K_d(\bar z, \eta', \eta'')$ we arrive at the lower bound on the integrated discontinuity ${\rm Disc}_t T_{2 \to 4}$.

While this argument bounds the discontinuity of $T_{2\to 4}$ in the unphysical kinematics of figure \ref{T24}.b, for ${36 m^2 \over 5} \leq t < 16 m^2$ from below, it does not tell us anything about $T_{2 \to 4}$ in the physical kinematics $t<0$. Let us however comment why one expects to have $T_{2 \to 4}$ of the same order also in the physical kinematics. Schematically, we can write the following representation of the $2\to4$ impact parameter amplitude
\be
T_{2 \to 4}(s,b) = (T_{2 \to 2})^2 e^{- b m} + \int\limits_{3 m}^\infty d M \rho(M) e^{-M b} .
\ee
We can rewrite this as follows
\be
| T_{2 \to 4}(s,b) - (T_{2 \to 2})^2 e^{- b m} | &\leq c_{\text{MP}}\, e^{- 3 m b}\ ,\qquad c_{\text{MP}} = \int\limits_{3 m}^\infty dM\, | \rho(M) | e^{-(M-3 m) b}\ ,
\ee
where $c_{\text{MP}}$ encodes the contribution of the multi-particle exchanges.

Without extra fine-tuning we expect that in a strongly coupled theory $T_{2 \to 2}, c_{\text{MP}} \sim\cO(1)$ and therefore the one-particle exchange to dominate for $b \gtrsim {1 \over m}$. This would make $T_{2 \to 4} \sim O(1)$. On the other hand, making $T_{2 \to 4} \ll 1$ given $T_{2 \to 2} \sim O(1)$ would require a very fine-tuned cancellation between the exchange of one-particle state and the multi-particle state as well as $c_{\text{MP}} \gg 1$. It seems quite possible that such a scenario is not consistent with multi-particle unitarity, but it is very hard to show it explicitly due to the complexity of the latter.

One scenario in which the regime of single particle dominance can be delayed to arbitrary large impact parameters is if the theory contains extended objects, such as strings. In such case we can choose $l_{\text{string}} \gg {1 \over m}$, and the one-particle exchange is expected to dominate only for $b \gtrsim l_{\text{string}}$. In this case however, the spectrum is expected to contain particles of mass ${1 \over l_{\text{string}}} \ll m$ which contradicts our assumption about $m$ being the lightest particle.

In some numerical applications that we discuss in more detail below, see e.g. \cite{Paulos:2017fhb}, it was observed that small $J$ partial wave converge very well.
Therefore we can set in the formula (\ref{eq:boundAmpunph}) $J_0 = 2$ and then use it in (\ref{eq:identityCr}). We can then evaluate the integral above to rigorously bound from below ${\rm Disc} T_{2\to 4}$ in the Mahoux-Martin kinematical region. In the numerical analysis when maximizing some couplings, one usually finds ${\rm Im} f_0(s) \sim 1$ therefore the formulas (\ref{eq:identityCr}), (\ref{eq:boundAmpunph}) will produce ${\rm Disc} T_{2 \to 4} \sim 1$. It is however still an open problem to translate this fact into a rigorous statement about $T_{2 \to 4}$ for physical kinematics.

Independently of the discussion above, in section \ref{sec:boundinelasticity} we bound inelasticity using an additional input about the discontinuity of the $2 \to 2$ amplitude.

\section{Threshold Expansion}\label{THboot}

\begin{figure}[t]
\centering
\def\svgwidth{16cm}
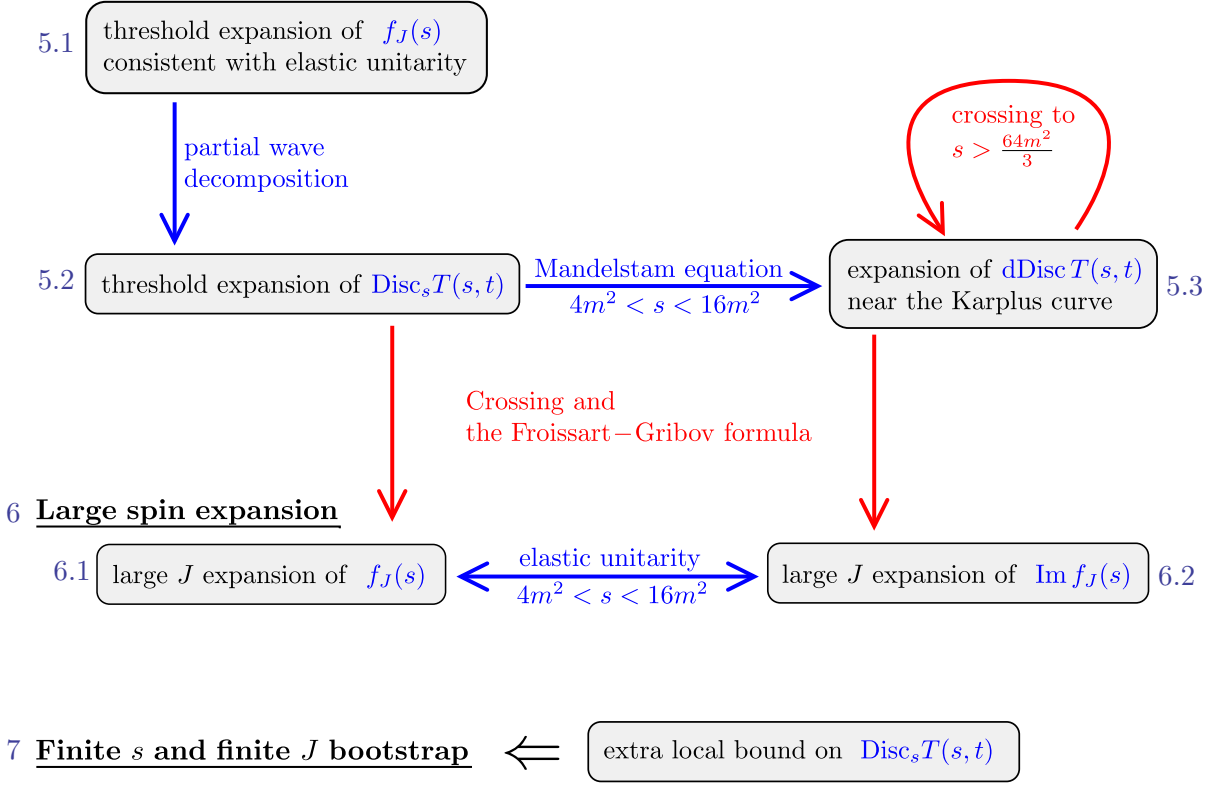
\caption{\small The logical structure of sections \ref{THboot}, \ref{LJboot} and \ref{sec:boundinelasticity}. We start in section \ref{THboot} by developing the idea of the threshold expansion. We solve elastic unitarity for $f_J(s)$ close to $s=4m^2$. This naturally leads to the threshold expansion both for the discontinuity of the amplitude $T_s(s,t)$ and its double discontinuity $\rho(s,t)$. In section \ref{LJboot} we use the Froissart-Gribov formula and crossing symmetry of $\rho(s,t)$ to translate the threshold expansion of discontinuity $T_s(s,t)$ into the large $J$ expansion of $f_J(s)$ and the threshold expansion of $\rho(s,t)$ into the large $J$ expansion of ${\rm Im}f_J(s)$ correspondingly. In section \ref{sec:boundinelasticity} we give an error bound on these expansion in terms of an extra phenomenological bound on the discontinuity of the amplitude.}
\label{thresholdfig}
\end{figure}

To further analytically constrain a nonperturbative amplitude a small parameter is needed. In this section, as well as the next one, we will study the expansions of the amplitude in two kinematical small parameters, as well as the relation between the two expansions.

One kinematical small parameter that always exists in a gapped theory is the 
energy distance from the two-particle thereshold in units of the mass gap, 
\beq
\sigma_s= {s \over 4 m^2} - 1 = {\vec p^2 \over m^2}\ ,
\eeq
where in the last equality we evaluated $\sigma_s$ in the center-of-mass frame (\ref{stu}).

The threshold expansion is, thus, an expansion in powers of $\sigma_s$. The small $\sigma_s$ expansion is  known in nuclear physics as  {\it the effective range expansion} \cite{Bethe:1949yr}. As we take $\sigma_s \ll 1$ or, equivalently, $|\vec p| \ll m$ scattering becomes non-relativistic and we can characterize it by some effective potential whose properties are captured by the threshold expansion parameters \cite{Reid:1968sq}.  In this section we 
use elastic unitarity, extended analyticity and crossing to argue that both the discontinuity and the double discontinuity of the scattering amplitude admit a natural threshold expansion. Importantly, the parameters that enter the threshold expansion are controlled by the low energy physics which is well-known from the experiments or lattice simulations, see e.g.  \cite{Colangelo:2001df}.

Whenever a theory has a conserved charge, there is a natural small parameter -- one over the charge of states that are being exchanged. In the case at hand, the only symmetry we assume is Lorentz symmetry. Correspondingly, the only available conserved charge is the spin. Indeed, in the next section we will show that the partial wave coefficients admit a systematic large $J$ expansion. Moreover, by combining the Froissart-Gribov formula with crossing we will relate the threshold expansion to the large spin expansion.

The logical structure and steps of 
this and following sections is summarized in figure \ref{thresholdfig}. We start by constructing a threshold expansion of the partial wave coefficients that is consistent with elastic unitarity.

\subsection{Threshold Expansion of $f_J(s)$}\label{Thresholdexp}

To expand the the partial wave coefficients close to the threshold in a way that is consistent with elastic unitarity, it is first convenient to solve the latter in a different fashion than (\ref{eq:elunPW}). After dividing by $|f_J(s)|^2$, the elastic unitarity condition (\ref{eq:EUfJ}) takes the form
\beq\label{eq:elunThr}
2{\rm Im }{1\over f_J(s)} = {1 \over i} \left( {1 \over f_{J}(s+i \eps)} - {1 \over f_{J}(s-i \eps)}  \right)=-{(s-4 m^2)^{d-3\over2}\over\sqrt s}\times\left\{\begin{array}{lc}1&4m^2<s<16m^2\\ 0&0<s<4m^2\end{array}\right.\ .
\eeq
Hence, $1/f_J$ has a branch cut at threshold for even $d$ and a logarithmic cut for odd $d$. The general solution to (\ref{eq:elunThr}) combined with real analaticity (\ref{eq:realanalytPW}) takes the form
\beq\label{eq:elasticunitarityFJS}
{1 \over f_J(s + i \eps) } =  b_J(\sigma_s) - {i \over 2} {(4 m^2\sigma_s)^{d-3\over2} \over \sqrt{s}}\times\left\{\begin{array}{ll}1&d\text{ even}\\
{i\over \pi}[\log\sigma_s-i\pi]&d\text{ odd}\end{array}\right.\ ,
\eeq
where $b_J(\sigma_s)$ is a real analytic and single-valued function in some finite neighborhood around the origin, except for potentially isolated singularity at $\sigma_s=0$. Note that in writing (\ref{eq:elasticunitarityFJS}) we used the continuity assumption for $f_J(s)$, see assumption 5 at the introduction and section \ref{sec:continuity}. 

As we have learned from the Froissart-Gribov representation (see (\ref{eq:thresholdFJFG})), as $\sigma_s\to0^+$ from above, $f_{J}(s)\to a_J\sigma_s^J$ (\ref{eq:scatteringlengths}). In terms of the function $b_J(\sigma)$, this behavior is
\be
\label{eq:equationforBJ}
b_{J}(\sigma)={m^{d-4}\over a_J \sigma_s^J}+\cO(\sigma_s^{1-J})
\ ,\qquad J\ge2\ ,
\ee
where we have assumed that $J_0(4m^2)<2$. For $J=0$ we do not expect the Froissart-Gribov representation to hold close to $\sigma_s=0$. Therefore we do not have a similar prediction. After factoring out the leading threshold singularity of $b_{J\ge2}(\sigma)$, it has a regular expansion 
\beq\label{eq:thresholdBJ}
b_{J\ge2}(\sigma_s)= {m^{d-4} \over \sigma_s^J} \left( {1 \over a_J} + \sum_{i=1}^{\infty} b_{J,i} \sigma_s^i \right)
\qquad\text{and}\qquad
b_0(\sigma_s)= {m^{d-4} \over \sigma_s^{\tilde J_0}} \sum_{i=0}^{\infty} b_{0,i}\, \sigma_s^i  \ ,
\eeq
where $\tilde J_0$ is an unconstrained integer at this point and $a_J$, $b_{J,i}$ are real coefficients. Reality of scattering lengths $a_J$ and effective ranges $b_{i,J}$ follows from real analyticity property of the partial waves coefficients (\ref{eq:realanalytPW}). For actual values of these parameters in various QCD processes see e.g. \cite{Nagels:1979xh}.

Similarly, using real analaticity of $b_J(s)$, we have the following equations for the imaginary part of the partial wave in the elastic unitarity region
\be\label{eq:elasticunitPWbJ}
{\rm Im} f_J(s) &= {(4m^2\sigma_s)^{d-3\over2} \over 2 \sqrt{s}}\[ b_J^2(\sigma_s) + {(4 m^2\sigma_s)^{d-3} \over 4s}\]^{-1}\qquad\qquad\qquad\qquad\qquad\qquad\quad\  d\text{ even}\ , \nn \\
{\rm Im} f_J(s) &={(4m^2\sigma_s)^{d-3\over2} \over 2 \sqrt{s}}\[ \left(b_J(\sigma_s) + {(4 m^2\sigma_s)^{d-3\over2} \over 2\pi\sqrt{s}} \log\sigma_s \right)^2 +  {(4 m^2\sigma_s)^{d-3} \over 4s} \]^{-1}\qquad d\text{ odd}\ .
\ee

To summarize, by plugging (\ref{eq:thresholdBJ}) into (\ref{eq:elasticunitarityFJS}) and (\ref{eq:elasticunitPWbJ}) we have constructed a threshold expansion of $f_J(s)$ and ${\rm Im}f_J(s)$ that automatically solves elastic unitarity for any real coefficients. 

Note that the most singular possible threshold behavior is completely fixed by elastic unitarity, with no free coefficient. It comes from $J=0$ and corresponds to setting $b_0(\sigma_s)=0$ in (\ref{eq:elasticunitarityFJS}), so that the partial wave is dominated by the universal term 
\beq\label{eq:asympf0}
{\rm Im} f_0 (s) =  {2^{5-d} m^{4-d} \over  \sigma_s^{{d-3 \over 2}} }\times\left\{\begin{array}{ll}1+\dots&\qquad d\text{ even} \\
{\pi^2 \over  \log^2 \sigma_s + \pi^2} +\dots&\qquad d\text{ odd}\end{array}\right. \ .
\eeq
In particular, this implies that the spin zero scattering length is infinite. 
In $d=3$ the threshold behavior (\ref{eq:asympf0}) is realized in massive $\phi^4$ theory \cite{Bros:1998tt}. In $d=4$ the asymptotic behavior (\ref{eq:asympf0})  appeared in the study of coupling maximization in \cite{Paulos:2017fhb} and corresponds to having a bound state at $s=4 m^2$.\footnote{By tuning the parameters of the theory it could be possible to reach this situation in QCD as well. We thank Mattia Bruno and Maxwell Hansen for the discussion.} For $J\geq 2\ge J_0(s)$ the singular behavior (\ref{eq:asympf0}) does not occur due to (\ref{eq:thresholdFJFG}).

Below we will also be interested in the case of finite ${\rm Im} f_0(s)$. 
In this case, the spin zero scattering length is finite, namely $b_0(s) = {m^{d-4} \over a_0}$, and we get the following leading behavior of ${\rm Im} f_0 (s)$
\be
\label{eq:finiteSCL0}
{\rm Im} f_0 (s) =2^{d-5}  m^{4-d}  a_0^2\, \sigma_s^{{d-3 \over 2}} +\dots\ ,\qquad d>3 \ .
\ee
As a final remark, note that the analysis of this section can also be generalized to non-integer spin, see appendix \ref{sec:nonintspinTh}.

\subsection{Threshold Expansion of Discontinuity}\label{THDisc}

For our purposes we will be interested in a 
closely related expansion, that of the discontinuity of the amplitude $T_s(s,t)$ for $t>4m^2$ and $\sigma_s \to 0$. As we will see,  the Froissart-Gribov formula (\ref{eq:FGform}) relates this expansion to the large $J$ behavior of the partial wave coefficients.

We proceed by considering the $s$-channel partial wave expansion for the discontinuity of the amplitude
\be
\label{eq:expagain}
T_s (s,t) 
=n_{J=0}^{(d)} {\rm Im} f_0 (s) + \sum_{J=2}^\infty n_J^{(d)} {\rm Im} f_J (s) P_J^{(d)}\left(1+{2 t \over s-4 m^2} \right) .
\ee
For $s$ close to $4m^2$, the sum over $J$ above converges inside the large $s$-channel Lehmann ellipse which is for $t< {16 m^2 s \over s-4 m^2}$. Given some fixed $t$, and considering the limit $s \to 4 m^2$ we stay within the convergence region. 

We can now plug (\ref{eq:elasticunitPWbJ}) into (\ref{eq:expagain}) and perform the threshold expansion under the sum. We get
\be
\label{eq:thresholdexpansion}
T_s (s,t) =n_{0}^{(d)} {\rm Im} f_0 (s) + n_{2}^{(d)} a_2^2 m^{4-d} \left({t \over m^2} \right)^2 {2^{d-7} (d-1) \over (d-2) } \sigma_s^{2+{d-3 \over 2}} + \dots  \ , 
\ee
where we can systematically expand $T_s (s,t)$ using the threshold expansion of partial waves (\ref{eq:thresholdBJ}) which is based on elastic unitarity. Note that the expansion parameter in even dimensions is simply $\sigma_s$, whereas in odd dimensions we have in addition powers of $\log \sigma_s$, as well as inverse powers ${1 \over \log \sigma_s}$, which we do not write here explicitly. 

Let us emphasize that in the argument above it was absolutely crucial to consider the discnontinuity of the amplitude $T_s (s,t)$ and not the amplitude $T(s,t)$ itself. Indeed, if we were to try repeating the argument above for $T(s,t)$ itself, we would run into the following problem. In (\ref{eq:expagain}) we have partial waves $f_J(s) \sim (s-4m^2)^J$ and Legendre polynomials behaving as $ P_J^{(d)}\left(1+{2 t \over s-4 m^2} \right) \sim (s-4m^2)^{-J}$. Therefore the expansion for $s \to 4 m^2$ requires re-summation of partial waves of all spins which is beyond our control.

The conclusion is that $T_s (s,t)$ admits a systematic threshold expansion in terms of the  solution to elastic unitarity in the $s$-channel. Moreover, the contribution of higher spin partial waves are suppressed by an additional factor of $\sigma_s^J$. We expect the threshold expansion converges as long as we stay below the leading Landau curve, namely for $t< {16 m^2 s \over s-4 m^2}$. Luckily for us, it is the discontinuity of the amplitude and not the amplitude itself that enters the Froissart-Gribov inversion formula. Therefore the results of this section can be readily put to use.

\subsection{Threshold Expansion for Double Spectral Density}
\label{sec:thresholdDSD}

We now turn to the study the double spectral density (\ref{dsd}) close to the leading Landau curve. 
We consider the Mandelstam equation (\ref{eq:doubleSD}) that is repeated here for convenience 
\be
\label{eq:doubleSDb}
\rho (s,t) &=  {(s - 4 m^2)^{d-3\over2} \over 4\pi^2 (4\pi)^{d-2} \sqrt{s} }  \int\limits_{z_1}^{\infty} d \eta'  \int\limits_{z_1}^{\infty} d \eta''\, \cT_t^{(+)}(s, \eta')\, \cT_t^{(-)}(s, \eta'')\, {\rm Disc}_z K(z, \eta', \eta'')\ ,
\ee
where $4 m^2 < s < 16 m^2$. The crucial observation is that when we are very close to the leading Landau curve $t = {16 m^2 s \over s-4m^2}$, the integral in (\ref{eq:doubleSDb}) is again controlled by the threshold expansion of $\cT^{(+)}_t(s,\eta')$ and $\cT_t^{(-)}(s,\eta'')$. The reason for this is that the region of integration starts at the threshold, $\eta',\eta''\ge z_1$, and ends at the boundary of support of the kernel, 
$\eta_+(\eta',\eta'') \leq z $. As $t$ approach the leading Landau curve, this region shrinks close to the threshold for both $\eta'$ and $\eta''$ integrals, see figure \ref{etarange}.

\begin{figure}[t]
\centering
\includegraphics[width=0.4\textwidth]{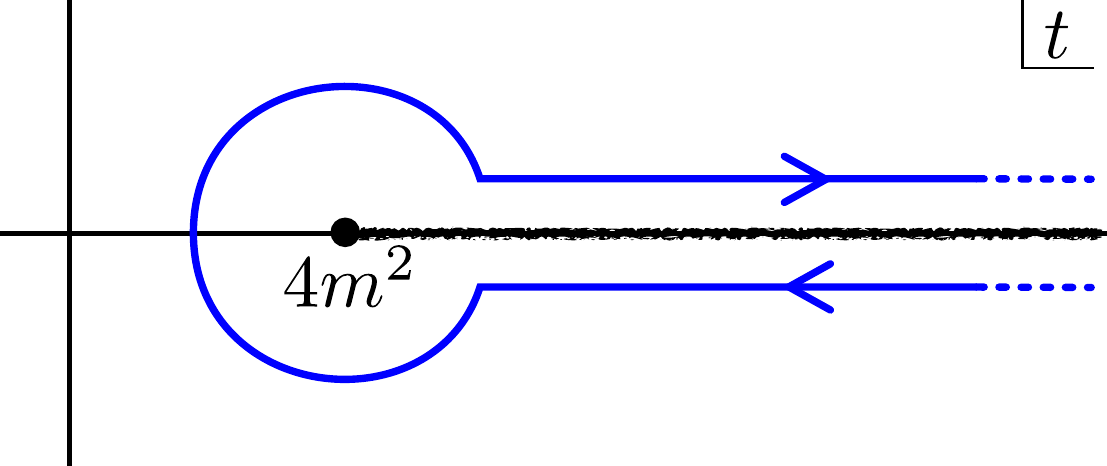}
\caption{\small The discontinuity of the Mandelstam equation (\ref{eq:doubleSD}) has been obtained by deforming the contour of integration in the finite integral (\ref{rhoTT}). Hence, the result cannot diverge. This means that an apparent divergence in (\ref{eq:deftilde}) that comes from the region of integration close to the threshold should be understood as a finite keyhole contour integral plotted here.}
\label{keyholefig}
\end{figure}

A convenient small parameter in the problem is the dimensionless distance from the Landau curve, see (\ref{Landau1})
\beq
\delta z\equiv z -(2 z_1^2 - 1)   \propto t - {16 m^2 s \over s - 4m^2}\ .
\eeq
We can now plug the threshold expansion of the discontinuity (\ref{eq:thresholdBJ}) into (\ref{eq:doubleSDb}) and expand the result in powers of $\delta z$. At any given order in that expansion only finite number of terms in the threshold expansion of the discontinuity contribute. For even spacetime dimension, the relevant integrals that appears in the expansion are 
\be
\label{eq:deftilde}
\tilde I_{n_1 ,n_2}^{(d)}(z) \equiv \int\limits_{z_1}^{\infty} d \eta' \int\limits_{z_1}^{\infty} d \eta''  \sigma_{t(\eta')}^{n_1 - {d-3 \over 2}}  \sigma_{t(\eta'')}^{n_2 - {d-3 \over 2}} {\rm Disc}_z K(z, \eta', \eta'')\ .
\ee
To leading order in $\delta z$ this integral is given by\footnote{To derive (\ref{eq:Iintthrrho}) it is useful to do the following change of integration variables close to $\delta z=0$
\be
\eta' &= z_1 + {\delta z \over 2 z_1} \alpha x\ ,\qquad \eta'' = z_1 + {\delta z \over 2 z_1} \alpha (1-x)\ ,
\ee
where the integration in (\ref{eq:doubleSDb}) for $\delta z \ll 1$ is restricted to $0 \leq x$ and $\alpha \leq 1$.}
\be
\label{eq:Iintthrrho}
\tilde I_{n_1,n_2}^{(d)}(z) =2 { \pi^{{d+1 \over 2}} (z_1-1)^{d-3-n_1-n_2} \over z_1^{n_1+n_2+1} (z_1^2-1)^{1/2}} \left( {\delta z \over 2} \right)^{n_1+n_2+{5-d \over 2}} {\Gamma(n_1+{5-d \over 2})\Gamma(n_2+{5-d \over 2}) \over \Gamma(n_1+n_2+{7-d \over 2})}\(1 +\cO(\delta z)\)\ .
\ee
For $n_1,n_2\le {d-5 \over 2}$ the integral (\ref{eq:deftilde}) diverges, but this apparent divergence is not physical. Namely, the integral in (\ref{eq:Iintthrrho}) originates from the contour integral that wraps around $\eta', \eta'' = z_1$. Therefore, the integral can be safely deformed to a keyhole contour around the dangerous region, see figure \ref{keyholefig}. 
The result of the keyhole integration is equivalent to analytically continuing (\ref{eq:Iintthrrho}) in $m,n$ (as long as the final result is finite). 

We see that indeed, higher powers in the threshold expansion result in higher powers of $\delta z$. For odd spacetime dimension, due to the presence of ${1 \over \log^2 \sigma_s + \pi^2}$ in (\ref{eq:elasticunitPWbJ}), the relevant integral is more complicated, but the power suppression in $\delta z$ is of course the same. We discuss it more in appendix \ref{sec:dsdOddD}.

The most singular possible behavior close to the Landau curve comes from the corresponding universal threshold behavior of ${\rm Im}\,f_0(t)$ in (\ref{eq:asympf0}) and corresponds to $n_1=n_2=0$
in (\ref{eq:deftilde}). In that case we find
\be
\rho(s,t) =  (2^{5-d} n_0^{(d)})^2 
\frac{(d-3) \pi ^{\frac{3-d}{2}}  m^{4-d} (z_1-1)^{\frac{d-3}{2}}}{32 (5-d) \cos \left(\frac{\pi  d}{2}\right) \Gamma \left(\frac{d-1}{2}\right)  z_1 (z_1+1)} \delta z^{\frac{5-d}{2}} + \dots  \ ,\qquad d\text{ even} \ .
\ee
On the other hand, for the case where the spin zero scattering length is finite (\ref{eq:finiteSCL0}), we have $n_1=n_2=d-3$
in (\ref{eq:deftilde}) and correspondingly the approach to the Landau curve is much softer
\be
\rho(s,t) = 
\frac{a_0^4 \pi^{\frac{1-d}{2}} m^{4-d} (z_1-1)^{\frac{9- 3d}{2} } z_1^{5-2 d} \Gamma \left(\frac{d-1}{2}\right)^2 (n_0^{(d)})^2}{512 (z_1+1) \Gamma \left(\frac{3 d - 5}{2}\right)} \delta z^{\frac{3 d - 7}{2}} + \dots  \ .
\ee

\subsection{Radius of Convergence}

Let us briefly discuss the radius of convergence of the threshold expansion introduced above. As usual the radius of convergence is controlled by the analytic properties of the function at hand.

Consider first $f_J(s)$. It has a normal threshold cut starting at $s=4m^2$ and multi-particle cuts at $s \geq 16 m^2$, as well as the $u$-channel cut for $s \leq 0$. In addition to that we can have bound states and resonances (which correspond to poles on the second sheet). In a given theory, we expect the singularity that is the closest to $s=4m^2$ to control the convergence radius of the threshold expansion of $f_J(s)$. 
In this paper we assume for simplicity that there are no bound states, however it should be easy to include them in the analysis. 

Consider next $T_s(s,t)$. In this case, the partial wave expansion (\ref{eq:expagain}) converges below the first Landau curve. For $4m^2<s<16m^2$ and $t>16m^2$, that is the regime of $\sigma_s < {16 m^2 \over t-16m^2}$. 
Moreover, if we are to first expand each ${\rm Im} f_J (s)$ close to the threshold under the sum, we will have to argue that the sum over $J$ and the threshold expansion commute. It is not clear to us how to do it. Instead below we adopt a different approach. We separate a few low spin partial waves in (\ref{eq:expagain}) and apply the threshold expansion to them. We then bound the sum over spins without doing the threshold expansion under the sum over spins.

\section{Large $J$ Expansion}\label{LJboot}

In this section we map the threshold expansion developed in the previous section to the large $J$ expansion of the partial waves coefficients $f_J(s)$.
\footnote{In $d=4$ this was pioneered by Dragt \cite{Dragt}, see also \cite{Streit}, and further developed by A.~W.~Martin in \cite{Martin:1969yw}. Here we consider a slightly more general form of the amplitude, as well as generalize the analysis to any spacetime dimension.}

The basic idea is very simple. The Froissart-Gribov formula (\ref{eq:FGform}) directly maps the threshold expansion of the discontinuity of the amplitude (\ref{eq:thresholdexpansion}) to the large $J$ expansion of the partial wave coefficients in the crossed channel.\footnote{This is analogous to the analytic bootstrap in CFTs \cite{Fitzpatrick:2012yx,Komargodski:2012ek,Caron-Huot:2017vep}.} Interestingly, once combined with crossing, the Froissart-Gribov formula allows us to map the low energy threshold expansion, that is dominated by the low spins, to the large spin behavior of the partial wave coefficients for any energy and in the same channel. 
In particular, it automatically predicts the amount of inelasticity at large $J$.

In this section we will use the results for the threshold expansion of the double spectral density and the discontinuity of the amplitude in the crossed $t$-channel as opposed to the $s$-channel threshold expansion of the previous section. We hope this will not cause any confusion. In sections \ref{sec:largespinF} and \ref{sec:imFcomp} we work out the leading large $J$ expressions for $f_J(s)$ and ${\rm Im} f_J(s)$ correspondingly. In section \ref{sec:Jexact} we compute the exact contribution of a given term in the threshold expansion of the amplitude to the partial waves. This includes infinitely many ${1 \over J}$ corrections to the results of \ref{sec:largespinF} and \ref{sec:imFcomp}.

Note that in this section we only concern ourselves with the threshold expansion of the amplitude close to the two-particle threshold. Of course, in addition there are contributions from multi-particle thresholds. These lead to further nonperturbative in ${1 \over J}$ corrections to the results of this section and are potentially important if we would like to discuss partial waves at finite $J$ which is the subject of the next section.

\subsection{Large Spin Expansion of $f_J(s)$}
\label{sec:largespinF}

We start by analyzing the Froissart-Gribov integral (\ref{eq:FGform}) in the large spin limit $J \gg 1$. For convinience, we quoting the integral here
\beq
\label{eq:FG2}
 f_J(s) ={2 \, {\cal N}_d\over\pi}\int\limits_{z_1}^\infty d z \,(z^2-1)^{{d-4 \over 2}}Q^{(d)}_J(z) T_t(s,t(z)) \ ,\qquad {\rm Re}[J]>J_0(s)\ .
\eeq
The large $J$ behavior of $Q_J^{(d)}(z)$ is given by\footnote{An efficient way to systematically expand $Q_J^{(d)}(z)$ to an arbitrary order in the large $J$, fixed $z$ expansion is to start from its representation in terms of the hypergeometric function \cite{largeQ}
\be\label{QexpA2}
Q^{(d)}_J(z)=2^{d-4}\sqrt\pi{\Gamma\(\tfrac{d-2}{2}\)\Gamma(J+1)\over\Gamma\(J+{d-1\over2}\)}{\lambda(z)^{-J}\over(\lambda(z)^2-1)^{d-3\over2}}\, _2F_1\left(1-\tfrac{d-3}{2},\tfrac{d-3}{2};J+\tfrac{d-1}{2};\tfrac{1}{1-\lambda(z)^2}\right)\ ,
\ee 
and then use the series representation for the hypergeometric function as $\, _2F_1(a,b;c;x)=\sum_{k=0}^\infty \frac{ (a)_k (b)_k}{k! (c)_k}x^k$. The spin, $J$, only enters in the Pochhammer $(c)_k=(J+{d-1\over2})_k$ and therefore the $k$'th term only contributes at order $1/J^k$ and higher. 
Note that for $d=3$ there are no ${1 \over J}$ corrections to (\ref{QexpA}). More generally, in odd $d$ the large $J$ properties of $Q^{(d)}_J(z)$ can be made manifest, see appendix \ref{sec:Q}.}
\be\label{QexpA}
Q^{(d)}_J(z)=2^{d-4} \sqrt{\pi}\, {\Gamma\(\tfrac{d -2}{2}\)\over J^{{d-3 \over 2}}} { \lambda(z)^{-J} \over  (\lambda(z)^2-1)^{d-3\over2} } \left(1 +  \cO(1/J) \right)\ ,
\ee
where 
\beq
\lambda(z) \equiv z+\sqrt{z^2 -1}=e^\theta\qquad\text{for}\qquad |z|>1\ ,\quad {\rm Re}\,\theta>0\ .
\eeq

Crucially, $Q^{(d)}_J(z)$ 
decays very fast for fixed $z>1$ at large $J$. The integral is therefore dominated by the region close to the threshold $z_1$. 
Explicitly, for $J \gg 1$ we have
\be
\label{eq:largeJqExp}
Q^{(d)}_J\left(z_* + {\sqrt{z_*^2-1} \delta z \over J} \right) = 2^{d-4} \sqrt{\pi}\, {\Gamma\(\tfrac{d -2}{2}\)\over J^{{d-3 \over 2}}} { \lambda(z_*)^{-J} \over  (\lambda(z_*)^2-1)^{d-3\over2} } e^{- \delta z}\(1 +\cO(1/J)\) \ ,
\ee
and therefore the integral in (\ref{eq:FG2}) is controlled by the region of the size $\sim {1 \over J}$ close to $z_*=z_1$. In that way, the large $J$ behavior is controlled by the threshold expansion of $T_t(s,t(z))$ established in the previous section in a manifest way.

Next, we explicitly plug the threshold expansion  into (\ref{eq:FG2}) and compute the leading large $J$ behavior of the partial wave coefficients. For that aim, it is convenient to express the integral as follows
\be\label{eq:largeJF}
f_J(s)\equiv{{\cal N}_d\over\sqrt\pi}{\Gamma\(\tfrac{d -2}{2}\)\over J^{{d-1 \over 2}}} { (\lambda(z_1)^2-1)^{{d-3\over2}} \over \lambda(z_1)^{J+d-3}  } \hat f_J(s) \ ,\qquad
\hat f_J(s)= \int\limits_0^\infty d \delta z\,  T_t(s,t(\delta z, J) )\,e^{- \delta z}\(1+\cO(1/J)\)\ ,
\ee
where we have\footnote{Note that $\delta z$ in (\ref{eq:sigmatlargeJ}) is different from the one used in the previous section. }
\be
\label{eq:sigmatlargeJ}
\sigma_t = {1 \over J} \left( {z_1 + 1 \over z_1 -1}  \right)^{1/2} \delta z\ .
\ee
Note that in deriving (\ref{eq:largeJF}) we took into account the integration measure $(z^2-1)^{{d-4 \over 2}}$ and the Jacobian from switching the integration variable to $\delta z$.

After performing the crossing transformation $s \leftrightarrow t$, we can now plug the leading threshold expansion expressions from the previous section (\ref{eq:thresholdexpansion}) to get the $J \gg 1$ limit of $f_J(s)$.
Let us start with the universal and most singular case (\ref{eq:asympf0}). For this case we get

\beq\label{fJlargeJ}
\hat f_J(s)= 2^{5-d} n_{0}^{(d)} m^{4-d}  J^{{d-3 \over 2}} \left({z_1 -1 \over z_1 + 1} \right)^{{d-3 \over 4}}\(1+\cO(1/J)\)\times\left\{
\begin{array}{ll}
\pi^2 g_{{d-3 \over 2}} \left(\log J   \sqrt{{z_1 - 1 \over z_1 +1}} \right)\,&\qquad d \text{ odd}\\
\quad \Gamma \left(\tfrac{5 -d}{2} \right) \quad&\qquad d \text{ even}
\end{array}\right.\ ,
\eeq
where the explicit form of the slowly-varying function $g_{{d-3 \over 2}}(\log x)$  is computed in appendix \ref{sec:keynoleint}.

If, on the other hand, we consider the situation with a finite spin zero scattering length (\ref{eq:finiteSCL0}), we get
\be
\hat f_J(s) &=2^{d-5} n_{0}^{(d)} a_0^2\,\Gamma \left(\tfrac{d-1}{2} \right) m^{4-d} {1 \over J^{{d-3 \over 2}} }  \left({z_1 + 1 \over z_1 - 1} \right)^{{d-3 \over 4}}\ \(1+\cO(1/J)\) .
\ee

It is clear that we can systematically include ${1 \over J}$ correction in this expansion. Note that in all cases, the leading large $J$ result for $f_J(s)$ is exponentially small and purely real. 
In the next section we compute the leading $J$ contribution to 
${\rm Im} f_J(s)$.

\subsection{Large Spin Expansion of ${\rm Im} f_J(s)$}
\label{sec:imFcomp}

We now repeat the analogues large $J$ expansion for ${\rm Im} f_J(s)$ in terms of the threshold expansion of the double spectral density. In this case, the Froissart-Gribov formula can be written as
\be
\label{eq:dsdPWb}
{\rm Im} f_J(s) &={2 {\cal N}_d \over \pi} \int\limits_{z_1}^\infty d z (z^2-1)^{{d-4 \over 2}}Q^{(d)}_J(z) \rho(s,t(z))\ ,\qquad {\rm Re}[J]>J_0(s)\ .
\ee
It is instructive to separate $s$ into three regions, $4 m^2<s<16 m^2$, ${64\over3}m^2>s>16m^2$ and $s>{64\over3}m^2$.

\subsubsection*{The Elastic Region $4m^2<s<16m^2$}

In this elastic strip the double spectral density is zero for $t<{16 m^2\,s \over s-4m^2}$ or, equivalently, for $z <2 z_1^2-1$, see figure \ref{fig:karplus1}. The large $J$ limit of ${\rm Im} f_J(s)$ is thus dominated by the expansion of double spectral density close to the leading Landau curve which we worked out in section \ref{sec:thresholdDSD}. Hence, we can simply use (\ref{eq:largeJqExp}) with $z_* = 2 z_1^2 - 1$, otherwise the consideration is identical to the one in the previous subsection. Explicitly, we have 
\be
\label{eq:largeJimF}
{\rm Im}f_J(s) &={{\cal N}_d\over\sqrt\pi} {\Gamma\(\tfrac{d -2}{2}\) \over J^{{d-1 \over 2}} } { ( [\lambda(2z_1^2-1)]^2-1)^{{d-3\over2}} \over [\lambda(2z_1^2-1)]^{J+d-3}  } {\rm Im} \hat f_J(s) \ ,\qquad 4m^2<s<16m^2\ , \nn \\
{\rm Im} \hat f_J(s) &= \int\limits_0^\infty d \delta z e^{- \delta z} \rho(s,t(\delta z, J) )\(1+\cO(1/J)\) \ ,
\ee
where 
\be
\label{eq:rholarge}
z=(2 z_1^2 - 1) + {2 z_1 \sqrt{z_1^2 -1} \over J} \delta z\ ,\qquad t(\delta z, J)= 8 m^2 \left( z_1+1 + {z_1 \over J} \left( {z_1+1 \over z_1 - 1} \right)^{1/2} \delta z \right) \ .
\ee
Therefore, upon making a substitute $\delta z \to {2 z_1 \sqrt{z_1^2 -1} \over J} \delta z$ in the formulas of section \ref{sec:thresholdDSD}, we can directly plug them into (\ref{eq:largeJimF}). The resulting large $J$ behavior with the universal threshold expansion (\ref{eq:asympf0}) takes the following form
\be\label{ImfJlargeJ}
{\rm Im} \hat f_J(s)=2^{-5{(d-3) \over 2}} \left( n_{0}^{(d)} \Gamma \left(\tfrac{5-d}{2}\right) \right)^2  m^{4-d} \pi^{{1 - d \over 2}} J^{{d-5 \over 2}}  z_1^{{3-d \over 2}} \left({z_1 -1 \over z_1 + 1} \right)^{d-1\over4} \(1+\cO(1/J)\)\ ,\qquad d \text{ even}\ ,
\ee
where again the case of $d$ odd should be considered separately using the results of appendix \ref{sec:keynoleint}.

Turning to the case with a finite spin zero scattering length (\ref{eq:finiteSCL0}), we get
\be
\label{eq:imfJlargeJscl}
{\rm Im} \hat f_J(s) &={2^{{3d-25 \over 2}}\( n_{0}^{(d)} a_0^2 \Gamma\left(\tfrac{d-1}{2} \right)\)^2 \over\pi^{d-1\over2} m^{d-4} J^{{3d-7 \over 2}} }   {1\over z_1^{d-3\over2}}  \left({z_1 + 1 \over z_1 - 1} \right)^{{3d-11 \over 4}} \(1+\cO(1/J)\)\ .
\ee

An alternative way to arrive at (\ref{ImfJlargeJ}) and (\ref{eq:imfJlargeJscl}) is to start from the large $J$ expansion of $f_J(s)$ discussed in the previous section (\ref{fJlargeJ}) and use elastic unitarity. Indeed, one can check that (\ref{fJlargeJ}) and (\ref{ImfJlargeJ}), (\ref{eq:imfJlargeJscl}) are consistent with elastic unitarity
\be
\label{eq:elunAA}
\log { 2{\rm Im }f_J(s) \over {(s-4 m^2)^{d-3\over2}\over\sqrt s}|f_J(s)|^2}=0\ ,\qquad4 m^2 < s< 16 m^2 \ .
\ee

\subsubsection*{The Inelastic Region $16m^2<s<{64\over3} m^2$}

This regime is in the multi-particle part of figure \ref{fig:karplus1}. To analyze it one should first identify the leading Landau curve in that segment, which is beyond the scope of this paper.

\subsubsection*{The Inelastic Region $s>{64\over3} m^2$}

As $s$ increases passed ${64m^2\over3}$, we know from crossing that the leading Landau curve is the one of the dual channel (the blue curve in figure \ref{fig:karplus1}). A remarkable consequence of this fact is that crossing allows us to measure the non-zero inelasticity at large spin $J$. 

As before the leading large $J$ behavior of ${\rm Im} f_J(s)$ is controlled by the leading Landau curve in the relevant kinematics. For $s>{64 \over 3} m^2$ the leading Landau curve is at $t=t_1(s)={4 m^2 s \over s-16 m^2}$, or equivalently at 
\be
\label{eq:zhatdef}
z=\tilde z_1 \equiv {4 - 3 z_1 + z_1^2 \over 5 - 3 z_1}  \ .
\ee
See blue curve on figure \ref{fig:karplus1}.

In this case we get that the Froissart-Gribov integral takes the same form as in (\ref{eq:dsdPWb}) with
\beq\label{eq:largeJimFinel}
z_1  \to\tilde z_1\ ,\qquad t(\delta z, J)=4 m^2 \({\tilde z_1 -1 \over z_1 -1} + {1 \over J}  {\sqrt{\tilde z_1^2 - 1} \over z_1 + 1} \delta z\)\qquad\text{and}\qquad s>{64 \over3 } m^2\ ,
\eeq
where $t(\delta z , J)$ is a definition of $\delta z$ that we will use below.

The double spectral density close to $\tilde z_1$ 
is given by the results of section \ref{sec:thresholdDSD} upon application of crossing $s \leftrightarrow t$.\footnote{By taking $s \to \infty$ with $x = {m^3 \over s^{3/2}} J$ kept fixed in (\ref{eq:largeJimF}), (\ref{eq:largeJimFinel}) one gets the $d$-dimensional version of the Haan-M{\"u}tter scaling law \cite{Haan:1974yy,Yndurain:1975py}.} 
Considering the amplitudes with the universal threshold expansion (\ref{eq:asympf0}) we get
\be
{\rm Im} \hat f_J(s)&={J^{{d-5 \over 2}}\(n_{0}^{(d)} \Gamma\(\frac{5-d}{2}\)\)^2 \over8(2m)^{d-4}(\tilde z_1^2 -1)^{{d-5\over 4}} \pi^{{d - 1 \over 2}}} { (z_1-1)^{{3d -13 \over 2}} \over (1+z_1)(3-z_1)(5-3 z_1)^{d-6}} \(1+\cO(1/J)\)\ ,\qquad d \text{ even}\ .
\ee

In odd $d$ the situation is more complicated due to the logarithmic nature of the threshold singularity. While the leading power dependence on $J$ can be easily computed, we do not give an explicit form for the dependence on $\log J$ that multiplies the leading power.\footnote{It should be possible to use the integral form for $\rho(s,t)$ given in appendix \ref{sec:dsdOddD} to efficiently evaluate the integral numerically for given $J$ but we do not pursue it here.}

For the case of 
finite spin zero scattering length (\ref{eq:finiteSCL0}), we get in any $d$
\be
{\rm Im} \hat f_J(s) &={\(n_{0}^{(d)} a_0^2\,\Gamma\({d-1 \over 2} \)\)^2\over 2^{d+7}m^{d-4} \pi^{d-1 \over2}} {(\tilde z_1^2 -1)^{{3d-7 \over 4}}\over J^{{3d-7 \over 2}} }
 {(5-3 z_1)^{3(d-2)} (z_1-1)^{{11 - 5 d \over 2}} \over (1+z_1)(3-z_1)^{2d-5} }\(1+\cO(1/J)\)\ .
\ee

We can use the results above for the leading large $J$ behavior of ${\rm Im} \hat f_J(s)$ for $s>{64 \over 3} m^2$ to estimate the amount of inelasticity that exists at given $s$ and $J$. Recall that if the scattering were purely elastic partial waves would satisfy $ 2{\rm Im }f_J(s) = {(s-4 m^2)^{d-3\over2}\over\sqrt s}|f_J(s)|^2 $, see (\ref{eq:elunAA}). 

\begin{figure}[H]
\centering
\includegraphics[width=0.75\textwidth]{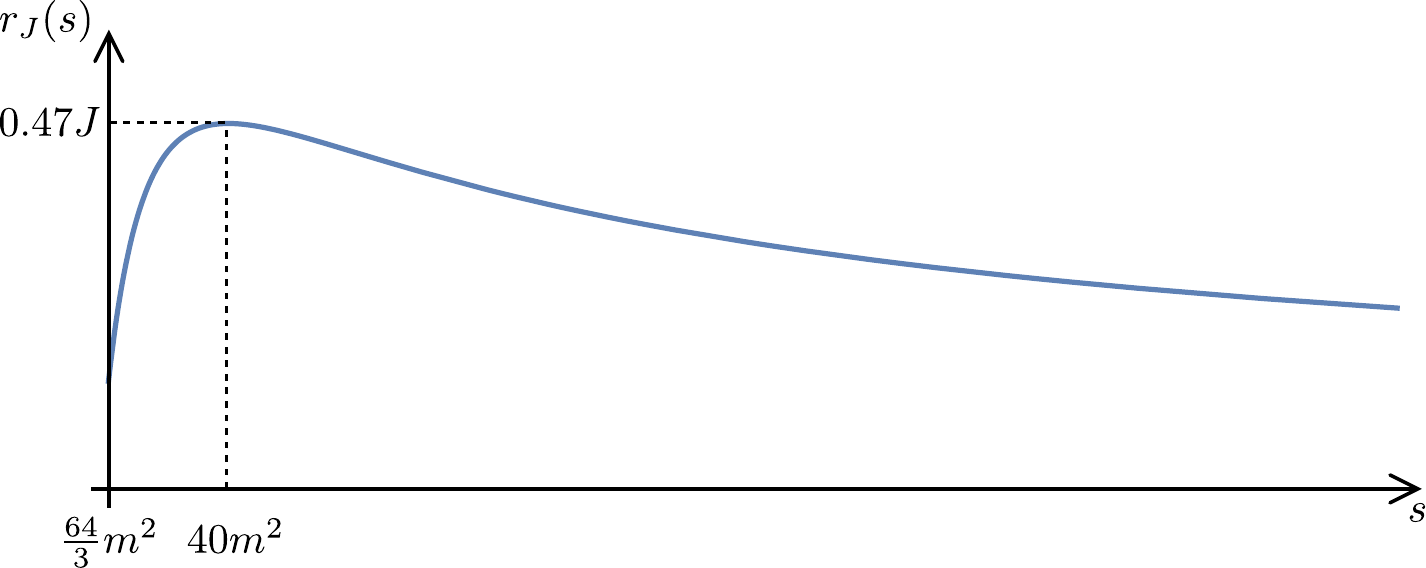}
\caption{\small The large spin inelastisity ratio $r_J(s)$ in (\ref{eq:ratiolead}) plotted for $s>{64\over 3}m^2$. The maximum value of the ratio is achieved at $s=40m^2$ for which we get $r_J(40m^2)\simeq 0.473 J(1 +\cO(1/J))$.}
\label{fig:inelast}
\end{figure}

Using the explicit results of this section we conclude that elastic unitarity together with crossing lead to the universal inelasticity ratio at large $J$
\be
\label{eq:ratiolead}
r_J(s) \equiv \log { 2{\rm Im }f_J(s) \over {(s-4 m^2)^{d-3\over2}\over\sqrt s}|f_J(s)|^2} =J \log {\lambda(z_1)^2 \over \lambda(\tilde z_1)} + O(1)\ ,\qquad s>{64 m^2\over 3}\ .
\ee
In figure \ref{fig:inelast} we plotted $r_J(s)$ as a function of $s/m^2$ for fixed $J$. We see that $r_J(s)$ approaches $0$ for $s=20m^2$ and $s = \infty$. It acquires its maximal value of about $0.47 J$ at $s=40m^2$.   

\subsection{Threshold Expansion in the $J$-space}
\label{sec:Jexact}

In this section we evaluate exactly the contribution of a given term in the threshold expansion to the partial waves.
This generalizes the results above which correspond to the large $J$ limit of the exact formulas that we present in the
current section.  For simplicity let us first focus on $d$ even, when the threshold behavior is of the square-root type.

Consider the threshold expansion of the discontinuity of the amplitude. In the regime where it converges close to the threshold we can write
\be
\label{eq:threxp}
T_t(s,t(z)) =
\sum_{n=0}^\infty c_n(s) \({z - z_1\over z_1-1}\)^{n+{3-d\over2}}=\sum_{n=0}^\infty c_n(s)\,\sigma_t^{n+{3-d\over2}}\ .
\ee

 In practice we truncate this expansion at any desired order and plug it into the Froissart-Gribov formula. Note that the Froissart-Gribov integral goes all the way to infinite $z$ where the expansion (\ref{eq:threxp}) is no longer valid. However, that regime of integration is exponentially suppressed in $J$ and therefore only affects the nonperturbative corrections at large $J$. 
It turns out that the relevant integral can be done exactly. It is given by
\be
\label{eq:masterintegral}
I_{n,J}^{(d)}(z_1) &\equiv {2 {\cal N}_d \over \pi} \int\limits_{z_1}^{\infty}\! dz(z^2-1)^{{d-4 \over 2}}\sigma_{t(z)}^{n+{3-d\over 2}}  Q_J^{(d)}(z)   =  {2^{{5\over2}}\Gamma(n+{5-d \over 2}) \over2^n(8\pi)^{d\over2} \Gamma({3 \over 2} + n) } { (z_1+1)^{n+{1 \over 2}}\over(z_1-1)^{{2-d\over2}}}Q_{J-n+{d-5 \over 2}}^{(2 n + 5)} (z_1)\ .
\ee
This formula is derived in appendix \ref{sec:derivation} and holds for arbitrary $J$, $d$ and $n$, not necessarily integer. If we expand both sides at large $J$, we reproduce the formulas of the previous section. In this way, we arrive at the explicit expression for the large $J$ expansion of $f_J(s)$ in terms of the coefficients of the threshold expansion in the crossed channel 
\be
\label{eq:resultF}
f_J(s)= \sum_{n=0}^\infty c_n I_{n,J}^{(d)}(z_1)+\text{[nonperturbative]}\ .
\ee
Using this expression and the large $J$ expansion of $I_{n,J}^{(d)}(z_1)$
\beq
I_{n,J}^{(d)}(z_1)={1\over J^{n+1}}\times{2\,\Gamma(n+{5-d\over2})
\over(8\pi)^{d-1\over2}\lambda(z_1)^{J+{d-3\over2}}}{(z_1+1)^{n+{1\over2}}\over(z_1-1)^{2-d\over2}}\({2\lambda(z_1)\over\lambda^2(z_1)-1}\)^{n+1}\(1+\cO(1/J)\)\ ,
\eeq
we see that higher orders in the sum are more suppressed in $J$. Hence, one can use this representation to explicitly and systematically obtain all the coefficients in the large $J$ expansion of $f_J(s)$, the leading $1/J$ result of course coincides with the previous analysis.

Similarly, we can write down the threshold expansion formula for 
$\rho(s,t)$. Based on the previous discussion we have
\be
\label{eq:rhoexp}
\rho(s,t(z)) = {1 \over (z-(2 z_1^2 - 1))^{{d-5 \over 2}} } \sum_{m=0}^\infty d_m(s) (z - (2 z_1^2 - 1) )^m\ .
\ee
The problem of finding $d_m(s)$ becomes completely algebraic after we note that (\ref{eq:rhoexp}) implies
\be
\label{eq:resultImF}
{\rm Im} f_J(s) &= \sum_{m=0}^\infty d_m(s) I_{m+1,J}^{(d)}(2 z_1^2 - 1)+\text{[nonperturbative]}\ ,
\ee
where we again performed the Froissart-Gribov integral exactly.

Elastic unitarity (\ref{eq:elunAA}) relates the two expansions, (\ref{eq:resultF}) and (\ref{eq:resultImF}). It then allows one to express the $d_m(s)$'s in terms of the $c_n(s)$'s. 
This is possible because the $d_m(s)$'s do not depend on $J$. Equivalently, we can use (\ref{eq:doubleSDb}) to directly map the threshold expansion of $T_t(s,t)$ to that of $\rho(s,t)$ near the Landau curve. The details of this expansion are presented in appendix \ref{sec:derivation}.

\section{Finite $J$ and Finite $s$}
\label{sec:boundinelasticity}

In practice, 
one is interested in making statements about partial waves at some finite (but potentially large) spin $J$ and finite energy $s$. Our ability to make such statements crucially depends on our ability to estimate an error in the Froissart-Gribov integral produced by the approximation to the amplitude. 
To that extent we can think of the discontinuity of the amplitude as follows
\be
T_t(s,t) = T^{{\rm approx}}_t(s,t) + T^{{\rm error}}_t(s,t)\ ,
\ee
where $T^{{\rm approx}}_t(s,t)$ is an approximation to $T_t(s,t)$ based on our knowledge about it. It can involve expansion coefficients close to various normal thresholds, information about resonances, or about the Regge limit. The more information is available to us, the less we can make $T^{{\rm error}}_t(s,t)$ and therefore the better is our knowledge of $f_J(s)$.

Assuming continuity of the amplitude we can try to bound the amplitude as follows. Consider a given $s$ and let us assume that $T_t(s,t) \sim t^{J_0(s)}$ at large $t$. Let us consider an integer $N> J_0(s)+{d - 3 \over 2}$ and write 
\be
\label{eq:threxpB}
T^{{\rm approx}}_t(s,t)=
\sum_{n=0}^{N-1} c_n(s) \({z - z_1\over z_1-1}\)^{n+{3-d\over2}}\ .
\ee

Continuity of the amplitude then implies that there exists $c_N(s)$ such that
\beq\label{eq:bound}
\big| T^{{\rm error}}_t(s,t)\big|<c_N(s) \({z - z_1\over z_1-1}\)^{N+{3-d\over2}} \ .
\eeq
\begin{figure}[t]
\centering
\includegraphics[width=0.85\textwidth]{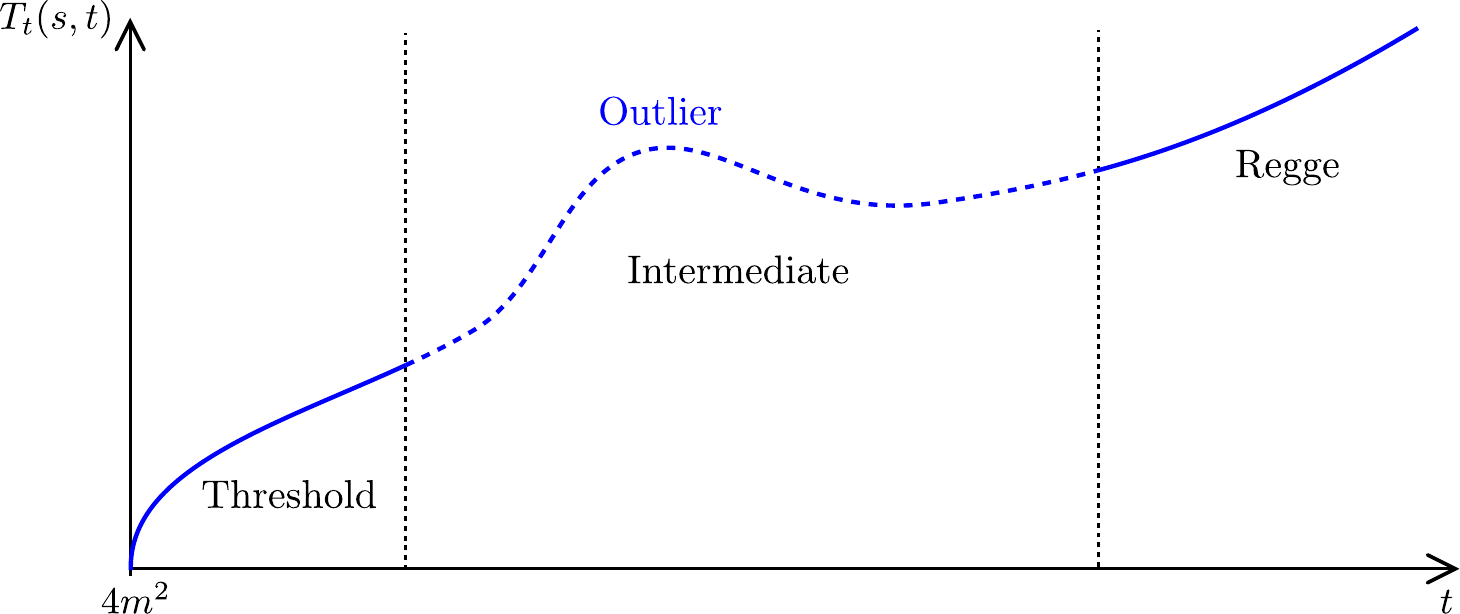}
\caption{\small The error in our approximation for the inelasticity depends on our knowledge about the discontinuity of the amplitude $T_t(s,t)$. Here, the discontinuity is plotted for different energy scales ($t$) at fixed $s$. Its structure in the large energy (Regge) and the low energy (threshold) regions are under relatively good control (solid blue line). At intermediate energy scales the amplitude may develop a bump. The problem of outliers is the problem of bounding the discontinuity in this regime, and hence bounding our error for the inelasticity.}\label{bump}
\end{figure}
Indeed, this bound matches the neglected term close to the threshold and is also consistent with the Regge behavior since $N> J_0(s)+{d - 3 \over 2}$. The minimal value of $c_{N}(s)$ depends on the behavior of the amplitude at intermediate $s$. In particular if at some fixed $t=t_0$ the discontinuity develops a ``bump'', see figure \ref{bump}, then $c_N(s)$ should be made large enough for (\ref{eq:bound}) to hold. A familiar example of such a bump is a resonance and it is due to a singularity on the second sheet.  More generally, we can call such a bump an outlier. Correspondingly, we  call the problem of bounding $c_N(s)$ ``the problem of outliers.''\footnote{For a similar discussion in the context of the CFT bootstrap see \cite{Kologlu:2019bco}.}

Note that for the integral on (\ref{eq:bound}) to converge at large $z$, we need to take $J>N+1$. Hence, a better choice of variables to encode our knowledge about the threshold behavior of $T_t(s,t)$ are ones that do not grow at large $z$. For example, we can replace $(z_1-1)\to(z-1)$ in (\ref{eq:threxpB}). 
Doing so is necessary if one wants to go to higher orders in the threshold expansion, but we will not pursue this in the present paper. The advantage of using just $(z-z_1)/(z_1-1)$ is that the Froissart-Gribov integral is known explicitly, see (\ref{eq:masterintegral}).

In a related manner, 
when deriving the Froissart-Gribov formula we do not have to close the contour all the way to infinity. We can instead keep arcs in the complex plane at some finite energy, see figure \ref{GFfig}.b. This is sometimes called the truncated Froissart-Gribov formula, see e.g. \cite{Yndurain:1975py,Martin:2017ndt}. The problem of deriving finite $s$ and $J$ formulae then requires bounding the contribution of the arcs. 
The advantage of this approach is that we do not have to assume extended analyticity all the way to $t = \infty$.

Let us proceed with $c_N(s)$ in (\ref{eq:bound}) being our phenomenological parameter and see how various quantities depend on it. First of all, we can immediately bound the error in $f_J(s)$. We have
\be\label{pwbound}
\left| f_J(s) -   \sum_{n=0}^{N-1} c_n I_{n,J}^{(d)}(z_1) \right| < c_N(s) I_{N,J}^{(d)}(z_1)\ .
\ee
where $I_{n,J}^{(d)}(z)$ is the integral (\ref{eq:masterintegral}). 

Next, we consider ${\rm Im} f_J(s)$. The relevant regime to study is $s>{64 m^2 \over 3}$, where we can estimate inelasticity using our knowledge about the threshold expansion in the crossed channel. In this case we can start from 
\be
{\rm Im} f_J(s) &= {2 {\cal N}_d \over \pi} \int\limits_{\tilde z_1}^\infty dz\, (z^2-1)^{d-4 \over 2} Q_J^{(d)}(z) \rho(s,t(z))\  ,
\ee
and split the $z$-integral in two regions, $4m^2<t<16 m^2$ and $t \geq 16 m^2$.

In the region $4m^2<t<16 m^2$ we can use the Mandelstam equation to compute $\rho(s,t)$. It will involve the terms in the threshold expansion that we worked out above together with an error term that is controlled by (\ref{eq:threxpB}) and (\ref{eq:bound}). For $t \geq 16 m^2$, since ${\rm Disc}_sT^{{\rm approx}}_t(s,t)=0$, we have
\be\label{eq:rhobound}
|\rho(s,t)|=|{\rm Disc}_sT^{{\rm error}}_t(s,t)| < |T^{{\rm error}}_t(s,t)|\ .
\ee
Together with (\ref{eq:threxpB}) and (\ref{eq:bound}), this can be used to estimate the bound on ${\rm Im}f_J(s)$. 

If one is not willing to make an assumption of the form (\ref{eq:bound}) about $T_t^\text{error}(s,t)$ then one can still derive a bound on inelasticity \cite{Martin:2017ndt}, albeit very weak and only asymptotic when $s \to \infty$.

\subsection{A Toy Model Example}
\label{sec:toymodel}

We now study a toy model example that is motivated by the numerical analysis of \cite{Paulos:2017fhb}. We consider four-dimensional spacetime ($d=4$) and take $T_t^\text{approx}(s,t)$ to be given by the first universal term in the threshold expansion (\ref{eq:asympf0})
\be
\label{eq:toymodel}
T_t^\text{approx}(s,t) =  32 \pi
\sqrt{z_1-1\over z-z_1}\ .
\ee
We assume that the Regge limit is bounded by $J_0(s)\le{1\over2}$ for real $s$ in some region. Hence, for this case we can take $N=1$ in (\ref{eq:bound}) and bound the error as
\be
\label{toyerror}
|T_t^\text{error}(s,t)|<\delta c(s)\times 32 \pi \sqrt{z-z_1\over z_1-1}\ .
\ee
where for convinience we have introduced the notation $\delta c(s)=c_1(s)/(32 \pi)$, (\ref{eq:bound}).

We will now go through the steps in figure \ref{thresholdfig} and apply them to the toy model (\ref{eq:toymodel}). We will use the bound on the error (\ref{toyerror}) to have a finite $J$ and finite $s$ bound on the error at each step.

We start with the error on the partial waves coefficients. Using the Froissart-Gribov formula, it is given by (\ref{pwbound})
\be\label{eq:errorfJ}
\left| f_J(s) - f_J^\text{approx}(s) \right|=\left| f_J(s) - 32 \pi I_{0,J}^{(d=4)}(z_1) \right| < 32 \pi \delta c(s) I_{1,J}^{(d=4)}(z_1)\ ,
\ee
Note that due to our assumption on the Regge trajectory, the Froissart-Gribov formula is applicable all the way down to $J \geq 2$ for any $s \geq 4m^2$.

To quantify the error in (\ref{eq:errorfJ}) we take the ratio between the error integral $I_{1,J}^{(d=4)}(z_1)$ to the approximation one $I_{0,J}^{(d=4)}(z_1)$
\beq\label{eq:ratiotest}
err_J(s)\equiv\delta c(s)\times {I_{1,J}^{(d=4)}(z_1) \over  I_{0,J}^{(d=4)}(z_1)}={\delta c(s)\over2J}\sqrt{z_1+1\over z_1-1}\(1+\cO(1/J)\)\ .
\eeq
The apprximation for the partial waves is good when this ratio is small. For example, we can focus on the point $s=40m^2$, where the large $J$ inelasticity ratio (\ref{eq:ratiolead}) is maximal.
\begin{figure}[t]
\centering
\includegraphics[width=0.55\textwidth]{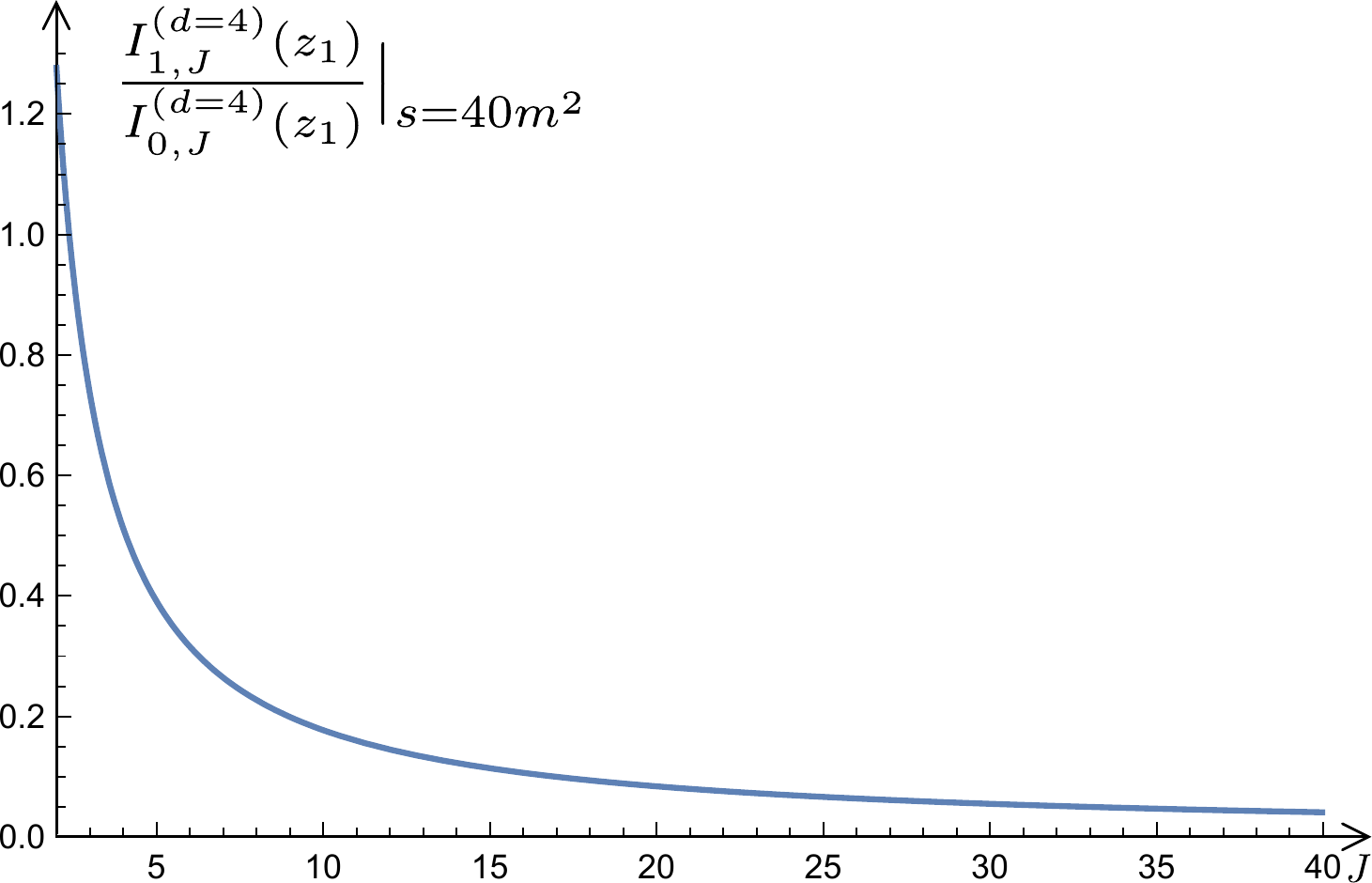}
\caption{\small We plot the ratio of the leading and subleading threshold integrals in (\ref{eq:ratiotest}) as a function of spin $J$ for $s=40m^2$. In particular, we get ${I_{1,J}^{(d=4)}(z_1) \over I_{0,J}^{(d=4)}(z_1)} |_{s=40 m^2 , J=4} \simeq 0.51$, and ${I_{1,J}^{(d=4)}(z_1) \over I_{0,J}^{(d=4)}(z_1)} |_{s=40 m^2 , J=20} \simeq 0.08$. }
\label{fig:ratioL}
\end{figure}
In figure \ref{fig:ratioL} we have plotted this ratio for $s=40m^2$ as a function of spin. For example, we have
\beq
err_{J=4}(40m^2)\simeq 0.51\,\delta c(40m^2)\ ,\qquad 
err_{J=20}(40m^2)\simeq 0.08\,\delta c(40m^2)\ .
\eeq

Next, we use the Mandelstam equation (\ref{eq:doubleSD}) to evaluate the double spectral density in the elastic strip and bound the error there. We get (see appendix \ref{sec:higherrho} for more details)
\beq
\label{eq:rhotoym}
\rho(s,t)={16\over\pi^2}\sqrt{s-4m^2\over s}\tilde{I}^{(4)}_{0,0}+\rho^\text{error}(s,t)\ ,\qquad s^\text{cross}_1(t)<s<16m^2\ ,
\eeq
where $\tilde{I}^{(d)}_{n_1,n_2}$ was defined in (\ref{eq:deftilde}), and recall that $s^\text{cross}_1(t)={4 m^2 t \over t-16 m^2}$.  
\be\label{eq:rhotoymodel}
| \rho^\text{error}(s,t) | \ <\  {16\over\pi^2}\sqrt{s-4m^2\over s}\(2 \delta c(s) \tilde{I}^{(4)}_{1,0} +\delta c^2(s)\tilde{I}^{(4)}_{1,1}\)\ .
\ee
In estimating the error we used the fact that both the Mandelstam kernel ${\rm Disc}_z K(z, \eta_1, \eta_2)$, and $T_t^\text{approx}$ are non-negative.

Next, we perform crossing transformation to the formulas above to estimate $\rho(s,t)$ at $s>{64m^2\over3}$. We plug (\ref{eq:rhotoym}) into the Froissart-Gribov formula and use it to evaluate ${\rm Im}f_J(s)$. As before, we focus on the point of maximal large $J$ inelasticity, $s = 40 m^2$. At this point, $\rho(40m^2,t)$ has only support for $t\geq t^\text{cross}_1(40m^2)={20 m^2 \over 3}$. We then consider the regions ${20 m^2 \over 3} \leq t < 16 m^2$ and $t>16 m^2$ separately
\be\label{GFsplit}
{\rm Im} f_J(s) &= {2 {\cal N}_d \over \pi}  \int\limits_{\tilde z_1}^{z_2} d z\, Q_J^{(d)}(z) \rho(s,t(z)) +  {2 {\cal N}_d \over \pi}\int\limits_{z_2}^{\infty} d z\, Q_J^{(d)}(z) \rho(s,t(z))\ ,
\ee
where $\tilde z_1$ was defined in (\ref{eq:zhatdef}) and  $z_2=1+{32m^2\over s-4m^2}$.

In the elastic $t$-strip the double spectral density is given by the crossing of (\ref{eq:rhotoymodel}).\footnote{Note that for $s=40m^2$ this region is inside the Mahoux-Martin positivity region (\ref{positiverho}).} To get a bound on the integral (\ref{GFsplit}) in that region, we replace $\delta c(t)$ by its maximal value there
\be
\delta c_{el} = {\rm max}_{{20 m^2 \over 3}<t<16 m^2} \delta c(t) \ .
\ee

Above the elastic strip we use (\ref{eq:rhobound}) 
($| \rho(s,t(z)) | \leq |T_t^\text{error}(s,t)|$) and the positivity of $Q_J^{(d)}(z)$. We do not have an explicit formula for the integral over this region but we can trivially compute it numerically for various values of $J$. 

In total, we find that for $J=20$ and $s=40m^2$
\beq
\left.\lambda_1^{2 J}(s) 2{\rm Im} f_{J}(s) \right|_{J=20,\,s=40m^2} = 88.53\pm\[16.6\,\delta c_{el} + 0.67\,\delta c_{el}^2 + 0.14\,\delta c(40m^2)\]\ .
\eeq
For these values we also get from (\ref{eq:errorfJ}) that
\beq
\left.\lambda_1^{2 J}(s)\sqrt{s-4 m^2\over s} | f_{J}(s)|^2\right|_{J=20,\,s=40m^2}\!\!\!\!\!\!\!\!\!\!=0.002\pm 3.5 \times 10^{-4} \delta c(40m^2) + 1.46 \times  10^{-5} (\delta c(40m^2))^2\ .
\eeq
Note that in this case the term $\sim (\delta c(40m^2))^2$ is sign-definite --- the consequence of using (\ref{eq:errorfJ}) to bound $| f_{J}(s)|^2$.

Similarly, for $J=2$ and $s=40m^2$ the result takes the form
\be
\left.\lambda_1^{2 J}(s) 2{\rm Im} f_{J}(s)\right|_{J=2,\,s=40m^2} &= 0.389 \pm \[ 0.19\,\delta c_{el} + 0.015 \delta c_{el}^2 + 2.18 \delta c(40m^2) \] \nn \\
\left.\lambda_1^{2 J}(s)\sqrt{s-4 m^2\over s}| f_{J}(s)|^2\right|_{J=2,\,s=40m^2} &= 0.14 \pm 0.358 \delta c(40m^2) + 0.228 (\delta c(40m^2))^2 \ .
\ee

Let us comment on the origin of the various terms. The error in $f_J(s)$ is given by (\ref{eq:errorfJ}). In the expression for $\lambda_1^{2 J} 2 {\rm Im} f_{J=20}(s=40m^2)$ the term linear in $\delta c_{el}$ comes from the error in the elastic region, and similarly the term linear in $\delta c(40m^2)$ comes from the integral over $t>16 m^2$. The $\delta c_{el}^2$ term comes from the error in $\rho(s,t)$ in the elastic region. 

To summarize, given a bound on our ignorance about $T_t(s,t)$, specified by $\delta c(s)$ for the case at hand, we can explicitly derive a lower bound on the amount of inelasticity that is present in the toy model. This is of course assuming that the underlying amplitude satisfies elastic unitarity. It would be also interesting if the analysis above can be improved using more refined error estimates. For example, a better approach would be to bound the error distributionally, see section \ref{sec:DistUnitNum} below.

\subsection{Elastic Unitarity and Coupling Maximization}

In \cite{Paulos:2017fhb} the numerical procedure outlined below in section \ref{Nreview} has been put forward and carried out 
for the problem of maximizing $T({4 m^2\over 3}, {4 m^2\over 3})$. It was observed that the low spin partial waves $f_J(s)$, with $J=0$ and $J=2$, converge very well as a function of $N_{\text{max}}$ which characterizes the number of terms used to approximate the amplitude, see section \ref{Nreview} for details. Moreover, it was found that that $f_{J=0,2}(s)$ tend to saturate elastic unitarity above $s\geq 4m^2$. 

It is interesting to apply the finite $J$ and $s$ analysis described above to this case. We find that the function produced by the numerics fits the toy model approximation we considered above in section \ref{sec:toymodel}. In particular, the relevant bounds on the the discontinuity of the amplitude $T_t(s,t)$ come from the threshold behavior of the amplitude
\beq\label{eq:extrabound}
\delta c(s) = {T_t(s,t) - 32 \pi\sqrt{z_1 - 1 \over z-z_1}\over 32 \pi\sqrt{z - z_1\over z_1 -1}}\Big|_{t = 4 m^2}\ ,\qquad \delta c_{el} =\delta c(16m^2)\ .
\eeq
The convergence properties of $\delta c(s)$ at general $s>4m^2$ are less clear as they probe the unphysical region. An interesting possibility to avoid the question of their convergence is to add the bounds on (\ref{eq:extrabound}) as extra conditions to the bootstrap algorithm.  
Note, that in theories that satisfy elastic unitarity $\delta c(s)$ does not depend on $s$ since in this case only $f_0(t)$ contributes. Therefore dependence of $\delta c(s)$ can itself be used as a probe of elastic unitarity.

\begin{figure}[t]
\centering
\includegraphics[width=0.8\textwidth]{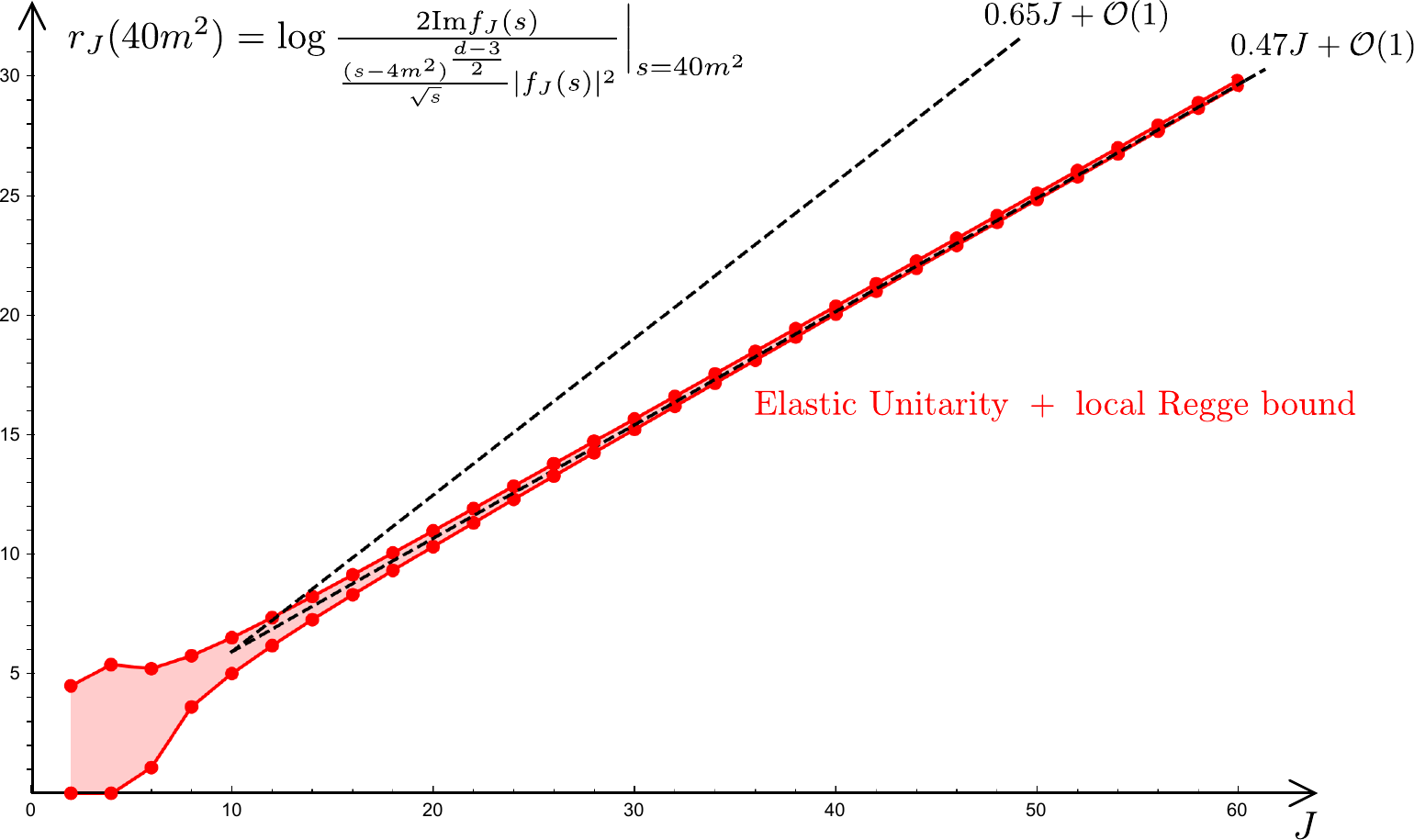}
\caption{\small Inelasticity ratio $r_J(s=40m^2)$, (\ref{eq:ratiolead}), 
as a function of spin $J$. The region between the red curves is the analytic prediction for a function that satisfies elastic unitarity and the local error bound (\ref{toyerror}) with parameters (\ref{eq:reggeparnum}) as taken from the numerics. 
For a function that satisfies elastic unitarity, the presence of the Steinmann shadow region where $\rho(s,t) = 0$, see figure \ref{fig:inelast}, gives $r_J(40m^2)\sim 0.47 J$ at large $J$. This is the asymptotic behavior of the red curves. On the other hand, a function that has $\rho(s,t) \neq 0$ in the Steinmann shadow region is expected to behave as $r_J(40m^2)\sim\log \lambda(z_1)|_{s=40 m^2}\sim 0.65 J$ this is what numerics produces for any finite $N_{{\rm max}}$.}
\label{fig:ratioLnum}
\end{figure}

Nevertheless, we proceed and consider numerics for $N_\text{max}=11$ and $J_\text{max}=36$ with the leading threshold behavior fixed to the universal behavior dictated by elastic unitarity. In this case we find from the numerics \cite{numerics}
\be\label{eq:reggeparnum}
\delta c_{el} = 0.67\ ,\qquad\delta c(40m^2) = 1.17\ .
\ee
In figure \ref{fig:ratioLnum} we plot the results for the inelasticity ratio $r_J(40m^2)$ (\ref{eq:ratiolead}) obtained in the toy model of the previous section with the parameters taken from the numerics (\ref{eq:reggeparnum}).

As we discuss below, there are several ways to improve the numerical procedure so that the plots would agree at finite $N_{max}$. 
We leave the exploration of these possibilities for future work \cite{numerics}.

\section{Analytical Methods and Numerical Bootstrap}
\label{sec:numboot}

In this section\footnote{We thank Madalena Lemos for collaboration on this section.} we discuss how some of the analytical methods described in the paper can be implemented in the numerical approach to the $S$-matrix bootstrap that has been put forward in \cite{Paulos:2017fhb}. We will report the actual results of the numerical explorations elsewhere \cite{numerics}.

\subsection{Review of the Numerical Framework of \cite{Paulos:2017fhb}}\label{Nreview}

Let us briefly review the setup of \cite{Paulos:2017fhb}. The basic idea is to write an ansatz for the expansion of the scattering amplitude which 
is linear in unknown real parameters $\alpha_{abc}$
\be
\label{eq:ansatzNum}
T(s,t) = \sum_{a,b,c=0}\left. \alpha_{a b c} \rho_s^a \rho_t^b \rho_u^c + {\rm extra}\right|_{u = 4m^2 -s -t}\ ,
\ee
where, the function $\rho_s \equiv {\sqrt{4m^2 - s_0} - \sqrt{4 m^2 + s} \over \sqrt{4m^2 - s_0} + \sqrt{4 m^2 + s}}$ maps the complex $s$-plane minus the $s$-channel cut to the unit circle and the point $s_0$ to the origin. 
Here, the extra terms may be added to make some particular properties of the amplitude manifest. Their presence or absence depends on the particular problem at hand. Crossing symmetry is imposed by demanding that the coefficients $\alpha_{a b c}$ are permutation-invariant. Finally, the relation $s+t+u=4m^2$ leads to a redundancy in the basis of coefficients that can be addressed systematically.

To approximate an amplitude using the ansatz (\ref{eq:ansatzNum}), the sum is truncated such that
\be
a+b+c \leq N_{{\rm max}}\ .
\ee
Given a finite $N_{{\rm max}}$, unitary in the form
\be
\label{eq:unitNum}
| S_J(s) | \leq 1\ ,\qquad s \geq 4 m^2\ ,\qquad J \in 2 \mathbb Z_+\ ,
\ee
is imposed 
over a finite grid of points and for spins that are truncated by some maximal value $J\le J_\text{max}(N_\text{max})$. As shown in \cite{Paulos:2017fhb}, remarkably unitarity in the form of (\ref{eq:unitNum}) can be restated as a semidefiniteness condition as follows. We write for physical $J$ and $s$
\be
S_J(s) = 1 + i \vec \alpha \cdot \vec f_J(s)\ ,
\ee
where $\hat f_J(s)$ are kinematical objects and all the dynamical information is in the coefficients $\vec \alpha$. The condition (\ref{eq:unitNum}) can be then rewritten as a semi-definitedness condition for the matrix
\be\label{semidef}
M \equiv \begin{pmatrix}
1 + \vec \alpha \cdot {\rm Re}\vec f_J(s) & 1 - \vec \alpha \cdot {\rm Im}\vec f_J(s)\\
1 - \vec \alpha \cdot {\rm Im}\vec f_J(s) & 1 - \vec \alpha \cdot {\rm Re}\vec f_J(s)
\end{pmatrix} 
\succcurlyeq 0\ .
\ee
At this point one can maximize numerically some quantity linear in the $\alpha$-parameters by imposing unitarity in the form (\ref{semidef}) over the chosen grip in $s$ and for $J\le J_\text{max}(N_\text{max})$. For example, in \cite{Paulos:2017fhb} the ``coupling", $T({4 m^2\over 3},{4 m^2\over 3})$, is maximized. 
If a certain maximization task reliably saturates as a function of $N_{{\rm max}}$ we stop the process and trivially extrapolate to $N_{{\rm max}} = \infty$ to get the actual bound on the space of physical $S$-matrices.

\subsection{Why and What Should be Improved
}\label{prosandcons}

The setup of \cite{Paulos:2017fhb} has several very important and desirable properties: 
\begin{itemize}
\item It is simple and practical. It is not too hard to implement, manipulate and obtain bounds.
\item The space of functions (\ref{eq:ansatzNum}) is complete inside the $\rho$ unit circle. Hence, any function analytic inside the circle (except, maybe a finite number of isolated poles which can be added explicitly) can be expanded in that way.
\item Crossing is trivialized and is satisfied exactly.
\item At any finite $N_{{\rm max}}$, (\ref{eq:ansatzNum}) satisfies maximal analaticity.
\end{itemize}
At the same time, there is some tension and potential issues in applying the procedure above for exploring physical amplitudes that we list below and comment on how one may improve on them:
\begin{itemize}
\item Physical amplitudes exhibit a set of normal multi-particle thresholds at $s>4m^2$. Hence, the $\rho$-expansion of such a function is not expected to converge point by point for real $s>4m^2$ ($|\rho_s|=1$). On the other hand, in the procedure above unitarity is imposed point-wise in exactly that regime. Thus, it is unclear if the space of functions that are probed in this procedure includes among them physical amplitudes, with multi-particle thresholds. Below we discuss two possible ways of addressing this point. One is to improve the ansatz and the other is to change the way in which unitarity is imposed.
\item The ansatz (\ref{eq:ansatzNum}) has restrictive behavior at $s=\infty$. In particular, the partial waves satisfy $\lim_{s \to \infty} S_J(s) = 1$. There is no reason to expect this to be a correct property of physical amplitudes. Hence, similar to multi-particle thresholds, the $\rho$-expansion is expected to be a poor approximation in this kinematical regime. One way to fix it is to explicitly add extra terms to the ansatz. 
Another way, which we discuss in more detail below, is to change the way in which unitarity is imposed.
\item Elastic unitarity is not satisfied. This is manifest for any finite $N_\text{max}$ -- elastic unitarity implies that $\rho(s,t) = 0$ in the elastic strips, below the first Landau curve. If one tries to impose it exactly on the truncated ansatz (\ref{eq:ansatzNum}) then clearly the only solution is $\alpha_{abc}=0$. One may still hope that elastic unitarity will emerge at large $N_\text{max}$. However, without extra constraints, we see no reason for that to happen. In practice, there is no conclusive evidence that the functions that emerge at the boundary of the allowed space satisfy elastic unitarity within the numerical error.

Imposing elastic unitarity is hard for the simple reason that this condition
\be
\label{eq:elunSN}
| S_J(s) |^2 = 1\ ,\qquad 4m^2 < s < 16 m^2\ ,\qquad J \in 2 \mathbb Z_+\ ,
\ee
is nonlinear in the unknown parameters $\alpha_{abc}$. Therefore, within the approach of \cite{Paulos:2017fhb}, we can only hope to impose elastic unitarity-type constraints. Namely, constraints that go beyond (\ref{eq:unitNum}), but still include physical amplitudes in the space of functions that satisfy them. This is important if eventually we want to explore the space of physical amplitude that in particular do satisfy (\ref{eq:elunSN}).

Below, we elaborate on several ways in which elastic unitarity can be pursued within the numerical approach of \cite{Paulos:2017fhb}. Importantly, each of them is still linear in $\alpha_{abc}$ and therefore possible to implement using the standard solvers.
\end{itemize}

\subsection{How the Numerical Framework Can Be Improved 
}

We now suggest a few ways of addressing the issues identified above within the framework of \cite{Paulos:2017fhb}.

\subsubsection{Distributional Unitarity}
\label{sec:DistUnitNum}

Here we address some of the issues identified above, namely the lack of point-wise convergence for real $s$, or $|\rho_s|=1$, and too restrictive behavior at $s=\infty$ by suggesting a different way in which unitarity can be imposed numerically.

The basic idea is that even though the truncated expansion of a function with multi-patricle thresholds does not converge at $|\rho|=1$, it can still converge to it as a distribution. Indeed, this is what is expected for the functions that are polynomially bounded as we approach $|\rho|=1$ or real Mandelstam variables $s$ and $t$. A relevant mathematical result that addresses this question is known as Vladimirov's theorem \cite{Vladimirov}. It was recently discussed in detail in the context of the conformal bootstrap in \cite{Kravchuk:2020scc} to which we refer the reader for technical details.

Let us consider negative $t$ needed to compute $S_J(s)$ starting from the amplitude. For such $t$, it is known that the amplitude is polynomially bounded as $|s| \to \infty$. It is also polynomially bounded for real $s$, namely as ${\rm Im} s \to 0$, see e.g. \cite{Epstein:1969bg}. These facts imply that we can apply Vladimirov's theorem to partial waves $S_J(s)$ and we expect that partial waves of physical amplitudes computed using the ansatz (\ref{eq:ansatzNum}) will converge for real $s$ distributionally, namely after integrating $S_J(s)$ against some smooth test function $g(s)$. Note that this includes also amplitudes that grow with $s$ or $t$, which can be modeled by ${(1 + \rho_t)^n \over (1 + \rho_s)^n}$ type terms. Expanding such terms around $\rho_t = \rho_s = 0$ leads to a series that converges distributionally for $|\rho_s|=1$ as explained in \cite{Kravchuk:2020scc}.

Therefore, if we treat the amplitude (and the partial wave coefficients) as a distribution, we can still approximate it by the ansatz (\ref{eq:ansatzNum}).
Given $N_{\text{max}}$, however, this approximation is expected to be good only for test functions that do not resolve the local in $s$ features of the amplitude/partial waves, or have support for very large $s$. In other words, given $N_{\text{max}}$ we can only hope to have a reliable approximation on average over big enough intervals of $s$, with the intervals becoming smaller as we increase $N_{\text{max}}$. 

Similarly, considering test functions localized at larger $s$ requires taking larger $N_{\text{max}}$. This is a consequence of the accumulation of the multi-particle thresholds at large $s$, as well as due to the potential growth of the amplitude as $s$ or $t$ become large. Both effects lead to poor distributional convergence of the ansatz (\ref{eq:ansatzNum}) at large $s$, or, equivalently, close to $\rho_s=-1$. By restricting the support of the test functions away from that region we can use the ansatz (\ref{eq:ansatzNum}) to probe functions that have multi-particle thresholds and grow in the Regge limit.

Given a real, non-negative function $g(s)$, unitarity on average takes the form
\be
\label{eq:distun}
\left|\, \int\limits_{4m^2}^\infty d s\, g(s) S_J(s) \right| \leq 1\ ,\qquad  \int\limits_{4m^2}^\infty ds\, g(s) = 1\ .
\ee
This can be still restated as a semi-definitedness condition and therefore can be implemented numerically using the same methods. In practice, given $N_{max}$ and expected properties of the physical amplitude there is a set of test functions $g(s, N_{\text{max}})$ that we can use. 

To understand what are the reasonable functions to be used let us note that on the boundary of the circle, truncation of the maximal power in $\rho^n$ is the same as truncating the Fourier harmonics of $\rho=e^{i\phi}$. Therefore if we truncate $n \leq N_{max}$ we cannot hope to resolve the amplitudes on scale $\delta \phi < {2 \pi \over N_{\max}}$. Through the map $\rho(s)$, this translates into a statement about the $s$-plane. We leave the detailed discussion of the test functions $g(s,N_\text{max})$ for the future.

Distributional convergence puts the numerical $S$-matrix bootstrap algorithm of \cite{Paulos:2017fhb} on a much more solid mathematical ground. It justifies imposing distributional unitarity (\ref{eq:distun}) for the truncated ansatz (\ref{eq:ansatzNum}). 

In practice, for certain types of problems it can happen that a more naive point-wise analysis of unitarity still leads to correct results. Based on the reasons explained above this is expected to work if the underlying amplitude does not exhibit multi-particle thresholds and does not grow in the Regge limit. This expectation agrees with the numerical results observed in \cite{Paulos:2017fhb}.

\subsubsection{Extended Basis}
\label{sec:extbasisnum}

Another way of addresing the issue of multi-particle thresholds is to extend the ansatz (\ref{eq:ansatzNum}). This ansatz makes the structure of the two-particle normal threshold manifest. An obvious extension of the basis which makes the structure of the multi-particle cuts manifest is to add to it any power of the functions
\be
\label{eq:higherthr}
\rho_{s}^{(n)} \equiv {\sqrt{(2n)^2m^2 - s_0} - \sqrt{(2n)^2 m^2 + s} \over \sqrt{(2n)^2 m^2 - s_0} + \sqrt{(2n)^2 m^2 + s}}\ .
\ee
Such an extension while making the structure of the multi-particle normal thresholds manifest still has the property that the double spectral density misses the regions where $\rho(s,t) = 0$ carved out
in the $(s,t)$-plane by the Landau curves.

Ideally, one would like to write down an ansatz which does not only make maximal analyticity manifest but also has a correct structure of the Landau curves. Such functions are naturally generated in perturbation theory. We can use them to write down functions that have expected behavior in the elastic strip. Let us take $\phi^4$ in $d=3$ and consider the following diagram
\be
&{\rm Box}(s,t,u)= {\rm Box}(s,t) + {\rm Box}(s,u) + {\rm Box}(t,u)\ ,\\
&\includegraphics[width=0.3\textwidth]{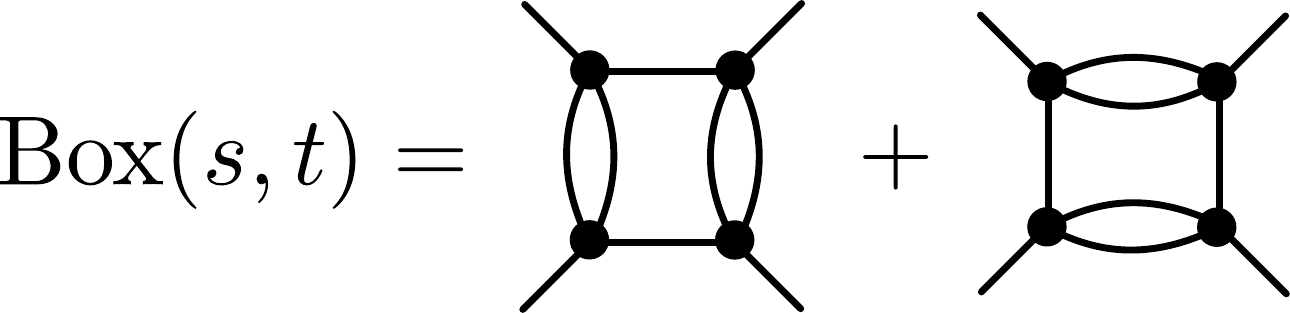}\ .
\ee
This function has the property that it is crossing symmetric and has the zero double discontinuity in the expected region. 

We can consider for example the following ansatz
\be\label{albasis}
T(s,t) &={\rm Box}(s,t,u)\times\sum_{a,b,c=0} \alpha_{a b c} (\rho_s^{(n)})^a (\rho_t^{(n)})^b (\rho_u^{(n)})^c |_{u = 4 m^2 - s -t} + {\rm extra}\ ,\qquad  n>2 \ .
\ee
By choosing $n>2$ we make sure that the normal threshold coming from $\rho^{(n)}_{s,t,u}$ starts after $16 m^2$ and the correct analytic structure inside the elastic strip comes from the sum of the $\phi^4$ diagrams. Hence, this ansatz has an advantage of having the built-in analytic structure consistent with elastic unitarity.

Similarly, other Landau curves can be manifestly incorporated by choosing different perturbative diagrams and dressing them by an appropriate $\rho$-ansatz. In this way we can hope to have an ansatz which has more structure of the actual scattering built in. At the same time, linearity in $\alpha_{abc}$ as well as crossing are still manifest. 

That said, without imposing extra constraints that result from elastic unitarity, there is no a priori reason for the numerics to turn on the $\alpha_{abc}$'s that are associated to sub-leading Landau curves. We turn to such conditions next.

\subsubsection{Elastic-type Unitarity at Non-integer $J$}

Another way of injecting constraints from elastic unitarity into the numerical bootstrap is by imposing unitarity for non-integer spins. As we explained in section \ref{sec:acJ}, the elastic unitarity condition (\ref{eq:elunSN}) can be analytically continued to complex $J$ as long as ${\rm Re} J > J_0(s)$, where $J_0(s)$ is the leading Regge trajectory. 

Let us consider real $J>J_0(s)$. Numerically, we can then impose the following elastic unitarity-type condition
\be
\label{eq:distunEl}
\Big| \int\limits_{4m^2}^{16 m^2}\!\! d s\, g(s) S_J(s) \Big| \leq 1\ ,\qquad  \int\limits_{4m^2}^{16 m^2}\!\! ds\, g(s) = 1\ ,\qquad  J>J_0(s).
\ee
Imposing this unitarity-type condition numerically for non-integer spin goes beyond (\ref{eq:unitNum}), while keeping the problem linear in the $\alpha$-coefficients. In this case, 
the partial waves $S_J(s)=1 + i {(s-4m^2)^{{d-3 \over 2}} \over \sqrt{s}} f_J(s)$ for non-integer $J$ are computed via the Froissart-Gribov integral (\ref{eq:FGform}).

While physical amplitudes will saturate this bound the condition above still includes them as a part of the solution however goes beyond the integer spin unitarity conditions. Importantly, it is still linear in $\alpha_{abc}$ and therefore can be easily implemented numerically. 

When imposed in that way, the elastic unitarity-type conditions (\ref{eq:distunEl}) are very similar to the original conditions (\ref{eq:unitNum}). Note that the Froissart-Gribov projection probes the regime of arbitrary large $t$ where the numerics convergence is slower. On the other hand, for integer spin one normally uses the partial wave projection (\ref{eq:pwprojection}) that only probes physical $t$'s.

\subsubsection{Lower Bound on Inelasticity}
\label{sec:inelastnum}

As we discussed in section \ref{sec:boundinelasticity} an additional knowledge of the behavior of the discontinuity of the amplitude $T_t(s,t)$ leads to a more refined prediction about the amount of inelasticity at finite $J$ and finite $s$. For example, we can put in some expectation about the Regge behavior and low energy data on the scattering length, as well as structure of bumps or resonances to get a relatively accurate estimate of $T_t(s,t)$. Similarly, it can happen that while running the numerical algorithm one observes that the gross features of $T_t(s,t)$ saturate quickly as one increases $N_{max}$. In this way one can get an accurate estimate of $c_N(s)$ in (\ref{eq:bound}). This in turn allows us to put a lower bound on the amount of inelasticity $r_J(s)$ (\ref{eq:ratiolead}) for finite $J$ and finite $s$.

The minimal amount of inelasticity can be easily implemented numerically. We can replace (\ref{semidef}) by 
a set of modified matrices
\be
\label{eq:modifsemidef}
M_{a}^{{\rm lower}} \equiv \begin{pmatrix}
1 + a_J(s) \vec \alpha \cdot {\rm Re}\vec f_J(s) & 1 - a_J(s)  \vec \alpha \cdot {\rm Im}\vec f_J(s)\\
1 - a_J(s)  \vec \alpha \cdot {\rm Im}\vec f_J(s) & 1 - a_J(s)  \vec \alpha \cdot {\rm Re}\vec f_J(s)
\end{pmatrix} 
\succcurlyeq 0\ ,\qquad a_J(s) \geq 1\ .
\ee
The corresponding modified positive semi-definetedness condition (\ref{eq:modifsemidef}) is equivalent to the inequality
\be
{2 \vec \alpha \cdot {\rm Im}\vec f_J(s) \over  |\vec \alpha \cdot {\rm Re}\vec f_J(s)|^2} \geq a_J(s) \geq 1\ ,
\ee
where in the original problem $a_J(s)=1$ and more generally $a_J(s)$ specifies the minimal amount of inelasticity at given $J$ and $s$.

From our discussion in section (\ref{sec:boundinelasticity}) and the large $J$ analysis we know that $a_J(s) \gg 1$ for large enough $J$ and $s>{64 m^2 \over 3}$
, see (\ref{eq:ratiolead}). Moreover, given a local bound on the discontinuity of the scattering amplitude of the type (\ref{eq:bound}), we can derive a set of improved elastic unitarity-type bounds (\ref{eq:modifsemidef}) at finite $J$ and finite $s$, see section \ref{sec:boundinelasticity}. Luckily, a bound of the type (\ref{eq:bound}) is again linear in $\alpha_{abc}$. Hence, the improved bound can be implemented numerically.

\subsection{Mahoux-Martin Positivity}

As we reviewed in section \ref{productionsec} elastic unitarity leads to the positivity property of the double spectral density in the so-called Mahoux-Martin region, see \ref{positiverho}. 

Positivity or more generally non-negativity of the double spectral density is obviously linear in $\alpha_{abc}$ and therefore is straightforward to implement numerically. We can think of this either using the improved basis described in section \ref{sec:extbasisnum} which directly implements the Steinmann shadow region where $\rho(s,t) = 0$. Alternatively, we can consider the original ansatz (\ref{eq:ansatzNum}) and impose the Mahoux-Martin type positivity constraints
\be
\label{eq:MMtypeNum}
\rho(s,t) \ge 0\ ,\qquad 4 m^2 < s < 16 m^2\ ,\qquad 4m^2 \le t \leq 4 m^2 {(3 s + 4 m^2)^2 \over (s - 4 m^2)^2}\ .
\ee
The condition (\ref{eq:MMtypeNum}) still includes physical amplitudes, which in the elastic strip $4 m^2 < s < 16 m^2$ satisfy the more restrictive conditions $\rho (s,4m^2 \le t < {16 m^2 s \over s - 4 m^2} )=0$ and $\rho (s,{16 m^2 s \over s - 4 m^2}< t \leq 4 m^2 {(3 s + 4 m^2)^2 \over (s - 4 m^2)^2} )>0$.

\section{Comments on CFTs}
\label{eq:commentCFT}

In this paper we assumed extended analyticity and we studied the structure of the amplitude for $s,t >0$, where the amplitude develops crossing-symmetric double spectral density $\rho(s,t)$. It is interesting to understand what are the analogous statements in CFTs.

Let us list a map between the $S$-matrix and CFT quantities:
\begin{center}
\begin{tabular}{ |c|c| } 
 \hline
 S-matrix elements & CFT correlators \\ 
 \hline
 $T(s,t)$ & ${\cal G} (u,v)$ \\
 \hline
 ${\rm Disc }\,T(s,t)$ & ${\rm dDisc}{\cal G} (u,v)$ \\

 \hline
 $\rho(s,t)$ & ${\rm qDisc} {\cal G} (u,v)$ \\ 
 \hline
  $f_J(s)$ & $c_J(\Delta)$ \\ 
 \hline
 Froissart-Gribov formula (\ref{eq:FGform}) & Lorentzian inversion formula \cite{Caron-Huot:2017vep,Simmons-Duffin:2017nub} \\ 
 \hline
 Dispersion relations (\ref{eq:dispersion}) & CFT Dispersion relations \cite{Carmi:2019cub} \\ 
\hline
Elastic unitarity&?\\
\hline
\end{tabular}
\end{center}
The double discontinuity ${\rm dDisc} {\cal G} $ and the quadruple discontinuity ${\rm qDisc} {\cal G} $ were introduced in \cite{Caron-Huot:2017vep}. It was shown in \cite{Caron-Huot:2017vep} that crossing symmetry of ${\rm qDisc} {\cal G}$, see also \cite{Simmons-Duffin:2016wlq,Meltzer:2019nbs}, readily implies the presence of multi-twist operators in the OPE. This is the CFT analog of the Aks theorem reviewed above.

Elastic unitarity is a consistency condition of the two-particle sector of the $S$-matrix. In generic CFTs, the analog of two-particle states in the dual AdS space are double-twist operators \cite{Fitzpatrick:2012yx,Komargodski:2012ek} that are defined at large spin $J$. In large $N$ CFTs the two-particle states in AdS correspond to double trace operators. Similarly, in AdS QFTs, see e.g. \cite{Paulos:2016fap}, we expect a natural set of operators corresponding to two-particle states to be present in the spectrum. However unitarity, as formulated in the CFT language through the OPE, does not admit a truncation analogous to elastic unitarity that emerges as we take the flat space limit of the theory.

Imposing that the twist spectrum structure of a CFT is the one coming from the light-cone bootstrap leads to the so-called Polyakov conditions, see e.g. \cite{Penedones:2019tng} for a recent discussion in the nonperturbative context. It is interesting to understand to what extent the consequences of imposing the Polyakov conditions in AdS are analogous to elastic unitarity in flat space.\footnote{We thank S. Caron-Huot and J. Penedones for discussion on this point.} 
Using this analogy, the exploration of the present paper suggests that an interplay between the Polyakov conditions and crossing symmetry of the quadruple discontinuity can lead to interesting results. It will be interesting to investigate it further.

Let us next comment on extended analyticity. In the context of amplitudes it implies in particular that the analytic structure of the discontinuity $T_s(s,t)$ is similar to the one of the scattering amplitudes, modulo interchanging normal thresholds to Landau curves.
In the context of CFTs it would require understanding analytic properties of ${\rm dDisc} {\cal G}$. In general this is a complicated problem since it requires going to the region of the $(u,v)$ space in which no OPE channel converges, see e.g. \cite{Qiao:2020bcs} for a detailed, recent discussion. It is however possible to make progress in 2d CFTs. Indeed, in this case thanks to the Virasoro symmetry \cite{Zamolodchikov:1985ie}, the OPE converges on an arbitrary sheet \cite{Maldacena:2015iua}. One finds that a statement analogous to Mandelstam analyticity indeed holds, namely the only singularities on an arbitrary sheet of ${\cal G}(u,v)$ are branch points at $u,v=0,\infty$. One does not expect an analogous statement in higher dimensional CFTs due to a more complicated structure of Lorentzian singularities of the correlator. However, it would be very interesting to investigate this analytic structure in more detail.

Finally, let us comment on the validity of the Mandelstam representation in CFTs. Recall, that to obtain Mandelstam representation in flat space we start with the usual dispersion relation (we ignore subtractions for simplicity)
\be
\label{eq:dispersion}
T(s,t) = \int\limits_{4 m^2}^{\infty} {d s' \over \pi} {T_{s}(s',t) \over s' - s} + \int\limits_{4 m^2}^\infty {d u' \over \pi} {T_{u}(u',t) \over u' - u}.
\ee
We then write the dispersion relation for the discontinuity of $T_{s}(s',t)$ 
\be
T_{s}(s,t) = \int {d t' \over \pi} {\rho(s,t') \over t' - t} +  \int {d u' \over \pi} {\rho(s,u') \over u' - u} ,
\ee
and plug in the formula above to get
\be
T(s,t) ={1 \over \pi^2} \int\limits_{4 m^2}^\infty {d s' d t' \  \rho(s',t') \over (s'-s)(t'-t)}+{1 \over \pi^2}\int\limits_{4 m^2}^\infty {d u' d t' \ \rho(u',t') \over (u'-u)(t'-t)} +{1 \over \pi^2} \int\limits_{4 m^2}^\infty {d s' d u' \  \rho(s',u') \over (s'-s)(u'-u)}.
\ee 
An important ingredient in this argument, apart from maximal analyticity, is polynomial boundedness of $T_{s}(s,t)$ for arbitrary $s$.\footnote{This is not expected to be a true property of nonperturbative amplitudes \cite{Mandelstam:1963cw}.}

Let us now see what is the analogous situation in CFTs. Let us first consider 2d CFTs where maximal analyticity follows from the Virasoro symmetry as described above. Let us recall what were the main ingredients in the derivation of CFT dispersion relations in  \cite{Carmi:2019cub}. There it was shown that given a single-valued ${\cal G}(u,v)$ analytic in the cut-plane and bounded in the Regge (and Euclidean OPE) limit one can write a dispersion relation.
What happens if as above we try to write dispersion relations for ${\rm dDisc}{\cal G}(u,v)$? Using the OPE, as described in \cite{Maldacena:2015iua}, one can clearly bound any limit of the correlator or ${\rm dDisc}{\cal G}(u,v)$ on any sheet.

Single-valuedness of ${\rm dDisc}{\cal G}(u,v)$ is however not obvious using the Virasoro block OPE \cite{Zamolodchikov:1985ie} and in general we do not expect it to hold. It is easy to check explicitly what happens in the case of minimal models. For the critical Ising model it is easy to see that ${\rm dDisc}{\cal G}(u,v)$ is single-valued. It therefore satisfies all the necessary properties to write dispersion relations \cite{Carmi:2019cub}. In this sense 2d Ising model correlators (somewhat trivially) admit Mandelstam representation. On the other hand, already in the tricritical Ising case single-valuedness does not hold, so we cannot apply the dispersion relations of \cite{Carmi:2019cub}.

In higher dimensions the situation is much more complicated due to absence of Virasoro symmetry. Here again there is no reason to expect single-valuedness of ${\rm dDisc} {\cal G}(u,v)$. On the other hand, single-valuedness is a true property of the double discontinuity in free field theories, which therefore also admit the CFT analog of Mandelstam representation. One can wonder if this property continues to hold for the theories with slightly broken higher spin symmetry, e.g. Chern-Simons vector models in $d=3$.

It would be also interesting to understand if there exists some other, more sophisticated way to think about writing an analog of Mandelstam representation in CFTs. As a different direction, thinking about some other versions of dispersion relations, see e.g. \cite{Bissi:2019kkx}, that do not rely on single-valuedness of the underlying correlator might very well be useful in certain applications.

\section{Conclusions}
\label{sec:conclusions}

One of the challenges of the modern conformal bootstrap is to efficiently combine analytical insights with the numerical methods to corner and solve physical theories \cite{Simmons-Duffin:2016wlq}. Analogously, in this paper we revisited analytical techniques for the nonperturbative $S$-matrix bootstrap. A natural next step for the $S$-matrix bootstrap program is to combine them with the existing \cite{Paulos:2017fhb} or future numerical methods to compute physical amplitudes.

Concretely, in this paper we studied the implications of elastic unitarity and extended analyticity for the relativistic, unitary, gapped $S$-matrix in $d \geq 3$. Our goal was to develop the analytical methods to constrain the nonperturbative scattering amplitude, which can be further used in the numerical bootstrap approaches.\footnote{Many remarkable structures were recently unraveled in the study of perturbative scattering amplitudes of both massless \cite{Benincasa:2007xk} and massive \cite{Arkani-Hamed:2017jhn} particles. In this paper we have focused on nonperturbative aspects of the two-to-two massive particle scattering. It would be interesting to see if any of these new insights can be put to use in the nonperturbative setting.} The analytic bootstrap was the subject of active investigation in the 60's. Most of our ideas and results, when restricted to $d=4$, are contained in some form in the old works of Dragt \cite{Dragt} and A.~W.~Martin \cite{Martin:1969yw}, as well as more recent work of Roy and A.~Martin \cite{Martin:2017ndt}. We believe however that there is some value in ``re-discovering'' these methods from the modern perspective and in pushing forward the current incarnation of the $S$-matrix bootstrap. 

As usual, if one wants to do analytic computations in a nonperturbative setting, a small parameter is needed. For a nonperturbative $S$-matrix, there are two expansions in two small kinematical parameters  -- the threshold expansion in ${s-4 m^2 \over 4m^2}$, and the large spin expansion in $1/J$. These two are related via the Froissart-Gribov formula (\ref{eq:FGform}). Ones combined with elastic unitarity and crossing symmetry, these two expansions lead to the bootstrap scheme outline in figure \ref{thresholdfig}. The upshot of this analysis is that one can start with the low energy, low spin data (the threshold expansion), and use it to bootstrap the amplitude away from this regime. We, however, do not restrict the low energy data. In this sense, the scheme is analogous to the analytic CFT bootstrap \cite{Fitzpatrick:2012yx,Komargodski:2012ek,Caron-Huot:2017vep}. 

While the analytic bootstrap methods reveal important structural properties of the amplitude, by themselves, they are not strong enough to ``solve'' the problem. Correspondingly, the low energy data that enters the threshold expansion and the bound on Regge are taken here as an unconstrained input for the analytic bootstrap scheme. 
Currently, the only known way of constraining these parameters systematically is using the numerical bootstrap techniques \cite{Paulos:2017fhb} or experiment \cite{Colangelo:2001df}. As we discussed in the present paper, the numerical methods should be improved by implementing the structure that originates from elastic unitarity and extended analyticity.  Indeed, it was observed in the numerical studies that the putative amplitude functions that saturate bounds tend to saturate unitarity. That seems in tension with the Aks theorem of section \ref{productionsec}. The problem with the latter is that it does not provide us with a finite energy lower bound on particle production that can be implemented numerically. Provided a local bound on the discontinuity of the amplitude however, one does get a finite lower bound on particle production. Moreover, it can be implemented numerically as an extra constraint. Hence, it is instructive to consider the $S$-matrix bootstrap in a given class of discontinuity bounded amplitudes. We discussed this in more detail in section \ref{sec:boundinelasticity}, where an explicit example is also given. In section \ref{sec:numboot} we suggest various ways in which inelasticity and other constraints that emerge from elastic unitarity can be implemented.

A related important question is to what extent physics in the elastic regions $4m^2 < (s \ \text{or} \  t) < 16 m^2$ studied in the present paper dominates the dynamics of the amplitude? In other words, under which conditions our ignorance of the multi-particle kinematics at $s,t>16 m^2$ leads to a small controllable error? As we have seen in section \ref{sec:boundinelasticity}, when considering the toy model, in the case where the low energy interaction is strong (infinite scattering length) and the Regge behavior is relatively soft, the elastic region strongly constrains the behavior of partial waves at finite $s$ and $J$. We can easily imagine a different situation, e.g. relevant for pion scattering, when the low energy interaction is weak. Based on our analysis in this case we do not have a reason to expect the physics of the partial waves to be dominated by the elastic region (unless the spin is very large). Correspondingly, in this case the dynamics in the multi-particle region is expected to be important. Bootstrapping such an $S$-matrix would then potentially require a more detailed understanding of the analytic constraints that result from the physics in the multi-particle region.

Let us briefly discuss a few future directions. First, it would be very interesting to extend the current numerical approaches to the $S$-matrix bootstrap by implementing structures that originate from elastic unitarity in one of the ways suggested in this paper. We will report on this in \cite{numerics}. Second, most of the explorations in this paper were bounded to the elastic strips, where one of the Mandelstam invariants is between the two- and the four-particle threshold energies, see figure \ref{fig:karplus1}. This region is particularly manageable because in one of the channels it is controlled by two-to-two amplitude only. It is an interesting and important task to explore the multi-particle region, where the energy is above the four-particle threshold in two of the channels. Finally, it would be interesting to explore the landscape $S$-matrices, other than $d=4$ QCD. Ideally, one would like to find an $S$-matrix in $d\ge3$ that may play the analogous rule to the one played by the Ising model in the conformal bootstrap. Whether such ``bootstrap-solvable" $S$-matrices in $d \geq 3$ exist or not is yet to be shown. If it exists, we expect its solution to teach us a lot about nonperturbative QFT in general. Implementing efficiently the structure of the amplitude that we discuss in the present paper would be an important step towards constructing an example. A natural candidate theory to explore is $\phi^4$ theory in $d=3$.

There are also a few technical avenues along which our work can be extended. One is relaxing the $\mathbb Z_2$ symmetry we assumed, which restricted the spin and the number of particles to be even. Another related extension is to include single-particle poles. Doing so will affect many of the details, but will not change the global picture. A more interesting generalization is to consider particles with spin, see e.g. \cite{deRham:2018qqo}. Similarly, it is an open problem to implement the known structure of the UV of the theory, say asymptotic freedom or the CFT data of the UV fixed point, into the $S$-matrix bootstrap.

Finally, one can wonder if there is anything to be learned from this analysis for the conformal bootstrap. In the latter case, the theory is gapless so naively there is no elastic unitarity. However, CFTs in $d>2$ have a twist gap, and multi-twist operators are mapped to the multi-particle states in the AdS dual theory. Therefore, it would be interesting to understand the AdS analog of the various aspects of the present paper more directly.

\section*{Acknowledgements}

Some of the results from this paper were presented by one of us at ICTP Bangalore, India, and The Third Mandelstam Theoretical Physics School and Workshop 2019 in Durban, South Africa.
We are grateful to Madalena Lemos for collaboration on section \ref{sec:numboot}. We are grateful to Paolo Benincasa, Mattia Bruno, Simon Caron-Huot, Barak Gabai, Rajesh Gopakumar, Victor Gorbenko, Maxwell Hansen, David McGady, Andr{\'e} Martin, Shiraz Minwalla, Jo\~{a}o Penedones, Ricardo Schiappa, Aninda Sinha, Piotr Tourkine, Gabriele Veneziano for useful discussions.

\appendix

\numberwithin{equation}{section}

\section{Derivation of the Mandelstam Kernel}\label{AppendixA}

In this appendix we compute the kernels $\mathcal{P}_d(z, z', z'')$ and $K_d (z, \eta', \eta'')$  defined in (\ref{angularkin}) and (\ref{MandelstamK}) correspondingly. 

\paragraph{The kernel $\mathcal{P}_d(z, z', z'')$} Recall that $z'$ and $z''$ are cosine of the angles between $\vec n$ in (\ref{angularkin}) and the vectors $\vec p_1$ and $\vec p_3$, (\ref{cosines}). They are related to the coordinates in the Sudakov decomposition of the unit vector $\vec n$
\be\label{Sudakov}
\vec n = \alpha {\vec p_1 \over |\vec p_1| } + \beta {\vec p_3 \over |\vec p_3| } + \vec n_{\perp}\ ,\qquad \vec n_{\perp} \cdot \vec p_1= \vec n_{\perp} \cdot \vec p_3 = 0
\ee
as
\beq\label{alphabeta}
z' 
= \alpha + \beta z\ ,\qquad z'' 
= \beta + \alpha z\ , \qquad \alpha = {z' - z z'' \over 1 - z^2}\ ,\qquad\beta = {z'' - z z' \over 1 - z^2}\ .
\eeq
In term of these coordinates, the angular integration in (\ref{angularkin}) reads
\be
\label{eq:meaapp}
\int d^{d-2} \Omega_{\vec n}& = 
2\int d^{d-1} \vec n\, \delta(\vec n^2 - 1)\cr
&=2 \sqrt{1-z^2} \int d \alpha\, d \beta\, d^{d-3} \vec n_{\perp} \delta( \vec n_{\perp}^2 + \alpha^2 + \beta^2 + 2 \alpha \beta z -1 ) \cr
&=2\sqrt{1-z^2} \int d \alpha\, d \beta\, {\Theta(1- \alpha^2-\beta^2-2 \alpha \beta z) \over(1 - \alpha^2 - \beta^2 - 2 \alpha \beta z)^{5-d\over2} }  \int  d^{d-3} \vec n_{\perp} \delta( \vec n_{\perp}^2 -1 ) , \cr
&=\sqrt{1-z^2} {\rm Vol}_{S^{d-4}} \int d \alpha\, d \beta\, {\Theta(1- \alpha^2-\beta^2-2 \alpha \beta z) \over (1 - \alpha^2 - \beta^2 - 2 \alpha \beta z)^{5-d\over2} }\ ,
\ee
where ${\rm Vol}_{S^{d-4}} =  {2 \pi^{(d-3)/2} \over \Gamma({d-3\over 2})}$. The above formula is only true in $d \geq 4$. In $d=3$ we have
\be
\label{eq:meaapp2}
\int d \Omega_{\vec n} =2 \int d^{2} \vec n\, \delta(\vec n^2 - 1) =2 \sqrt{1-z^2} \int d \alpha\, d \beta\, \delta(\alpha^2 + \beta^2 + 2 \alpha \beta z - 1 )\ ,
\ee
which can be also obtained as a distributional limit from (\ref{eq:meaapp}) when $d \to 3$.
By plugging the relation (\ref{alphabeta}) into (\ref{eq:meaapp}) and (\ref{eq:meaapp2}), we arrive at (\ref{eq:dlips}).

\paragraph{The Mandelstam kernel $K_d (z, \eta', \eta'')$} Instead of plugging the explicit form of $\cP_d(z,z',z'')$ into the definition (\ref{MandelstamK}), we have find it simpler to compute the Mandelstam kernel directly using the Sudakov decomposition (\ref{Sudakov}). For $|\eta'|,|\eta''|>1$ we have
\be\label{eq:mandelstamapp}
K_d (z, \eta', \eta'') 
&=\int{ d^{d-2}\Omega_{\vec n} \over (\eta' - z')(\eta'' - z'')}= \int  {d^{d-2} \Omega_{\vec n} \over(\eta' - \alpha - z \beta)(\eta'' - \beta - z \alpha)  }\\
&=\sqrt{1-z^2} {\rm Vol}_{S^{d-4}} \int {\Theta(1- \alpha^2-\beta^2-2 \alpha \beta z) \over (1 - \alpha^2 - \beta^2 - 2 \alpha \beta z)^{5-d\over2} }{ d \alpha\, d \beta\over(\eta' - \alpha - z \beta)(\eta'' - \beta - z \alpha)  }\nn\ ,
\ee
where in the second step we have used (\ref{alphabeta}) with $d\ge4$ and in the third we have used (\ref{eq:meaapp}).

Next, we shift $\alpha \to \alpha - z \beta$ and after it we rescale $\beta \to {\beta \over \sqrt{1 - z^2}}$. In this way we get in $d>3$
\be
K_d (z, \eta', \eta'') &= {\rm Vol}_{S^{d-4}} \int d \alpha\, d \beta\,   {\Theta(1- \alpha^2-\beta^2) \over (1 - \alpha^2 - \beta^2)^{5-d\over2} } {1 \over(\eta'-\alpha)(\eta''-\sqrt{1-z^2}\beta-z \alpha)} \cr
&={\rm Vol}_{S^{d-4}}\int\limits_0^1 d r  \int\limits_0^{2 \pi} d \phi {r \over  (1-r^2)^{5-d\over2}} {1 \over(\eta' - r \cos \phi)(\eta'' - r \cos (\phi+\theta))}\ ,
\ee
where $re^{i\phi}=\alpha-i\beta$. The integral over $\phi$ gives
\be\label{phiint}
\int\limits_0^{2 \pi}\!{d\phi\over(\eta'-r\cos \phi)(\eta''-r\cos(\phi+\theta))}
= {2 \pi \over \eta' \eta''+ \sqrt{(\eta'^2-r^2)(\eta''^2-r^2)} - z r^2  } \[{\eta' \over \sqrt{\eta'^2-r^2}} + {\eta'' \over \sqrt{\eta''-r^2}}\]\ .
\ee
for $|\eta'|,|\eta''|>1$. Otherwise, we analytically continue (\ref{phiint}). Next, we change the $r$ integration variable to
\beq
\eta\equiv{1\over r^2}\(\eta' \eta''+ \sqrt{(\eta'^2-r^2)(\eta''^2-r^2)}\)\ .
\eeq
In that way we arrive at
\beq\label{Kdge4}
K_{d \geq 4} (z, \eta', \eta'')=2\pi {\rm Vol}_{S^{d-4}}\int\limits_{\eta_+}^\infty{d\eta\over \eta - z} {(\eta^2 - 1)^{4-d \over 2}\over (\eta - \eta_+)^{5-d\over2} (\eta - \eta_-)^{5-d\over2}}\ ,\qquad |\eta'|,\,|\eta''|>1\ ,
\eeq
where $\eta_\pm$ are defined in (\ref{etapm}).

Similarly, for $d=3$ we have
\be\label{K3d}
K_3(z, \eta', \eta'') 
=&2\sqrt{1-z^2} \int d \alpha\, d \beta{ \delta(\alpha^2 + \beta^2 + 2 \alpha \beta z - 1 )\over(\eta' - \alpha - z \beta)(\eta'' - 2\beta - z \alpha)  }\\
=&2\int d \alpha\, d\beta\,{\delta(\alpha^2+\beta^2-1) \over (\eta'-\alpha)(\eta''-\sqrt{1-z^2}\beta-z \alpha)}=\int\limits_0^{2 \pi}{d \phi\over(\eta' - r \cos \phi)(\eta'' - r \cos (\phi+\theta))}\nn\\
=&{2\pi\over\eta_+-z}\({\eta'\over\sqrt{\eta'^2-1}}+{\eta''\over\sqrt{\eta''^2-1}}\)\ ,\qquad|\eta'|,\,|\eta''|>1 \ .\nn
\ee

Finally, the Mandelstam kernel with $|\eta'|<1$ or $|\eta''|<1$ is obtained from (\ref{Kdge4}) and (\ref{K3d}) by analytically continuation.

\section{Useful Identities for Gegenbauer $P$- and $Q$-functions}\label{Idappendix} 

As discussed in the main text, due to the $SO(1,d-1)$ symmetry, the elastic unitarity kernel and the Mandelstam kernel are diagonal in spin. They take the form \eqref{eq:PPP} and \eqref{eq:PQQ} correspondingly. In this appendix we derive these forms together with the related integrated expression \eqref{eq:Atkinson}.

%Here we prove the identities on products of Gegenbauer functions, which connect elastic unitarity in terms of the amplitude  eq. \eqref{eq:elauni1} with the formulation in terms of the partial waves, eq. \eqref{eq:EUfJ}. Namely, expressions \eqref{eq:PPP}, \eqref{eq:PQQ} and \eqref{eq:Atkinson}. We note the 
The literature on the properties and identities of the Gegenbauer functions is extensive \cite{HTF}, starting with the Gegenbauer addition formula which dates back to 1893 to Gegenbauer himself \cite{Gegenbauer}. Better suited for us is the integrated form of this identity \cite{HTF} which in our conventions reads
\be
2^{4-d} {\Gamma(d-3) \over \Gamma^2({d-3 \over 2})} \int\limits_{-1}^1dz\, P^{(d)}_J\left(z_1 z_2 + z\sqrt{1 -z_1^2} \sqrt{1-z_2^2} \right) (1 - z^2)^{d - 5 \over 2}= P^{(d)}_J(z_1) P^{(d)}_J(z_2)\ .
\ee
Perhaps the cleanest way to derive the above formula is to use group theoretic techniques \cite{Arfken}. If $z_1$ and $z_2$ are the cosines of the polar angles of unit vectors $\mathbf{n_1}$ and $\mathbf{n_2}$  and $z$ is the cosine of the azimuthal angle difference between $\mathbf{n_1}$ and $\mathbf{n_2}$, then $z_1 z_2 + z\sqrt{1 -z_1^2} \sqrt{1-z_2^2}  = \mathbf{n_1} \cdot \mathbf{n_2}$. One can then apply a rotation to make $\mathbf{n_2}$ aligned along the $z$-axis, as the vector product is invariant under this transformation. The relation between $P^{(d)}_J\left(z_1 z_2 + x \sqrt{1 -z_1^2} \sqrt{1-z_2^2} \right)$ and  $P^{(d)}_J(z_2)$ will then involve the $SO(d-1)$ matrix representation of this rotation, the Wigner D-matrix \cite{Wigner}, which for a specific entry is given up to a factor by $P^{(d)}_J(z_1)$. The integration in $z$ will select this entry and project out the others, so we end up with a closed equation between Gegenbauer polynomials.
\par
Let us change variable to
\be
y = z_1 z_2 + z \sqrt{1-z_1^2} \sqrt{1-z_2^2}
\ee
with integration limits
\be
z_1 z_2 - \sqrt{1-z_1^2} \sqrt{1-z_2^2} < y < z_1 z_2 + \sqrt{1-z_1^2} \sqrt{1-z_2^2},
\ee
or equivalently,
\be
1 - z_1^2 - z_2^2 - y^2 + 2yz_1z_2 > 0.
\ee
The Gegenbauer addition formula then becomes
\be
\label{eq:gaf}
{1\over2}\pi^{2 -d \over 2} \Gamma\left(\tfrac{d - 2}{2}\right) \int\limits_{-1}^1 dy\, \mathcal{P}_d(y,z_1,z_2)(1-y^2)^{d-4\over2} P^{(d)}_J(y)= (1-z_1^2)^{d-4\over2} P^{(d)}_J(z_1) \, (1-z_2^2)^{d-4\over2} P^{(d)}_J(z_2)\ ,
\ee
with $\mathcal{P}_d$ given in \eqref{eq:dlips}.
\par
Multiplying the above by $n_J^{(d)} P^{(d)}_J(z)$ and summing over $J$ allows usage of the Gegenbauer completeness relation
\be
\label{eq:PP}
\sum_{J=0}^\infty n_J^{(d)} P^{(d)}_J(y) P^{(d)}_J(z) = {2 \over \mathcal{N}_d} (1 - z^2)^{4-d\over2} \delta(y - z)\ ,
\ee
with $n_J^{(d)}$ given by \eqref{eq:nJd}. We then arrive at \eqref{eq:PPP}.
%\be
%{\cal P}_d(z,z_1,z_2) =(4\pi)^{d-2}{\cal N}_d^2\, (1 - z_1^2)^{d - 4 \over 2} (1 - z_2^2)^{d - 4 \over 2}\sum_{J=0}^\infty n_J^{(d)} P_J^{(d)}(z) P_J^{(d)}(z_1) P_J^{(d)}(z_2)\, . \nn \tag{\ref{eq:PPP}}
%\ee

When $z \to 1$, we can use \eqref{eq:PP} and the kernel localizes to
\be
\label{eq:PPlocal}
{\cal P}_d(1,z_1,z_2) = 2 (4 \pi)^{d-2} {\cal N}_d (1 - z_1^2)^{d - 4 \over 2} \delta(z_1 - z_2)\ .
\ee
\par
We can get a similar identity to \eqref{eq:PPP} for the Mandelstam kernel,
\be
\label{eq:MandelstamKP}
K_d (z, \eta_1, \eta_2) &\equiv \int\limits_{-1}^1 dz_1 \int\limits_{-1}^1 dz_2\, {\mathcal{P}_d(z,z_1,z_2) \over (\eta_1 - z_1)(\eta_2 - z_2)} \ ,
\ee
by using the definition of the Gegenbauer function of the second kind, eq. \eqref{eq:Qfunc}, to get \eqref{eq:PQQ}.
%\be
%K_d (z, \eta_1, \eta_2) =4 (4 \pi)^{d-2} {\cal N}_d^2 (\eta_1^2-1)^{d - 4 \over 2} (\eta_2^2-1)^{d - 4 \over 2}\sum_{J=0}^\infty n_J^{(d)} P_J^{(d)}(z) Q_J^{(d)}(\eta_1) Q_J^{(d)}(\eta_2). \nn \tag{\ref{eq:PQQ}}
%\ee

We can get a hint at the analytic structure of the Mandelstam kernel from \eqref{eq:PQQ}. In terms of $\eta'$ and $\eta''$ the kernel shares the $[-1,1]$ branch cut of $ Q_J^{(d)}$. The situation is more interesting in the $z$ plane. Given that $P_J^{(d)}(z)$ is a polynomial in $z$, analytic everywhere, $K_d (z, \eta', \eta'')$ can only be singular whenever $z$ is such that the sum in $J$ no longer converges. Indeed, when $J \to \infty$ we have $P_J^{(d)}(z) \sim \lambda(z)^J$ and $Q_J^{(d)}(z) \sim \lambda(z)^{-J}$ (see appendix \ref{sec:Q}), 
and the series diverges when 
\be
\lambda(z) = \lambda(\eta') \lambda(\eta'), \, \, \text{              or          } \, \,z = \eta_+ \, ,
\ee
which signals the singularity of the kernel as deduced in appendix A.
\par
Representation \eqref{eq:PQQ} makes the symmetries of the kernel manifest. 
In particular, $K_d(z, \eta', \eta'')$ is symmetric in its last two arguments, and further obeys 
\be
K_d (-z, \eta', \eta'') = - K_d (z, -\eta', \eta'') = - K_d (z, \eta', - \eta'')\ ,
\ee
where we used $Q_J^{(d)}(-z) = (-1)^{J+d-3} Q_J^{(d)}(z)$ and $P_J^{(d)}(-z) = (-1)^J P_J^{(d)}(z)$. This symmetry of the kernel is responsible for the $t-u$ symmetry of the double spectral density \eqref{eq:doubleSD}.
\par
Finally, let us derive \eqref{eq:Atkinson}. We start with integer $J$ and $|\eta_1|, |\eta_2| > 1$. We take \eqref{eq:gaf} and apply to it $\iint\limits_{-1}^1  {dz_1 dz_2 \over (\eta_1 - z_1)(\eta_2 - z_2)}$. In this way we get 
\be
\label{eq:178}
{1\over8}\pi^{2 -d \over 2}\Gamma\left(\tfrac{d - 2}{2}\right) \int\limits_{-1}^1dy\, K_d(y,\eta_1,\eta_2)(1-y^2)^{d-4\over2} P^{(d)}_J(y)=(\eta_1^2-1)^{d-4\over2} Q^{(d)}_J(\eta_1) \, (\eta_2^2-1)^{d-4\over2} Q^{(d)}_J(\eta_2)\ ,
\ee
where we used the definition of $K_d$ in terms of $\mathcal{P}_d$ \eqref{eq:MandelstamKP} and definition of $Q^{(d)}_J$ in terms of $P^{(d)}_J$ \eqref{eq:Qfunc}. Noting that $P^{(d)}_J$ is related to the discontinuity of $Q^{(d)}_J$, see \eqref{eq:Qdisc}, we write the integral as an anticlockwise contour around $[-1,1]$ 
\be
\label{eq:236}
\int\limits_{-1}^1dy\,  K_d(y,\eta_1,\eta_2)  (1-y^2)^{d-4\over2} P^{(d)}_J(y)=  {1 \over  \pi i} \oint\limits_{[-1,1]}d\eta\, K_d(\eta,\eta_1,\eta_2)  (\eta^2-1)^{d-4\over2} Q^{(d)}_J(\eta) \ ,
\ee
where we used the fact that for $|\eta_1|, |\eta_2| > 1$ the kernel is analytic in a finite region around $[-1,1]$. We now want to blow up this contour to infinity to pick up the branch cut of the Mandelstam kernel \eqref{Kdge4}. At infinity we have $K_d(\eta \to \infty,\eta_1,\eta_2) \sim {\log \eta \over \eta}$ for $d > 3$ and $\sim {1 \over \eta}$ for $d = 3$. Given that $Q_J(\eta \to \infty) \sim \eta^{3-d-J}$ we get that the integrand goes like $\sim \eta^{-J-2}$, which gives a null contribution to the arc at infinity for $\mathrm{Re} J > -1$. In this way we arrive at
\be
\label{eq:237}
{1 \over  \pi i} \oint\limits_{(-1,1)}d\eta\, K_d(\eta,\eta_1,\eta_2)  (\eta^2-1)^{d-4\over2}Q^{(d)}_J(\eta)  = {2 \over \pi} \int\limits_{\eta_+}^\infty d\eta\, {\rm Disc}_\eta K_d(\eta, \eta_1, \eta_2) (\eta^2-1)^{d-4\over2} Q^{(d)}_J(\eta)\ .
\ee
Note that in the main text we included $\theta(\eta-\eta_+)$ in the discontinuity of the kernel, see \eqref{eq:kernelpositivity}. In the formula above, which is valid for complex $(\eta_1, \eta_2)$, the variable $\eta$ is integrated from $\eta_+$ to $\infty$ and we can simply use 
\be
{\rm Disc}_\eta K_d(\eta, \eta_1, \eta_2) = {4\pi^{d+1\over2} \over \Gamma({d-3\over 2})} {(\eta^2 - 1)^{4-d \over 2}  \over (\eta-\eta_-)^{5-d\over2}(\eta-\eta_+)^{5-d\over2}} .
\ee
Plugging \eqref{eq:237} and \eqref{eq:236} back into \eqref{eq:178} yields \eqref{eq:Atkinson}.
\be
\int\limits_{\eta_+}^\infty d \eta (\eta^2-1)^{{d-4 \over 2}} Q_J^{(d)}(\eta) {\rm Disc}_\eta K_d(\eta, \eta_1, \eta_2) ={4 \pi^{d/2} \over \Gamma({d-2 \over 2})}(\eta_1^2-1)^{d-4\over2} Q^{(d)}_J(\eta_1) \, (\eta_2^2-1)^{d-4\over2} Q^{(d)}_J(\eta_2)\ . \tag{\ref{eq:Atkinson}}
\ee
This is valid for complex $\eta_1, \eta_2$ satisfying $|\eta_1|, |\eta_2|>1$ and integer $J$. The integral is taken along the path that does not cross any cuts of the integrand, e.g. ${\rm arg}[\eta]={\rm arg}[\eta_+]$.

The above equation can be continued in spin. Indeed, note that both sides of (\ref{eq:Atkinson}) are manifestly analytic in spin $J$ for $\mathrm{Re}[J] > - 1$ and coincide for positive integer $J$. To argue that they coincide for any $J$ we also need to check the growth at infinity. One can check that both sides of (\ref{eq:Atkinson}) have a large $J$ asymptotic $\lambda(\eta_+)^{-J}$ and therefore the conditions of Carlson's theorem are satisfied.

\section{The $Q_J^{(d)}(z)$ Large $J$ Expansion}
\label{sec:Q}
The large $J$ expansion of the $Q$-function is given in (\ref{QexpA}) and (\ref{QexpA2}). The aim of this appendix is to argue that there are no nonperturbative corrections to this expansion of the form $\lambda(z)^{-\alpha J}$, with $\alpha>1$. This fact is used in section \ref{sec:Karplus} to derive the Landau curves in the elastic region. We consider separately the cases when the spacetime dimension is even and when it is odd.

\paragraph{Odd dimensions}\label{Qappendix}
In odd spacetime dimensions the $Q$-function (\ref{eq:Qfunc}) takes a simple form
\beq\label{Qodd}
Q_J^{(d\text{ odd})}(z)={2^{d-4}\sqrt\pi\,\Gamma({d-2\over2})\over\lambda(z)^J(\lambda(z)^2-1)^{d-4}}{\Gamma(J+1)\over\Gamma(J+d-3)}{\mathbb P}_{d-5\over2}(J,\lambda(z)^2-1)
\eeq
Where ${\mathbb P}_n(J,x)$ is a polynomial of degree $n$ in $x$ and $J$. For example, we have
\beq
{\mathbb P}_{-1}={1 \over Jx}\ ,\qquad{\mathbb P}_{0}=1\ ,\qquad {\mathbb P}_{1}=x(J+3)+2\ ,\qquad {\mathbb P}_{2}=x^2(J+5)(J+4)+6x(J+5)+12\ .
\eeq
This form makes the large $J$ expansion trivial.

\paragraph{Even dimensions} 

For even spacetime dimensions, in analogues to (\ref{Qodd}), the $Q$-function takes the form
\beq\label{Qeven}
Q_J^{(d\text{ even})}(z)={(-1)^{d\over2}\over2}P_J^{(d)}(z)\log{z+1\over z-1}+\({z^2\over z^2-1}\)^{d-4\over2}{\mathbb K}^{(d)}_{J-1}(z)\ ,
\eeq
where ${\mathbb K}^{(d)}_n(x)$ is a polynom of degree $n$.\footnote{This form can be derived by expanding $(1-z'^2)^{d-4\over2}P_J^{(d)}(z')$ in (\ref{eq:Qfunc}) in powers of $(z'-z)$. An explicit way of fixing ${\mathbb K}_{J-1}^{(d)}$ from $P_J^{(d)}$ is be demanding that $Q_J^{(d)}$ decays as in (\ref{Qasimp}). From (\ref{Qeven}) we see that the branch-cut discontinuity in odd $d$ is replaced by a logarithmic one for even $d$.} Because the degree of the polynom depends on $J$, this form is not so useful for understanding the large $J$ expansion. 

Instead, we consider the exact integral (\ref{eq:masterintegral}) for any integer $n$. We observe that on the right hand side of that equation we have again a $Q$-function, but now in odd spacetime dimension instead of an even one. By expanding both sides of that equation at large $J$ (and fixed $n$) one can map between the coefficients of the large $J$ expansion of the $Q$-functions in even and odd spacetime dimensions. Now supposed $Q^{(d\text{ even})}$ had an $\lambda(z)^{-\alpha J}$-type correction. Assuming no cancellations, such a correction would result in an analogous correction in the expansion of $Q^{(d\text{ odd})}(z_1)$ in the right hand side of (\ref{eq:masterintegral}). From the above however it is clear that corrections of this type are absent.

\section{Gribov's Theorem}
\label{sec:gribovstheorem}

It is possible to use elastic unitarity condition continued in spin  $J$ (\ref{eq:generalFJ}) to constrain the high-energy asymptotic of the discontinuity of the amplitude \cite{Gribov:1961ex,Gribov:1961fm}.

Consider the following ansatz for the discontinuity of the amplitude
\be
\label{eq:discTG}
\lim_{t \to \infty} T_t(s,t) = B(s,\log t)\, t^{\alpha(s)}\ ,\qquad  \alpha(s) \in \mathbb{R}\ ,\qquad  4 m^2 < s < 16 m^2 .  
\ee
where $B(s,\log t)$ is a slowly varying function of $t$ that grows slower than a power, e.g. $B(s,\log t) \sim (\log t)^{q(s)}$.

{\bf Gribov's Theorem:} Let us assume the high energy behavior of the discontinuity (\ref{eq:discTG}) in the elastic region $4 m^2 < s< 16 m^2$. Elastic unitarity then implies that
\be
\label{eq:grint}
\int^{\infty} d \log t \ B(s,\log t) < \infty \ .
\ee
Historically, Gribov's theorem excluded the classical picture of diffraction from a black body $T_t^{(+)}(s,t) = B(s) t$ in QFT.

The easiest way to prove Gribov's theorem is to note that if the integral (\ref{eq:grint}) diverges $f_J(s)$ develops a singularity on the real axis at $J = \alpha(s)$. Taking $J = \alpha(s) + \eps$ where $0<\eps \ll 1$ and real, we get from the Froissart-Gribov formula
\be
f_{\alpha(s)+\eps}(s) \sim \int^\infty d x \ B(s,x)e^{- \eps x}.
\ee
Elastic unitarity close to the leading Regge singularity $J = \alpha(s)$ then takes the schematic form
\be
\label{eq:gribovC}
\int^\infty d x \ {\rm Im} B(s,x)e^{- \eps x} &\propto \left|\int^\infty d x \ B(s,x)e^{- \eps x} \right|^2 \nn \\
&= \left( \int^\infty d x \ {\rm Re} B(s,x)e^{- \eps x} \right)^2 + \left( \int^\infty d x \ {\rm Im} B(s,x)e^{- \eps x} \right)^2 ,
\ee
which can be only consistent if (\ref{eq:grint}) holds. Indeed, otherwise we get that the singularity in the RHS of (\ref{eq:gribovC}) does not match the singularity in the LHS of (\ref{eq:gribovC}).

A simple, and physically natural, way out of the contradiction is to assume that $\alpha(s) \in \mathbb{C}$. Indeed, consider a model, where the leading Regge trajectory is given by a single Regge pole
\be
f_J(s) = {\beta(s) \over J - \alpha(s)}+ \dots  ,\qquad  {\rm Im}[\alpha(s)] \neq 0\ ,\qquad  4m^2<s<16m^2 \ .
\ee
This corresponds to the discontinuity of the amplitude that takes the form $T_t(s,t) = \beta(s) t^{\alpha(s)}$.

Let us now impose elastic unitarity (\ref{eq:generalFJ}) at $J = \alpha^* (s)$. We get that the solution is
\be
\label{eq:solutionEU}
\beta(s) =  {2 \sqrt s \over (s - 4 m^2)^{{d-3 \over 2}}} {\rm Im}\alpha(s) \ .
\ee

In $d>3$ the consideration above tacitly assumed that ${\rm Re} \alpha(s)> -1$. This is related to the fact that in $d>3$, as can be seen explicitly from (\ref{eq:Qfunc}), $Q^{(d)}_J(z)$ develop a pole at $J=-1$ and we run into the same problem as above where the singularities do not match in the elastic unitarity equation.

It would be interesting to understand better properties of the full nonperturbative leading Regge trajectory in the complex $s$ plane, see e.g. discussion in \cite{Kupsch:2008hq} for some common assumptions. The properties of the leading Regge trajectory in the planar theory are relatively well-understood, see e.g. \cite{Brower:2006ea,Caron-Huot:2016icg}.

\section{Karplus Curves from the Mandelstam Equation}
\label{sec:Mandelstam_Karplus}

The functional shape of the Landau curves in the elastic strip can be derived by impose the consistency between the positions of the thresholds of $T_t(s,t)$ (\ref{thresholds}) and those of $\rho(s,t)$ (\ref{thresholds2}) with elastic unitarity. In section \ref{sec:Karplus} we have done so by imposing elastic unitarity at the level of the partial waves. In this appendix we use the Mandelstam equation instead to generate the same curves.

The Mandelstam equation \eqref{eq:doubleSD} can be written as
\be\label{eq:rhoK}
\rho(s,t)= \mathrm{K}[\cT_t^{(+)},\cT_t^{(-)}](s,z_s(t))\ ,
\ee
where we have introduced the functional  
\be\label{eq:MandelstamK}
\mathrm{K}[A,B](s,z) \equiv {(s - 4 m^2)^{d-3\over2} \over 4\pi^2 (4\pi)^{d-2} \sqrt{s} }\int\limits_{z_1}^{\infty} d \eta'  \int\limits_{z_1}^{\infty} d \eta''\,A(s, \eta') B(s, \eta'')\, {\rm Disc}_z K(z, \eta', \eta'')\ .
\ee
Permutation symmetry of the Mandelstam kernel, $K_z(z, \eta', \eta'') = K_z(z, \eta'', \eta')$, implies symmetry of the functional $\mathrm{K}[A,B] = \mathrm{K}[B,A]$. Due to real analyticity $\[\cT_t^{(+)}\]^* = \cT_t^{(-)}$, it then follows from \eqref{eq:rhoK} that $\rho(s,t)$ is real when $s,t$ are real and positive. 

The structure of $T_t(s,t)$ in physical ($t$-channel) kinematics is given in (\ref{thresholds}).
As we continue $s$ out of that region, $T_t^{2 \to 2n}(s,t)$ may develop a discontinuity at the Karplus curves. Given that $\rho(s,t) \equiv \mathrm{Disc}_s T_t(s,t) = \im_s T_t(s,t)$, we can separate $T_t(s,t)$ into real and imaginary parts as
\be
\label{eq:Ttdecomp}
T_t^{(\pm)}(s,t) = R(s,t) \pm i \rho(s,t)\ ,
\ee
where for physical $s$, $\rho(s,t)$ vanishes and $R(s,t)$ is simply given by the normal thresholds of \eqref{thresholds}. It can be written schematically as
\be
\label{eq:firstDisc}
R(s,t) \sim \sum_{n=1}^{\infty} \Theta(z-z|_{t=2nm^2})\ ,\qquad 4-t < s < 0\ .
\ee

By plugging \eqref{eq:Ttdecomp} into \eqref{eq:rhoK} we get that 
\be\label{eq:unirhoK}
\rho = \mathrm{K}[R,R] + \mathrm{K}[\rho,\rho]\ .
\ee
We now take $R$ to just be given by \eqref{eq:firstDisc} as a seed to \eqref{eq:unirhoK} and iterate this equation to find the minimal consistent set of Landau curves in (\ref{thresholds2}) that is consistent with this equation.

Note that for $s > 4m^2$, $R$ can have additional  discontinuities in addition to normal thresholds. Hence, at each step we correct $R$ with the additional thresholds of $\rho$ that are generated through the iteration of \eqref{eq:unirhoK}. In the end we check the iteration procedure converges to a closed set of Landau curves.

The first contribution to $\rho$ comes from inserting \eqref{eq:firstDisc} into \eqref{eq:unirhoK}
\be\label{eq:karplus1}
\mathrm{K}[R,R](s,z) &\sim \sum_{n,m=1}^\infty \int\limits_1^\infty d\eta' d\eta'' \Theta(\eta' - z_n) \Theta(\eta'' - z_m) \Theta\(z - \eta^+(\eta',\eta'')\)    \nn \\
&\sim \sum_{n,m=1}^\infty \Theta\(z - \eta^+(z_n,z_m)\) = \sum_{n,m}^\infty \Theta\(\lambda(z) - \lambda(z_n) \lambda(z_m) \)\ .
\ee
We see that normal thresholds generate a first set of Landau curves for $\rho(s,t)$ given by (\ref{nmLandauc}) 
via the $\mathrm{K}[R,R]$ term. 
Including the \eqref{eq:karplus1} Karplus curves into $R$ requires, from consistency with \eqref{eq:unirhoK}, an additional set of ``cubic" curves,
\be
\mathrm{K}[R,\mathrm{K}[R,R]](s,z) &\sim \sum_{n,m,l=1}^\infty \int\limits_1^\infty d\lambda' d\lambda'' \Theta(\lambda' - \lambda(z_n)) \Theta(\lambda'' - \lambda(z_m) \lambda(z_l)) \nn \\
& \sim \sum_{n,m,l=1}^\infty \Theta\(\lambda(z) - \lambda(z_n) \lambda(z_m) \lambda(z_l) \),
\ee
and also ``quartic" curves,
\be
K[K[R,R],K[R,R]](s,z) &\sim \sum_{n,m,l,k=1}^\infty \int\limits_1^\infty d\lambda' d\lambda'' \Theta(\lambda' - \lambda(z_n) \lambda(z_m)) \Theta(\lambda'' - \lambda(z_k) \lambda(z_l) ) \Theta\(\lambda(z) - \lambda' \lambda'' \)   \nn \\
&\sim \sum_{n,m,l,k=1}^\infty \Theta\(\lambda(z) - \lambda(z_n) \lambda(z_m) \lambda(z_k) \lambda(z_l) \).
\ee
Iterating further, we find that a natural set of Landau curves consistent with (\ref{eq:unirhoK}) and the existence of normal thresholds is 
\be
\label{eq:Rmaximal}
R\(s,t(z)\) \sim \sum_{L=1}^\infty \, \sum_{\{n_1, \dots, n_{L}\}}^\infty \Theta(\lambda(z) - \lambda(z_{n_1}) \cdots \lambda(z_{n_{L}}) ) \ ,
\ee
and
\be
\label{eq:rhomaximal}
\rho\(s,t(z)\) \sim \sum_{L=2}^\infty \, \sum_{\{n_1, \dots, n_{L}\}}^\infty \Theta(\lambda(z) - \lambda(z_{n_1}) \cdots \lambda(z_{n_{L}}) ) \ ,
\ee
where importantly $4 m^2 < s < 16 m^2$. Note that the difference between the supports of \eqref{eq:Rmaximal} and \eqref{eq:rhomaximal} is precisely \eqref{eq:firstDisc}.

\par
Below we present the curves that asymptote to $t = 16m^2, \,36m^2 ,\, 64 m^2$.
\be
\label{eq:20}
t_{\{2,0,\dots\}} = {16 m^2 s \over s - 4m^2}\ ,\qquad t_{\{3,0,\dots\}} = {36m^2(s + {4 m^2 \over 3})^2 \over (s - 4m^2)^2}\ ,\qquad  t_{\{4,0,\dots\}} = {64m^2s(s+4 m^2)^2 \over (s - 4m^2)^3} , 
\ee
\be
\label{eq:11}
t_{\{1,1,0,\dots\}} = {20m^2 s + 16m^2\sqrt{s(s+12m^2)} + 48m^4 \over s - 4m^2}\ ,\qquad t_{\{0,2,0,\dots\}} = {64m^2(s+12m^2) \over s - 4m^2} , 
\ee
\be
\label{eq:101}
t_{\{1,0,1,0,\dots\}} &= {128m^4 + 40m^2 s + 24m^2 \sqrt{s(s+32m^2)} \over s - 4m^2}, \nn \\
t_{\{2,1,0,\dots\}} &= \frac{16m^2 \left(s+2 m \sqrt{s}+2 m \sqrt{s+12m^2}+\sqrt{s (s+12m^2)}+8m^2\right)^2}{\left(\sqrt{s}-2m\right)^2 \left(s+8m \sqrt{s+12m^2}+28m^2\right)} \ .
\ee
The curves are plotted below.
\begin{figure}[H]
  \centering
    \includegraphics[width=0.65\textwidth]{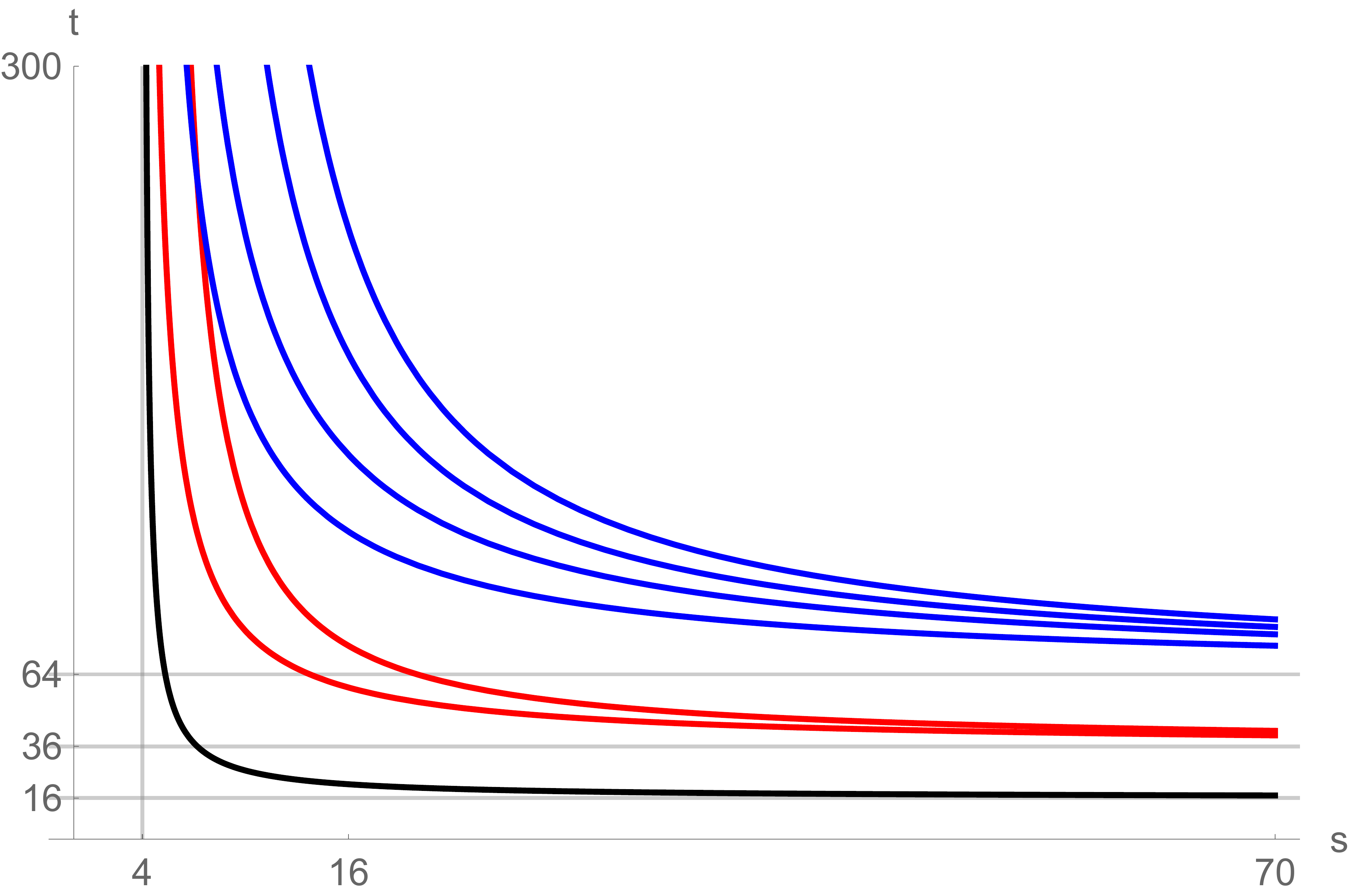}
  \caption{\small Plot of the Landau curves given by equations \eqref{eq:20} to \eqref{eq:101} for $m=1$. The curves are organized by color according to the asymptote at $t \to \infty$. Black is the leading curve $t_{\{2,0,0,\dots\}}$. Red curves obey $t_{\{1,1,0,\dots\}} < t_{\{3,0,\dots\}}$. Blue curves obey $t_{\{1,0,1,0,\dots\}} < t_{\{0,2,0,\dots\}} < t_{\{2,1,0,\dots\}} < t_{\{4,0,\dots\}}$.}
\label{fig:karplus2}
\end{figure}

\section{Threshold Expansion for Non-Integer $J$}
\label{sec:nonintspinTh}

It is interesting to ask about the continuation of the formula (\ref{eq:elasticunitarityFJS}) for the solution to elastic unitarity to non-integer $J$. For a related discussion, see e.g. \cite{ThresholdRegge}.

The starting point is the observation that the Froissart-Gribov formula can be written in the following form
\be
\label{eq:FGJte}
{f_J(s) \over (s-4 m^2)^J} &=2 {(16\pi)^{2-d\over2}\over\Gamma\({d-2\over2}\)} \int\limits_{z_1}^\infty {d z \over \pi} \,(z^2-1)^{{d-4 \over 2}} {Q^{(d)}_J(z) \over (s-4 m^2)^J } T_t(s,t(z)) \ 
\ee
admits a simple analytic continuation to $s<4 m^2$ for real $J$.

Indeed, $Q^{(d)}_J(z)$ for non-integer $J$ have a branch point at $s=4m^2$ or, equivalently, $z=\infty$. On the other hand, ${Q^{(d)}_J(z) \over (s-4 m^2)^J }$ has only branch cut for $z \in [-1,1]$ and satisfies
\be
(z^2-1)^{{d-4 \over 2}} {Q^{(d)}_J(z) \over (s-4 m^2)^J } = - ((-z)^2-1)^{{d-4 \over 2}} {Q^{(d)}_J(-z) \over (4 m^2-s)^J } ,
\ee
for $|z|>1$ as can be easily seen from (\ref{eq:Qfunc}).

Switching in the RHS of (\ref{eq:FGJte}) to the integral $\int\limits_{4m^2}^{\infty} dt$ we can continue $f_J(s)$ to $s<4 m^2$. Together with the fact that $T_t(s,t(z))$ is positive and real for $0<s< 4 m^2$ we conclude that ${f_J(s) \over (s-4 m^2)^J}$ is real analytic function of $s$ with a branch cut starting at $s=4m^2$. Moreover, ${f_J(s) \over (s-4 m^2)^J}$ is positive and real for $0<s<4m^2$.

Let us now impose continued in spin elastic unitarity (\ref{eq:generalFJ}). It is convenient to rewrite it as follows
\be
{1 \over i} \left( {(s-4 m^2)^J \over f_{J}(s+i \eps)} - {(s-4 m^2)^J \over f_{J}(s-i \eps)}  \right)=-{(s-4 m^2)^{J+{d-3\over 2}}\over\sqrt s} .
\ee

The general solution to it takes the form
\be
\label{eq:elasticunitarityFJSb}
f_J(s) &= {(s-4 m^2)^J\over \tilde b_J(s) + {e^{- i \pi (J+{d-3 \over 2})} \over \sin \pi (J+{d-3 \over 2})} {(s-4m^2)^{J+{d-3\over2}} \over 2 \sqrt{s}} }\ ,\qquad  s>4 m^2 , 
\ee
where $\tilde b_J(s)$ are analytic around $s=4m^2$ and $b_J(4) \sim a_J > 0$ for $J \geq 2$.

Few comments are in order. Let us first discuss how the formula above reduces to (\ref{eq:elasticunitarityFJS}) when $J$ is even integer. In even $d$ it is trivial upon identifying $\tilde b_{J=2 k}(s) = (s-4m^2)^{2k} b_{2k}(s)$. In odd $d$ the situation is more subtle because in this case ${e^{- i \pi (J+{d-3 \over 2})} \over \sin \pi (J+{d-3 \over 2})} = {e^{- i \pi J} \over \sin \pi J}$ which develops a pole for even $J$. The residue of this pole ${(s-4m^2)^{J+{d-3\over2}} \over 2 \sqrt{s}}$ however is analytic around $s=4m^2$ and therefore it can be canceled by $\tilde b_J(s)$ if we impose that
\be
\lim_{J \to 2 k} \tilde b_J(s) = - {1 \over J - 2 k } {(s-4 m^2)^{2k+{d-3 \over 2}} \over 2 \pi} + (s-4 m^2)^{2k}b_{2 k}(s) + O(J-2k)\ ,\qquad  J> {\rm Re}\alpha(s) ,  ~~~ d \ {\rm odd}, 
\ee
the $J^0$ term then correctly reproduces (\ref{eq:elasticunitarityFJS}).

Secondly, note that when $|J| \to \infty$ the ratio  ${e^{- i \pi (J+{d-3\over2})} \over \sin \pi (J+{d-3\over2})}$ is polynomially bounded which is consistent with the expected behavior of $J_J(s)$ at infinity. This poses a potential problem in even $d$ for half-integer $J$ and in odd $J$ integer $J$. 

Let us consider even $d$ first. In this case ${e^{- i \pi (J+{d-3 \over 2})} \over \sin \pi (J+{d-3 \over 2})} = - i {e^{- i \pi J} \over \cos \pi J}$ develops a pole at odd $J$ which corresponds to zero of $f_J(s)$.
Therefore unless there is a cancellation between the two terms in the denominator of (\ref{eq:elasticunitarityFJS}) we have that $f_{{1 \over 2} +  {\mathbb Z}} = 0$. If this is the case via Carlson theorem we then conclude that $f_J (s) = 0$ identically. Therefore, an infinite number of poles should cancel with the corresponding poles in $\tilde b_J(s)$. In this way we get
\be
\lim_{J \to k-{1 \over 2}} \tilde b_{J}(s) =-  {1 \over J- (k-{1 \over 2})} {(s-4 m^2)^{k+{d-4 \over 2}} \over 2 \pi} \ + \dots \ ,\qquad  J> {\rm Re}\alpha(s) ,  ~~~ d \ {\rm even} .
\ee
Note again that in even $d$ $(s-4 m^2)^{k+{d-4 \over 2}}$ is analytic at $s=4m^2$ which is consistent with the predicted property of $\tilde b_{J}(s)$.

Finally, in odd $d$ for odd integer $J$ we get the following cancellation condition
\be
\lim_{J \to 2 k - 1} \tilde b_J(s) = - {1 \over J - (2 k - 1) } {(s-4 m^2)^{2k+{d-5 \over 2}} \over 2 \pi} + \dots \ ,\qquad  J> {\rm Re}\alpha(s)  ~~~ d \ {\rm odd}. 
\ee

\section{Threshold expansion in $J$-space: Technical Details}
\label{sec:derivation}

In this appendix we collect various results and technical details that are relevant to the inversion of the threshold expansion using the Froissart-Gribov formula performed in section \ref{sec:Jexact}.

We start with the derivation of (\ref{eq:masterintegral}). Consider first $d$ to be even. We would like to evaluate the following integral
\be
\label{eq:froissartgr}
I_{n,J}^{(d)}(z_1) &\equiv {2 {\cal N}_d \over \pi} \int\limits_{z_1}^{\infty} d z (z^2-1)^{{d-4 \over 2}} Q_J^{(d)}(z) {(z_1-1)^{{d-3 \over 2} - n} \over (z - z_1)^{{d-3 \over 2} - n} } = {2 {\cal N}_d \over \pi} (z_1-1)^{{d-3 \over 2} - n} I(d,n,z_1) \nn , \\
I(d,n,z_1) &\equiv \int\limits_{z_1}^{\infty} d z (z^2-1)^{{d-4 \over 2}} Q_J^{(d)}(z) {1 \over (z - z_1)^{{d-3 \over 2} - n}} \ ,\qquad  n \in \mathbb{Z} \ \ ,\qquad  n \geq 0 \ .
\ee
We would like to do the integral for general $J$. The strategy is to do the integral first for integer $J$ exactly and then analytically continue it to arbitrary $J$.

As a first step we note that the integrand can be interepreted as a discontinuity of some simple function
\be
{\rm Disc}_t {1 \over \sin \pi ({d-3 \over 2} - n) } {1 \over (z_1 - z)^{{d-3 \over 2} - n} }  = {1 \over (z - z_1)^{{d-3 \over 2} - n}} .
\ee 
For integer $n$ and $d$ this only holds for even $d$, where the power is half-integer and therefore we have a square-root type discontinuity.

Therefore if we interpret the integrand as a discontinuity $T_t (s,t)={1 \over (z - z_1)^{{d-3 \over 2} - n}}$ of the amplitude  $T(s,t) ={1 \over \sin \pi ({d-3 \over 2} - n) } {1 \over (z_1 - z)^{{d-3 \over 2} - n} }$ then (\ref{eq:froissartgr}) is nothing but the Froissart-Gribov integral for this amplitude! In this way we can immediately rewrite it as follows
\be
\label{eq:rewrote}
I(d,n,z_1)  = {\pi \over 2} {(-1)^{n} \over \sin \pi {d-3 \over 2} } \int\limits_{-1}^{1} d z (z^2-1)^{{d-4 \over 2}} P_J^{(d)}(z) {1 \over (z_1 - z)^{{d-3 \over 2} - n} } .
\ee 
Here we used that $J$ is integer and dropped the contour at infinity which requires $J>n-{d-3 \over 2}$.

We note that the integral (\ref{eq:froissartgr}) satisfies a very simple recursion relation based on the identity
\be
\label{eq:recu}
\pa_{z_1} I(d,n,z_1)  =- (n-{d-3 \over 2}) I(d,n-1,z_1).
\ee
Therefore we can first compute the integral for $n=0$ and then use this differential equation to compute the integral for $n>0$.

Let us now find explicitly $I(d,0,z_1)$. To do this let us do the following change of variable
\be
{1 \over t} = \lambda(z_1) \equiv z_1 + \sqrt{z_1^2 - 1}. 
\ee 
We can then write
\be
{1 \over (z_1 - z)^{{d-3 \over 2}} } &= {1 \over (2 t)^{{3-d \over 2}}} {1 \over (1 - 2 t z + t^2)^{{d-3 \over 2}}} = {1 \over (2 t)^{{d-3 \over 2}}}  \sum_{J=0}^{\infty}  C_J^{({d-3 \over 2})}(z)  t^{J} \nn \\
&= {1 \over (2 t)^{{3-d \over 2}}}  \sum_{J=0}^{\infty}  {\Gamma(d-3) \Gamma(J+1) \over \Gamma(d+J-3)} P_J^{(d)}(z)  t^{J} ,
\ee
where we used the relation between $P_J^{(d)}(z)$ and the Gegenbauer polynomials $C_J^{({d-3 \over 2})}(z)$
\be
P_J^{(d)}(z) = {\Gamma(d+J-3) \over \Gamma(d-3) \Gamma(J+1)} C_J^{({d-3 \over 2})}(z) .
\ee
Using the orthogonality property of $P_J^{(d)}(z)$ we immediately get
\be
\label{eq:resultZe}
I(d,0,z_1)  &= {\pi \over 2} {1 \over \sin \pi {d-3 \over 2} } {1 \over (2 t)^{{3-d \over 2}}}  {2 \over {\cal N}_d n_J^{(d)}} {\Gamma(d+J-3) \over \Gamma(d-3) \Gamma(J+1)} t^J = \nn \\
&={\pi \over \sin {\pi d \over 2}} {2^{{3 d -11 \over 2}} \Gamma({d-2 \over 2})^2 \over (2 J + d-3) \Gamma(d-3)}  \lambda(z_1)^{-(J+{d-3 \over 2})} .
\ee
The rest we can get trivially using (\ref{eq:recu}). Note that the final result holds in any $d$ and for any $J$.

The basic integral is the following
\be
\int d z_1 \lambda(z_1)^{-c} = {1 \over 2} \left( {\lambda^{-1-c} \over c+1} - {\lambda^{1-c} \over c-1} \right) .
\ee
Therefore, it is clear that we get the following result for the integral
\be
I(d,n,z_1) ={\pi \over \sin {\pi d \over 2}} {2^{{3 d -11 \over 2}} \Gamma({d-2 \over 2})^2 \over (2 J + d-3) \Gamma(d-3)} \lambda(z_1)^{n-(J+{d-3 \over 2})} \sum_{k=0}^n c_{k,n} \lambda(z_1)^{-2 k} ,
\ee
where $c_{k,n}$ can be explicitly found.

We have $c_{0,0}=1$. It is then easy to check that we have the following result
\be
\label{eq:resultFinalA}
I(d,n,z_1) = & \;{2^{{3 d -13 \over 2}} \Gamma({d-2 \over 2})^2  \Gamma({d-3 \over 2}) \Gamma(n + {5 - d \over 2}) \over \Gamma(d-3)}  \lambda(z_1)^{n-(J+{d-3 \over 2})}  \nn \\
& \qquad \qquad \times 2^{-n}\sum_{k=0}^n {n \choose k} {(-1)^k \over \prod_{j=0}^n (J + k - j + {d - 3 \over 2})  }\lambda(z_1)^{-2 k}.
\ee
\par
Also note that when $J \to \infty$, the denominator $\sim J^{n+1}$ factors out and the sum becomes Newton's binomial. We get
\be
\lim_{J \to \infty} I(d,n,z_1) = {2^{{3 d -13 \over 2}} \Gamma({d-2 \over 2})^2  \Gamma({d-3 \over 2}) \Gamma(n + {5 - d \over 2}) \over J^{n+1} \Gamma(d-3)} (z_1^2 - 1)^{n \over 2} \lambda(z_1)^{-(J+{d-3 \over 2})} 
\ee
where we used $\lambda(z_1) - \lambda^{-1}(z_1) = 2 z_1^2 - 2$.
\par
From the above we conclude
\be
\lim_{J \to \infty} { I(d,n+1,z_1) \over I(d,n,z_1)} =  \sqrt{z_1^2 - 1} \(n+ {5 - d \over 2}\) {1 \over J}   .
\ee
This is the essence of why a threshold expansion maps to a systematic large $J$ expansion through the Froissart-Gribov integral: Consecutive terms in the threshold expansion are roughly suppressed by $\sim {1 \over J}$ with respect to one another. 
\par
Note that the original integral $I(d,0,z_1) $ looks divergent in even $d\geq 6$, whereas the final result (\ref{eq:resultZe}) is finite. The resolution to this apparent contradiction is that when we changed the contour and wrote the integral as (\ref{eq:rewrote}) we implicitly used that the original integral was defined via the keyhole prescription. Therefore, the result (\ref{eq:resultFinalA}) is correct.

Formula (\ref{eq:resultFinalA}) can be rewritten as follows
\be
\label{eq:resultFinalB}
I(d,n,z_1) &=- {1 \over \pi}{2^{{3 d -12 \over 2}} \Gamma({d-2 \over 2})^2  \Gamma({d-3 \over 2}) \Gamma(n + {5 - d \over 2}) \over \Gamma(d-3)} \lambda(z_1)^{n-(J+{d-3 \over 2})} 2^{-n} \cos {\pi (d+2 J) \over 2} \tilde I \nn \\
\tilde I &\equiv \sum_{k=0}^n {n \choose k} (-1)^k \Gamma({3-d \over 2} - J - k) \Gamma({d-3 \over 2} +J + k - n)\lambda(z_1)^{-2 k} .
\ee

We now note that we can write
\be
\tilde I = \Gamma({3-d-2J \over 2}) \Gamma({2J - 2n +d -3 \over 2}) \ _2 F_1 ({d-3 \over 2} +J-n, - n, J+{d-1 \over 2}, {1 \over \lambda_1(z)^2}) ,
\ee
which we can now analytically continue away from integer $n$. Using the formula (\ref{QexpA}) for $Q_J^{(d)}(z)$ we finally get (\ref{eq:masterintegral}) which can be now checked to hold for arbitrary $J$, $d$ and $n$.

\subsubsection{Odd $d$}

In odd dimensions we are also interested in the following integral
\be
I_{log}(d,n,z_1,c_i) =  \int\limits_{z_1}^{\infty} d z (z^2-1)^{{d-4 \over 2}} Q_J^{(d)}(z) {(z-z_1)^{n-{d-3 \over 2}} \over \log^2 (z - z_1) + c_1 \log (z-z_1) + c_2 } .
\ee
This comes from plugging the threshold expansion into the Froissart-Gribov integral. In this case we note that
\be
(\pa_n^2 + c_1 \pa_n + c_2 ) I_{log}(d,n,z_1,c_i) =I(d,n,z_1) ,
\ee
where $I(d,n,z_1)$ was computed in the subsection above. It is easy to write a general solution to this differential equation, which should suffice for doing the integrals numerically for given $J$. We have not pursued this further in the present paper.

\subsection{Higher Order Corrections to $\rho(s,t)$}
\label{sec:higherrho}

Here we present some details for computing the double spectral density that comes from plugging $c_{n_1} c_{n_2} {(\eta' - z_1)^{n_1 - {d-3 \over 2}} \over (z_1 - 1)^{n_1 - {d-3 \over 2}} }  { (\eta'' - z_1)^{n_2 - {d-3 \over 2}} \over (z_1 - 1)^{n_2 - {d-3 \over 2}}}$ for the square of discontinuity of the amplitude in the Mandelstam equation (\ref{eq:doubleSD}).
We denote the result of this integration by $\rho_{n_1, n_2} (s,t(z))$. 

It admits the following expansion close to the leading Landau curve
\be
\rho(s,t(z))=\sum_{n_1,n_2}c_{n_1}c_{n_2}\rho_{n_1, n_2} (s,t(z))\ ,\qquad \rho_{n_1, n_2} (s,t(z)) \equiv {(s - 4 m^2)^{d-3\over2} \over 4\pi^2 (4\pi)^{d-2} \sqrt{s} } \tilde I_{n_1,n_2}^{(d)}(z)\ ,
\ee
where $\tilde I_{n_1,n_2}^{(d)}(z)$ was defined in (\ref{eq:deftilde}).
This can be expanded close to the leading Landau curve
\be
\label{eq:thrExpRho}
\rho_{n_1,n_2}(s,t(z) ) ={1 \over (z_1-1)^{n_1+n_2-(d-3)}} {(z - (2 z_1^2 - 1) )^{n_1+n_2} \over (z-(2 z_1^2 - 1))^{{d-5 \over 2}} } \sum_{m=0}^\infty d_{n_1,n_2;m}(s) (z - (2 z_1^2 - 1) )^m\  ,
\ee
where  $d_{n_1,n_2;m}(s)$ are the coefficients that we would like to compute.

Instead of computing the integral (\ref{eq:doubleSD}) we can use (\ref{eq:froissartgr}) to get ${\rm Im}f_J(s)$ and then extract $d_{n_1,n_2;m}(s)$ by imposing elastic unitarity order by order in ${1 \over J}$. The result takes the following form 
\be
d_{n_1,n_2;0}(s) =2^{{d-3 \over 2}} {{\cal N}_d \over \sqrt{\pi} } {(s-4 m^2)^{d-3\over2}\over\sqrt s} {\lambda_1^{2+n_1 + n_2} (\lambda_1^{2}- 1)^{n_1 + n_2} \over (\lambda_1^4- 1)^{n_1+n_2+1}} { \Gamma({d-2 \over 2}) \Gamma(n_1+{5-d \over 2})  \Gamma(n_2+{5-d \over 2}) \over \Gamma(n_1+n_2 +{7 - d \over 2})}\ ,
\ee
where $\lambda_1 \equiv \lambda(z_1)$.

Proceeding to higher orders we get
\be
\label{eq:resultrho}
{d_{n_1,n_2;m}(s) \over d_{n_1, n_2; 0}(s)} = {\Gamma(n_1+1+m) \over \Gamma(n_1+1)}  {\Gamma(n_2+1+m) \over \Gamma(n_2+1)}  \ { \lambda_1^{2m} \over (\lambda_1^4-1)^{2m} } {\Gamma(n_1+n_2 +{7 - d \over 2}) \over \Gamma(n_1+n_2 +{7 - d \over 2} + m)} {(-1)^m 2^m \over \Gamma(m+1)} {\rm poly}_m \ ,
\ee
where ${\rm poly}_m$ does not depend on $d$ and takes the form
\be
{\rm poly}_m = \sum_{i=0}^{2m} \lambda_1^{2 i} c_{i,m} .
\ee

We quote here some results on the properties of the polynomial ${\rm poly}_m$
\be
c_{0,m} &= c_{2m,m} = 1 \ ,\qquad c_{i,m} = c_{2m-i,m} \ , \nn \\
c_{1,m} & = - m \left( {n_2 \over n_1+1} + {n_1 \over n_2 +1} \right) \ , \nn \\
c_{2,m} &=m^2 +{m(m-1) \over 2} \left( {n_2 (n_2-1) \over (n_1+1)(n_1+2)} + {n_1 (n_1-1) \over (n_2+1)(n_2+2)} \right)
\ee

We can also write down an explicit result for $n_1 = n_2 = 0$
\be
c_{i,m} &={1+(-1)^i \over 2} {m \choose {i \over 2} }^2 ,
\ee
so that
\be
{\rm poly}_m = \ _2 F_1 (-m,-m,1,\lambda_1^4) \ ,\qquad  n_1=n_2 = 0 .
\ee
For $n_1 = n_2 =1$ and $n_1 =0, n_2 =1$  we get
\be
c_{i,m} &=(-1)^i {m \choose [{i \over 2}]} {m \choose [{i+1 \over 2}]} \nn \\
&=(-1)^i \left( {1+(-1)^i \over 2} {m \choose {i \over 2} }^2 + {1-(-1)^i \over 2} {m \choose {i-1 \over 2} } {m \choose {i+1 \over 2} }   \right) \ ,
\ee
where $[x]$ stands for the integer part of $x$. Similarly one can write an explicit result for ${\rm poly}_m$ in terms of hypergeometric functions.

Note that the threshold expansion of the double spectral density (\ref{eq:thrExpRho}) does not reflect the behavior of $\rho_{n_1,n_2}(s,t(z) )$ at large $z \gg 1$. Indeed, one can check that it takes the form $\rho_{n_1,n_2}(s,t(z) ) \sim z^{{\rm max}[n_1, n_2]-{d-3 \over 2}}(1+ \delta_{n_1,n_2} \log z)$.\footnote{The fact that $n_1 = n_2$ term acquires an extra $\log z$ is closely related to Gribov's theorem which constrains the possible leading Regge behavior of the scattering amplitude in the elastic region, see appendix \ref{sec:gribovstheorem}.}

A.~W.~Martin \cite{Martin:1969yw} found an elegant closed expression for $\rho_{n_1,n_2}(s,t(z) )$ in $d=4$. It takes the following form
\be
\label{eq:martInt}
\rho^{(d=4)}_{n_1,n_2}(s,t) &={d_{n_1, n_2 ;0} \over (z_1-1)^{n_1+n_2-1}} {(\delta z)^{n_1+n_2+{1 \over 2}}  \over \left( 1 +{2 \delta z \l_1^2 \over (\l_1^2 - 1)^2} \right)^{{1 \over 2}} } \nn \\
&\times \sum_{p,q,r=0}^{\infty} {\Gamma(n_1+n_2 +{3 \over 2}) \Gamma(n_2+2p+r+{1 \over 2}) \Gamma(n_1+2q+r+{1 \over 2}) \over p! q! r! \Gamma(n_1+{1 \over 2}) \Gamma(n_2+{1 \over 2}) \Gamma(n_1+n_2+p+q+r+{3 \over 2})} \\
&\qquad\times
\left({\delta z \over 4(z-1)z_1^2}\right)^{p+q}\left( {-z \delta z \over 2  (z-1) z_1^2 } \right)^r\ , \nn
\ee
where $\delta z\equiv z - (2 z_1^2 - 1)$. The advantage of the formula above is that by performing the sum over $r$ it makes the large $z$ limit of $\rho(s,t)$ manifest. 

Similarly, in $d=3$ the correction to the double spectral density takes the following form
\be
\label{eq:doubleSDb3d}
\rho^{(d=3)}_{n_1,n_2} &= {1 \over 8 \pi \sqrt{s} } \int\limits_{{\rm arccosh} \ z_1 }^{{\rm arccosh} \ z - {\rm arccosh} \ z_1 } d \theta_1 {(\cosh \theta_1 - z_1)^{n_1} \over (z_1-1)^{n_1} } {(\cosh ({\rm arccosh} \ z -  \theta_1) - z_1)^{n_2} \over (z_1-1)^{n_2} } \ ,
\ee
which can be explicitly evaluated for given $n_1$, $n_2$. 

The representations above are particularly useful if one would like to perform computations at finite $J$ and finite $s$, see section \ref{sec:boundinelasticity}. Indeed, plugging the threshold expansion formula in the Froissart-Gribov integral becomes dangerous at high enough order because the integral goes all the way to $z \to \infty$. Martin's formula (\ref{eq:martInt}) does not have this problem after performing the $r$ resummation. The same holds true for the $d=3$ result (\ref{eq:doubleSDb3d}). These formulae make both the threshold and the large $\delta z$ behavior of $\rho(s,t)$ manifest. We discuss the generalization of the formulae above to other dimensions below.

\subsubsection{Mandelstam Integral for $\rho(s,t)$}

In solving elastic unitarity (\ref{eq:doubleSDb}) within the threshold expansion we sometimes want to compute 
the following integral 
\be
\label{eq:doubleSDc}
J_{n_1,n_2}^{(d)} (s,t) &\equiv (z^2-1)^{{d-4 \over 2}} \int\limits_{z_1}^{\infty} {d \eta' \over \pi} \int\limits_{z_1}^{\infty} {d \eta'' \over \pi} (\eta' - z_1)^{n_1-{d-3 \over 2}} (\eta'' - z_1)^{n_2-{d-3 \over 2}} {\rm Disc}_z K(z, \eta', \eta'')\ .
\ee

Above we discussed the results in case of $d=3$ and $d=4$ as well as the threshold expansion in general $d$. Here we would like to note that in other dimensions the integral can be evaluated recursively by noting that
\be
\pa_{\delta z} J_{n_1,n_2}^{(d)} (\delta z) = -(d-5)\left( (1-z_1^2 - \delta z) J_{n_1-1,n_2-1}^{(d-2)} + J_{n_1,n_2}^{(d-2)} +z_1 (J_{n_1,n_2-1}^{(d-2)}+ J_{n_1-1,n_2}^{(d-2)}) \right)\ .
\ee
This recursion can be used in even $d$ and in odd $d \geq 7$. The case of $d=5$ can be treated explicitly similarly to the case of $d=3$ in the previous section.

\subsection{More General Integral}

Above we obtained the following result
\be
\label{eq:froissartgrB}
I(d,n,z_1) &= \int\limits_{z_1}^{\infty} d z (z^2-1)^{{d-4 \over 2}}  {Q_J^{(d)}(z) \over (z - z_1)^{{d-3 \over 2} - n}} = {\Gamma({d\over 2}-1) \Gamma(n+{5-d \over 2}) \over 2^{n-{d-5 \over 2}}\Gamma({3 \over 2} + n) }  (z_1^2-1)^{n+{1 \over 2}} Q_{J-n+{d-5 \over 2}}^{(2 n + 5)} (z_1) \ . 
\ee
Let us consider a slightly more general integral
\be
I(d,n,m,z_1) \equiv \int\limits_{z_1}^{\infty} d z (z^2-1)^{{d-4 \over 2}} Q_J^{(d)}(z) {1 \over (z - z_1)^{{d-3 \over 2} }} {(z-z_1)^n \over (z-1)^m}\ ,\qquad n,m \geq 0 \  .
\ee
The advantage of this integral is that for $m=n$ this a natural threshold expansion in terms of ${z-z_1 \over z-1}$ which does not grow for $z \to \infty$. For $m=0$ we get (\ref{eq:froissartgrB}).

We can rewrite this integral as follows
\be
{1 \over (z-1)^m} &= {1 \over (z-z_1 + [z_1-1])^m} \\
&= {1 \over \Gamma(m)} \int\limits_{- i \infty}^{i \infty} {d \alpha \over 2 \pi i}  \Gamma(m+\alpha) \Gamma(- \alpha) {(z_1-1)^{\alpha} \over (z - z_1)^{m+\alpha}} \ ,\qquad  -m < {\rm Re} \ \alpha < 0\ .\nn 
\ee

In this way we immediately get
\be
I(d,n,m,z_1) = {1 \over \Gamma(m)} \int\limits_{- i \infty}^{i \infty} {d \alpha \over 2 \pi i}  \Gamma(m+\alpha) \Gamma(- \alpha) {I(d,n-m-\alpha,z_1)  \over (z_1 - 1)^{m+\alpha}}\ ,
\ee
where the large $z$ convergence requires ${\rm Re} J > n-m- {\rm Re}\ \alpha -{d-3 \over 2}$. Next we can use the Mellin representation for the hypergeometric function
\be
\ _2 F_1 (a,b,c,z) = {\Gamma(c) \over \Gamma(a) \Gamma(b)} \int {d s \over 2 \pi i} {\Gamma(a+s) \Gamma(b+s) \Gamma(-s) \over \Gamma(c+s)} (-z)^s\ .
\ee
To use straight contour we would like to have ${\rm Re}[a,b,c]>0$. For the case above this becomes
\be
n-m < \alpha < 1 - m + n\ .
\ee
Therefore to apply the formula for $n=m$ we need to deform the contour across $\alpha = 0$ pole and pick the residue. 
We then perform the $\alpha$ integration. 

Let us for simplicity present the result for $d=4$ and $m=n$
\be
{I(4,n,n,z_1) \over \sqrt{2} \pi \lambda_1^{-J-{1 \over 2}}} =& {1 \over 2 J + 1}
- \sqrt{ {\l_1 - 1 \over \l_1 + 1} } \sum_{k=0}^{\infty} {1 \over (\l_1^2 -1)^{k}} {\Gamma(1+J) \Gamma(n+{1 \over 2}) \Gamma(k+{1 \over 2}) \over \Gamma({1 \over 2}-k) \Gamma(k+1)\Gamma(J+k+{3 \over 2}) \Gamma(n)}\\
&\qquad\qquad\qquad\qquad\qquad\times\ _3 F_2 (1+J,k+{1 \over 2}, n+{1 \over 2}; {3 \over 2}, {1 \over 2 }-k, {\l_1 - 1 \over \l_1 + 1}) \nn \\
&+{2 \over \sqrt{\pi}} {\l_1 - 1 \over \l_1 + 1} \sum_{k=0}^{\infty} \left( {4 \over (1+\l_1)^2} \right)^{2k} {\Gamma(k+n+1) \Gamma(k+{1 \over 2}) \over \Gamma(n) \Gamma(2k+3)}  \nn \\
&\qquad\qquad\qquad\qquad\qquad\times \ _3 F_2 ({3 \over 2}+J +k, 2 k+1, n+k+1; k+{3 \over 2}, k+2, {\l_1 - 1 \over \l_1 + 1})\ .\nn
\ee
For given $n$ this can be quite easily expanded at large $J$. Important property of this expansion is that higher terms in $k$ have an extra suppression in ${1 \over J}$. Using this formula one can in principle repeat the analysis of section \ref{LJboot} up to an arbitrary high order in ${z-z_1 \over z -1 }$ without spoiling the Regge behavior of the amplitude.

\section{Keyhole Integrals in Odd $d$}
\label{sec:keynoleint}

In this appendix we collect some of the useful integrals in odd $d$. They appear both in the large $J$ expansion of partial waves, and in the threshold expansion of the double
spectral density. A key difference compared to even $d$ is appearance of powers of both logarithm and inverse logarithm of the threshold expansion parameter $\sigma_t$.

\subsection{Partial Wave}

In the discussion of the large $J$ expansion we encountered the integral (\ref{eq:largeJF}), and in odd $d$ we introduced a function $g_n(J)$ in (\ref{fJlargeJ}) that controls the large $J$ asymptotic behavior of partial waves. Let us compute it explicitly.

For the universal threshold asymptotic in odd $d$, see (\ref{eq:asympf0}), the relevant integral takes the following form
\be
\label{eq:basisintegral}
g_n(\log J) ={1 \over 2 i} \oint\limits_{{\rm keyhole}} {d z \over z^n} {1 \over \log {z \over J} - i \pi} e^{- z}\ ,\qquad  n = 0,1, 2 , \dots  \ . 
\ee
In our problem $n = {d-3 \over 2}$. The keyhole contour is depicted in figure \ref{keyholefig} and it naturally appears when deriving the Froissart-Gribov formula.

For $n=0,1$ ($d=3$ and $d=5$) it is not necessary to keep the keyhole since the integral (\ref{eq:basisintegral}) converges and we can simply write
\be
\label{eq:zerooneg}
g_{n=0,1}(\log J) = \int\limits_0^\infty {d z \over z^n} {1 \over \log^2{z \over J} + \pi^2} e^{- z} \ .
\ee
To compute $n \geq 2$ it is convenient to slightly modify the integral and use the following recursion relation
\be
\label{eq:diffeqg}
g_n(\log J,\alpha) &= {1 \over 2 i} \oint\limits_{{\rm keyhole}}   {d z \over  z^n} { 1 \over \log {z \over J} - i \pi} e^{- \alpha z}\ , \nn \\
\pa_\alpha g_{n}(\log J,\alpha) &=- g_{n-1}(J,\alpha) \ ,
\ee
with the starting point given by $g_0(\log J,\alpha)$ which does not require regularization and can be efficiently computed numerically, see (\ref{eq:zerooneg}). We also note that 
\be
\lim_{J \to \infty} g_n (\log J, \alpha) = 0\ ,
\ee
which allows us to fix the integration constant in the differential equation (\ref{eq:diffeqg}). The original integral (\ref{eq:zerooneg}) is recovered by setting $\alpha = 1$.

Let us start with $n=0$. It is convenient to rewrite the $n=0$ integral as follows
\be
\label{eq:zeroresult}
g_0 (J) &= \Gamma(1 - \pa_{\log J}) {1 \over (\log J)^2 +  \pi^2}  = {1 \over \log^2 J } + \dots \  .
\ee
A slight advantage of writing $ \Gamma(1 - \pa_{\log J})$ is that to generate the large $J$ expansion we can treat $\pa_{\log J}$ in the argument of gamma-function as a small parameter.

To derive (\ref{eq:zeroresult}) we can write more generally
\be
\label{eq:laplaceG}
\Gamma(1 - \pa_{\log J})  g(\log J) &\equiv \int\limits_0^{\infty} d t e^{-t} t^{- \pa_{\log J}} g(\log J) = \int\limits_0^{\infty} d t e^{-t} g(\log {J \over t})=J {\cal L}[g(-\log t)](J)\ ,
\ee
where we used that $e^{-a \partial_{x}} g(x) = g(x-a)$.

By solving the recursion we then then get
\be
\label{eq:resoddF}
g_1 (J) &= \Gamma(1 - \pa_{\log J}) {i \over 2 \pi} \log {1-{i \pi \over \log J}  \over 1+{i \pi \over \log J}} = {1 \over \log J } + \dots\ , \nn \\
g_n(J) &={(-1)^{n-1} \over \Gamma(n) \log J} + \dots  \ .
\ee

More generally, we can write
\be
g_n(\log J) = \Gamma(1 - \pa_{\log J}) \hat g_n (\log J) = J {\cal L}[\hat g_n(-\log t)](J)\ ,
\ee
where a few of the $\hat g_n$'s were listed above in (\ref{eq:resoddF}). Note that (\ref{eq:resoddF}) can be easily computed numerically for finite $J$ using (\ref{eq:laplaceG}). This is an advantage compared to the original integral (\ref{eq:basisintegral}) which requires a keyhole regularization.

The leading large $J$ asymptotic of partial waves in odd $d$ therefore takes the form
\be
\hat f_J(s) &= 2^{5-d} n_{0}^{(d)} m^{4-d}  J^{{d-3 \over 2}} \left({z_1 -1 \over z_1 + 1} \right)^{{d-3 \over 4}} \pi^2 g_{{d-3 \over 2}} \left(\log J   \sqrt{{z_1 - 1 \over z_1 +1}} \right) \nn \\
&=2^{5-d} n_{0}^{(d)} m^{4-d}  J^{{d-3 \over 2}} \left({z_1 -1 \over z_1 + 1} \right)^{{d-3 \over 4}} \pi^2 J {\cal L}[\hat g_{{d-3 \over 2}}(-\log \sqrt{{z_1 + 1 \over z_1-1}}  \delta z)](J)\ .
\ee

\subsection{Double Spectral Density}
\label{sec:dsdOddD}

We next consider the problem of computing of $\rho(s,t)$ close to the threshold in odd $d$ for the universal threshold asymptotic (\ref{eq:asympf0}). The idea is to use the result of the previous subsection together with elastic unitarity.

Recall that due to elastic unitarity for $4 m^2 < s< 16 m^2$ we have, see (\ref{eq:largeJimF}) and (\ref{eq:rholarge}), 
\be
{\rm Im} \hat f_J(s) = 2^{-{d+5 \over 2}} J^{{1-d \over 2}} m^{d-4} \pi^{{1-d \over 2}} z_1^{{3-d \over 2}} \left( {z_1 - 1 \over z_1 + 1} \right)^{{5-d \over 4}} | \hat f_J(s)  |^2 \ .
\ee

Using the results of the previous subsection we can immediately read off the large $J$ expansion of ${\rm Im} \hat f_J(s)$. The latter is related to $\rho(s,t)$ as follows, see (\ref{eq:largeJimF}),
\be
\label{eq:rhointegral}
{\rm Im} \hat f_J(s) &= J {\cal L}[ \rho(s, t(\delta z))] (J) = J \int\limits_0^\infty d \delta z e^{- J \delta z} \rho(s,t(\delta z) )\ , \nn \\
t(\delta z, J) &= 8 m^2 \left( z_1+1 + z_1 \left( {z_1+1 \over z_1 - 1} \right)^{1/2} \delta z \right)\ ,
\ee
where the integral in the first line should be defined via the keyhole contour whenever it is divergent. For the universal threshold behavior in odd $d$ this happens for $d \geq 9$.

Let us limit ourselves to the situations when the integral (\ref{eq:rhointegral}) does not require the keyhole regularization, namely $d<9$. In this case the elastc unitarity takes the form
\be
{\cal L}[ \rho(s, t(\delta z) )] (J) = 2^{-{5 \over 2}(d-3)} m^{d-4} (n_0^{(d)})^2 \pi^{{9-d \over 2}} z_1^{{3-d \over 2}} \left( {z_1 - 1 \over z_1 + 1} \right)^{{d-1 \over 4}} J^{{d-3 \over 2}} \left( {\cal L}[\hat g_{{d-3 \over 2}}(-\log \sqrt{{z_1 + 1 \over z_1-1}}  \delta z)](J) \right)^2 \ .
\ee

Using the basic properties of the Laplace transform we can rewrite this as follows
\be
\label{eq:eqforrho}
 \rho(s, t(\delta z) ) &= 2^{-{5 \over 2}(d-3)} m^{d-4} (n_0^{(d)})^2 \pi^{{9-d \over 2}} z_1^{{3-d \over 2}} \left( {z_1 - 1 \over z_1 + 1} \right)^{{d-1 \over 4}} \nn \\
&(-1)^{{d-3 \over 2}}\pa_{\delta z}^{{d-3 \over 2}} \int\limits_0^{\delta z} d \delta z' \hat g_{{d-3 \over 2}} (-\log \sqrt{{z_1 + 1 \over z_1-1}}  \delta z') \hat g_{{d-3 \over 2}} (-\log \sqrt{{z_1 + 1 \over z_1-1}}  [\delta z - \delta z'])\ .
\ee
Together with the result of the previous subsection it allows us to compute the leading threshold contribution to the double spectral density.

Let us analyze in a little bit more detail the case of $d=3$. In this case we get
\be
\rho(s, t(\delta z) ) &=(n_{0}^{(d)})^2 m \pi^3 \left( {z_1 - 1 \over z_1 + 1} \right)^{{1 \over 2}}  \int\limits_0^{\delta z} d \delta z' {1 \over [ \log \left( {z_1 + 1 \over z_1 - 1} \right)^{{1 \over 2}} \delta z' ]^2 + \pi^2} {1 \over [ \log \left( {z_1 + 1 \over z_1 - 1} \right)^{{1 \over 2}} (\delta z - \delta z') ]^2 + \pi^2} \nn \\
 &=(n_{0}^{(d)})^2 m \pi^3  \left( {z_1 - 1 \over z_1 + 1} \right)^{{1 \over 2}}  \delta z \int\limits_0^{1} d x {1 \over [ \log \left( {z_1 + 1 \over z_1 - 1} \right)^{{1 \over 2}} \delta z x  ]^2 + \pi^2} {1 \over [ \log \left( {z_1 + 1 \over z_1 - 1} \right)^{{1 \over 2}} \delta z (1-x)  ]^2 + \pi^2} \nn \\
 &=(n_{0}^{(d)})^2 m \pi^3  \left( {z_1 - 1 \over z_1 + 1} \right)^{{1 \over 2}}  {\delta z \over \left( \log  \left( {z_1 + 1 \over z_1 - 1} \right)^{{1 \over 2}} \delta z  \right)^4 } \left( 1 + {4 \over  \log  \left( {z_1 + 1 \over z_1 - 1} \right)^{{1 \over 2}} \delta z} + {20 - {8 \over 3} \pi^2 \over  \log^2  \left( {z_1 + 1 \over z_1 - 1} \right)^{{1 \over 2}} \delta z} + \dots   \right)\ .
\ee
One can easily check that the leading order result agrees with the formulas in the main body of the paper.

\end{document}